\documentclass[12pt]{elsarticle}
\usepackage[utf8x]{inputenc}
\usepackage{graphicx,epsfig,color,booktabs}
\usepackage{amssymb,amsmath,a4wide,url}

\journal{Physics Reports}

\def\cL{{\cal L}}

\def\cF{{\cal F}}

\newcommand{\req}[1]{Eq.~(\ref{#1})}
\newcommand{\abs}[1]{\lvert #1\rvert}    % absolute value
\newcommand{\norm}[1]{\lVert#1\rVert}    % norm
\newcommand{\AM}{\mathsf{A}}             % adjacency matrix
\newcommand{\LM}{\mathsf{L}}             % Laplacian matrix
\newcommand{\PM}{\mathsf{P}}             % transition matrix
\newcommand{\EE}[1]{\mathrm{E}[#1]}      % expected value
\newcommand{\dd}{\mathrm{d}}             % differential
\newcommand{\eg}{\emph{e.g.}}            % e.g.
\newcommand{\ie}{\emph{i.e.}}            % i.e.
\newcommand{\mR}{\mathsf{R}}             % matrix with ratings
\newcommand{\AUC}{\mathrm{AUC}}

\newcommand{\fig}[1]{Fig.~\ref{#1}}

\newcommand{\vek}[1]{\boldsymbol{#1}}

\newcommand{\figg}[2]{\includegraphics[scale=#1]{#2}}\def\use{{i}}
\def\ite{{\alpha}}
\def\hid{{k}}
\def\data{{r}}
\def\dataM{{R}}
\def\toti{{M}}
\def\totu{{N}}

\def\toth{{K}}
\def\dira{{\boldsymbol a}}
\def\dirb{{\boldsymbol b}}
\def\diraa{{a}}
\def\dirbb{{b}}
\def\btheta{{\boldsymbol\theta}}
\def\bphi{{\boldsymbol\phi}}
\def\svdU{{W}}
\def\svdV{{V}}
\def\svdu{{w}}
\def\svdv{{v}}
\def\score{{r}}
\def\thetau{{\theta^{(\use)}}}
\def\thetauk{{\theta^{(\use)}_{\hid}}}
\def\phik{{\phi^{(\hid)}}}
\def\phiki{{\phi^{(\hid)}_{\ite}}}
\def\co{{\mu}}
\begin{document}

\begin{frontmatter}
\title{Recommender Systems}
\author[hz,fr,cd]{Linyuan L\"u}
\author[fr]{Mat\'u\v s Medo}
\author[fr,uk]{Chi Ho Yeung}
\author[hz,fr,cd]{Yi-Cheng Zhang\corref{cor1}}
\ead{yi-cheng.zhang@unifr.ch}
\author[hz,fr,cd]{Zi-Ke Zhang}
\author[hz,fr,cd,bj]{Tao Zhou}
\cortext[cor1]{Corresponding author}
\address[hz]{Institute of Information Economy, Hangzhou Normal University, Hangzhou, 310036, PR China}
\address[fr]{Department of Physics, University of Fribourg, Fribourg, CH-1700, Switzerland}
\address[cd]{Web Sciences Center, University of Electronic Science and Technology of China, Chengdu, 610054, PR China}
\address[uk]{The Nonlinearity and Complexity Research Group, Aston University, Birmingham B4 7ET, United Kingdom}
\address[bj]{Beijing Computational Science Research Center, Beijing, 100084, PR China}
\begin{abstract}
The ongoing rapid expansion of the Internet greatly increases the necessity of effective recommender systems for filtering the abundant information. Extensive research for recommender systems is conducted by a broad range of communities including social and computer scientists, physicists, and interdisciplinary researchers. Despite substantial theoretical and practical achievements, unification and comparison of different approaches are lacking, which impedes further advances. In this article, we review recent developments in recommender systems and discuss the major challenges. We compare and evaluate available algorithms and examine their roles in the future developments. In addition to algorithms, physical aspects are described to illustrate macroscopic behavior of recommender systems. Potential impacts and future directions are discussed. We emphasize that recommendation has a great scientific depth and combines diverse research fields which makes it of interests for physicists as well as interdisciplinary researchers.
\end{abstract}
\begin{keyword}
recommender systems, information filtering, networks
\end{keyword}
\end{frontmatter}

\newpage
\tableofcontents
\pagenumbering{arabic}
\newpage

\section{Introduction}
Thanks to computers and computer networks, our society is undergoing rapid transformation in almost all aspects. We buy online, gather information by search engines and live a significant part of our social life on the Internet. The fact that many of our actions and interactions are nowadays stored electronically gives researchers the opportunity to study socio-economical and techno-social systems at much better level of detail. Traditional ``soft sciences'', such as sociology or economics, have their fast-growing branches relying on the study of these newly available massive data sets \cite{Watts2007ZT,Vespignani2009ZT}. Physicists, with their long experience with data-driven research, have joined this trend and contributed to many fields such as finance \cite{ManSta00,BouPot09}, network theory \cite{Albert2002,Dorogovtsev2002,Newman2003,Boccaletti2006,Newman2006} and social dynamics \cite{Castellano2009} which are outside their traditional realm. The study of recommender systems and information filtering in general is no exception with the interest of physicists steadily increasing over the past decade. The task of recommender systems is to turn data on users and their preferences into predictions of users' possible future likes and interests. The study of recommender systems is at crossroads of science and socio-economic life and its huge potential was first noticed by web entrepreneurs in the forefront of the information revolution. While being originally a field dominated by computer scientists, recommendation calls for contributions from various directions and is now a topic of interest also for mathematicians, physicists, and psychologists. For instance, it is not a coincidence that an approach based on what psychologists know about human behavior scored high in a recent recommendation contest organized by the commercial company Netflix \cite{wired}.

When computing recommendations for a particular user, the very basic approach is to select the objects favored by other users that are similar to the target user. Even this simple approach can be realized in a multitude of ways---this is because the field of recommendation lacks general ``first principles'' from which one could deduce the right way to recommend. For example, how best to measure user similarity and assess its uncertainty? How to aggregate divergent opinions from various users? How to handle users for whom little information is available? Should all data be trusted equally or can one detect reckless or intentionally misleading opinions? These and similar issues arise also when methods more sophisticated than those based on user similarity are used. Fortunately, there exist a number of real data sets that can be used to measure and compare performance of individual methods. In consequence, similarly to physics, it is the experiment what decides which recommendation approach is good and which is not.

It would be very misleading to think that recommender systems are studied only because suitable data sets are available. While the availability of data is important for empirical evaluation of recommendation methods, the main driving force comes from practice: electronic systems give us too much choice to handle by ourselves. The interest from industry is hardly surprising---an early book on the nascent field of recommendation, \emph{Net Worth} by John Hagel III and Marc Singer \cite{Hagel3}, clearly pointed out the enormous economic impact of ``info-mediaries'' who can greatly enhance individual consumers' information capabilities. Most e-commerce web sites now offer various forms of recommendation---ranging from simply showing the most popular items or suggesting other products by the same producer to complicated data mining techniques. People soon realized that there is no unique best recommendation method. Rather, depending on the context and density of the available data, different methods adapting to particular applications are most likely to succeed. Hence there is no panacea, and the best one can do is to understand the underlying premises and recommender mechanisms, then one can tackle many diverse application problems from the real life examples. This is also reflected in this review where we do not try to highlight any ultimate approach to recommendation. Instead, we review the basic ideas, methods and tools with particular emphasis on physics-rooted approaches.

The motivation for writing this review is multifold. Firstly, while extensive reviews of recommender systems by computer scientists already exists \cite{Herlocker2004,Ricci2011,AdoTuz2005}, the view of physicists is different from that of computer scientists by using more the complex networks approach and adapting various classical physics processes (such as diffusion) for information filtering. We thus believe that this review with its structure and emphasis on respective topics can provide a novel point of view. Secondly, the past decade has already seen a growing interest of physicists in recommender systems and we hope that this review can be a useful source for them by describing the state of the art in language which is more familiar to the physics community. Finally, the interdisciplinary approach presented here might provide new insights and solutions for open problems and challenges in the active field of information filtering.

This review is organized as follows. To better motivate the problem, In Section~\ref{sec:applications} we begin with a discussion of real applications of recommender systems. Next, in Section~\ref{sec:definitions} we introduce basic concepts---such as complex networks, recommender systems, and metrics for their evaluation---that form a basis for all subsequent exposition. Then we proceed to description of recommendation algorithms where traditional approaches (similarity-based methods in Section~\ref{sec:sim-based} and dimensionality reduction techniques in Section~\ref{sec:dim_red}) are followed by network-based approaches which have their origin in the random walk process well known to all physicists (in Section~\ref{sec:diffusion}). Methods based on external information, such as social relationships (in Section~\ref{sec:social}), keywords or time stamps (in Section~\ref{sec:meta}), are also included. We conclude with a brief evaluation of methods' performance in Section~\ref{sec:performance} and a discussion on the outlook of the field in Section~\ref{sec:outlook}.

                     % written, could be extended
\section{Real Applications of Recommender Systems}
\label{sec:applications}
Thanks to the ever-decreasing costs of data storage and processing, recommender systems gradually spread to most areas of our lives. Sellers carefully watch our purchases to recommend us other goods and enhance their sales, social web sites analyze our contacts to help us connect with new friends and get hooked with the site, and online radio stations remember skipped songs to serve us better in the future (see more examples in Table~\ref{tab:services}). In general, whenever there is plenty of diverse products and customers are not alike, personalized recommendation may help to deliver the right content to the right person. This is particularly the case for those Internet-based companies that try to make use of the so-called long-tail \cite{Anderson2006} of goods which are rarely purchased yet due to their multitude they can yield considerable profits (sometimes they are referred to as ``worst-sellers''). For example on Amazon, between 20 to 40 percent of sales is due to products that do not belong to the shop's 100\,000 most saled products \cite{BryHuSmi03}. A recommender system may hence have significant impact on a company's revenues: for example, $60\%$ of DVDs rented by Netflix are selected based on personalized recommendations.\footnote{As presented by Jon Sanders (Recommendation Systems Engineering, Netflix) during the talk ``Research Challenges in Recommenders'' at the 3rd ACM Conference on Recommender Systems (2009).}

As discussed in~\cite{SchKoRi01}, recommender systems not only help decide which products to offer to an individual customer, they also increase cross-sell by suggesting additional products to the customers and improve consumer loyalty because consumers tend to return to the sites that best serve their needs (see \cite{CheWuYo04} for an empirical analysis of the impact of recommendations and consumer feedback on sales at Amazon.com).

Since no recommendation method serves best all customers, major sites are usually equipped with several distinct recommendation techniques ranging from simple popularity-based recommendations to sophisticated techniques many of which we shall encounter in the following sections. Further, new companies emerge (see, for example, string.com) which aim at collecting all sorts of user behavior (ranging from pages visited on the web and music listened on a personal player to ``liking'' or purchasing items) and using it to provide personalized recommendations of different goods or services.

\begin{table}
\centering
\begin{tabular}{ll}
\hline\hline
Site           & What is recommended\\
\hline
Amazon         & books/other products\\
Facebook       & friends\\
WeFollow       & friends\\ 
MovieLens      & movies\\
Nanocrowd      & movies\\
Jinni          & movies\\
Findory        & news\\
Digg           & news\\
Zite           & news\\
Meehive        & news\\   
Netflix        & DVDs\\
CDNOW          & CDs/DVDs\\
eHarmony       & dates\\
Chemistry      & dates\\
True.com       & dates\\
Perfectmatch   & dates\\
CareerBuilder  & jobs\\
Monster        & jobs\\
Pandora        & music\\
Mufin          & music\\ 
StumbleUpon    & web sites\\
\hline\hline
\end{tabular}
\caption{Popular sites using recommender systems. Besides, there are also some companies devoting themselves to recommendation techniques,  such as Baifendian (\textsf{www.baifendian.com}), Baynote (\textsf{www.baynote.com}), ChoiceStream (\textsf{www.choicestream.com}), Goodrec (\textsf{www.goodrec.com}), and others.}
\label{tab:services}
\end{table}

\subsection{Netflix Prize}
\label{sec:netflix_prize}
In October 2006, the online DVD rental company Netflix released a dataset containing approximately 100 million anonymous movie ratings and challenged researchers and practitioners to develop recommender systems that could beat the accuracy of the company's recommendation system, Cinematch \cite{BenLan07}. Atlhough the released data set represented only a small fraction of the company's rating data, thanks to its size and quality it fast became a standard in the data mining and machine learning community. The data set contained ratings in the integer scale from 1 to 5 which were accompanied by dates. For each movie, title and year of release were provided. No information about users was given. Submitted predictions were evaluated by their root mean squared error (RMSE) on a qualifying data set containing over 2,817,131 unknown ratings. Out of 20,000 registered teams, 2,000 teams submitted at least one answer set. On 21 September 2009, the grand prize of \$1,000,000 was awarded to a team that overperformed the Cinematch's accuracy by 10\%. At the time when the contest was closed, there were two teams that achieved the same precision. The prize was awarded to the team that submitted their results 20 minutes earlier than the other one. (See \cite{wired} for a popular account on how the participants struggled with the challenge.)

There are several lessons that we have learned in this competition \cite{BeKo07}. Firstly, the company gained publicity and a superior recommendation system that is supposed to improve user satisfaction. Secondly, ensemble methods showed their potential of improving accuracy of the predictions.\footnote{The ensemble methods deal with the selection and organization of many individual algorithms to achieve better prediction accuracy. In fact, the winning team, called BellKor's Pragmatic Chaos, was a combined team of BellKor \cite{Koren2009}, Pragmatic Theory \cite{Piotte2009} and BigChaos \cite{Toscher2009} (of course, it was not a simple combination but a sophisticated design), and each of them consists of many individual algorithms. For example, the Pragmatic Theory solution considered 453 individual algorithms.}
Thirdly, we saw that accuracy improvements are increasingly demanding when RMSE drops below a certain level. Finally, despite the company's effort, anonymity of its users was not sufficiently ensured \cite{NaraShm08}. As a result, Netflix was sued by one of its users and decided to cancel a planned second competition.

\subsection{Major challenges}
\label{sec:challenges}
Researchers in the field of recommender systems face several challenges which pose danger for the use and performance of their algorithms. Here we mention only the major ones:

\begin{enumerate}
\item \emph{Data sparsity.} Since the pool of available items is often exceedingly large (major online bookstores offer several millions of books, for example), overlap between two users is often very small or none. Further, even when the average number of evaluations per user/item are high, they are distributed among the users/items very unevenly (usually they follow a power-law distribution~\cite{Newman05}) and hence majority of users/items may have expressed/received only a few ratings. Hence, an effective recommender algorithm must take the data sparsity into account \cite{Huang2004ZT}.

\item \emph{Scalability.} While the data is mostly sparse, for major sites it includes millions of users and items. It is therefore essential to consider the computational cost issues and search for recommender algorithms that are either little demanding or easy to parallelize (or both). Another possible solution is based on using incremental versions of the algorithms where, as the data grows, recommendations are not recomputed globally (using the whole data) but incrementally (by slightly adjusting previous recommendations according to the newly arrived data) \cite{Sarwar2002ZT,Jin2009ZT}. This incremental approach is similar to perturbation techniques that are widely used in physics and mathematics \cite{Holmes95}.

\item \emph{Cold start.} When new users enter the system, there is usually insufficient information to produce recommendation for them. The usual solutions of this problem are based on using hybrid recommender techniques (see Section \ref{sec:hybrid}) combining content and collaborative data \cite{Schein02,Lam08} and sometimes they are accompanied by asking for some base information (such as age, location and preferred genres) from the users. Another way is to identify individual users in different web services. For example, Baifendian developped a technique that could track individual users' activities in several e-commerce sites, so that for a cold-start user in site $A$, we could make recommendation according to her records in sites $B$, $C$, $D$, etc.

\item \emph{Diversity vs. accuracy.} When the task is to recommend items which are likely to be appreciated by a particular user, it is usually most effective to recommend popular and highly rated items. Such recommendation, however, has very little value for the users because popular objects are easy to find (often they are even hard to avoid) without a recommender system. A good list of recommended items hence should contain also less obvious items that are unlikely to be reached by the users themselves \cite{McNee06}. Approaches to this problem include direct enhancement of the recommendation list's diversity \cite{Smyth01,Ziegler05,Hurley2011} and the use of hybrid recommendation methods \cite{Zhou_PNAS}.

\item \emph{Vulnerability to attacks.} Due to their importance in e-commerce applications, recommender systems are likely targets of malicious attacks trying to unjustly promote or inhibit some items \cite{MoBuBhWi07}. There is a wide scale of tools preventing this kind of behavior, ranging from blocking the malicious evaluations from entering the system to sophisticated resistant recommendation techniques \cite{Lam2006}. However, this is not a easy task since the strategies of attackers also get more and more advanced as the developing of preventing tools. As an example, Burke \emph{et al.} \cite{Burke2011} introduced eight attacking strategies, which are further divided into four classes: basic attack, low-acknowledge attack, nuke attack and informed attack. 

\item \emph{The value of time.} While real users have interests with widely diverse time scales (for example, short term interests related to a planned trip and long term interests related to the place of living or political preferences), most recommendation algorithms neglect the time stamps of evaluations. It is an ongoing line of research whether and how value of old opinions should decay with time and what are the typical temporary patterns in user evaluations and item relevance \cite{MinHan05,XiangL2010}.

\item \emph{Evaluation of recommendations.} While we have plenty of distinct metrics (see Section \ref{sec:metrics}), how to choose the ones best corresponding to the given situation and task is still an open question. Comparisons of different recommender algorithms are also problematic because different algorithms may simply solve different tasks. Finally, the overall user experience with a given recommendation system---which includes user's satisfaction with the recommendations and user's trust in the system---is difficult to measure in ``offline'' evaluation. Empirical user studies thus still represent a welcome source of feedback on recommender systems.

\item \emph{User interface.} It has been shown that to facilitate users' acceptance of recommendations, the recommendations need to be transparent \cite{Sinha2002,Cooke2002}: users appreciate when it is clear why a particular item has been recommended to them. Another issue is that since the list of potentially interesting items may be very long, it needs to be presented in a simple way and it should be easy to navigate through it to browse different recommendations which are often obtained by distinct approaches.
\end{enumerate}

Besides the above long-standing challenges, many novel issues appear recently. Thanks to the development of methodlogy in related branches of science, especially the new tools in network analysis, scientists started to consider the effecrs of network structure on recommendation and how to make use of known structural features to improve recommendation. For example, Huang \emph{et al.} \cite{Huang2007} analyzed the consumer-product networks and proposed an improved recommendation algorithm preferring edges that enhance the local clustering property, and Sahebi \emph{et al.} \cite{Sahebi} designed an improved algorithm making use of the community structure. Progress and propagation of new techniques also bring new challenges. For example, the GPS equipped mobile phones have become mainstream and the Internet access is ubiquitous, hence the location-based recommendation is now fesaible and increasingly significant.\footnote{Websites like Foursquare, Gowalla, Google Latitude, Facebook, Jiapang, and others already provide location-based services and show that many people want to share their location information and get location-based recommendations.}
Accurate recommendation asks for both the high predictability of human movements \cite{Song2010,Cho2011} and quantitative way to define similarities between locations and people \cite{Zheng2010,Clements2011}. Lastly, intelligent recommender systems should take into account the different behavioral patterns of different people. For example, new users tend to visit very popular items and select similar items, while old users usually have more specific interests \cite{Shang2010,ZhangCJ2012}, and users behave much differently between low-risk (e.g., collecting bookmarks, downloading music, etc.) and high-risk (e.g., buying a computer, renting a house, etc.) activities \cite{Vig2011,ChenLi2012}. 
              % checked, done
\section{Definitions of Subjects and Problems}
\label{sec:definitions}
We briefly review in this chapter basic concepts that are useful in the study of recommender systems.

\subsection{Networks}
Network analysis is a versatile tool in uncovering the organization principles of many complex systems \cite{Albert2002,Dorogovtsev2002,Newman2003,Boccaletti2006,Newman2006}. A network is a set of elements (called \emph{nodes} or \emph{vertices}) with connections (called \emph{edges} or \emph{links}) between them. Many social, biological, technological and information systems can be described as networks with nodes representing individuals or organizations and edges capturing their interactions. The study of networks, referred to as \emph{graph theory} in mathematical literature, has a long history that begins with the classical \emph{K\"onigsberg bridge problem} solved by Euler in 18th century \cite{Euler1736}. Mathematically speaking, a network $G$ is an ordered pair of disjoint sets $(V,E)$ where $V$ is the set of nodes and the set of edges, $E$, is a subset of $V\times V$ \cite{Bollobas1998}. In an \emph{undirected network}, an edge joining nodes $x$ and $y$ is denoted by $x\leftrightarrow y$, and $x\leftrightarrow y$ and $y\leftrightarrow x$ mean exactly the same edge. In a \emph{directed network}, edges are ordered pairs of nodes and an edge from $x$ to $y$ is denoted by $x\rightarrow y$. Edges $x\rightarrow y$ and $y\rightarrow x$ are distinct and may be present simultaneously. Unless stated otherwise, we assume that a network does not contain a \emph{self-loop} (an ``edge'' joining a node to itself) or a \emph{multi-edge} (several ``edges'' joining the same pair of nodes). In a \emph{multinetwork} both loops and multi-edges are allowed.

In an undirected network $G(V,E)$, two nodes $x$ and $y$ are said to be adjacent to each other if $x\leftrightarrow y \in E$. The set of nodes adjacent to a node $x$, the \emph{neighborhood} of $x$, is denoted by $\Gamma_x$. Degree of node $x$ is defined as $k_x=\lvert\Gamma_x\rvert$. The degree distribution, $P(k)$, is defined as the probability that a randomly selected node is of degree $k$. In a \emph{regular network}, every node has the same degree $k_0$ and thus $P(k)=\delta_{k,k_0}$. In the classical Erd\"os-R\'enyi random network \cite{Erdos1959} where each pair of nodes is connected by an edge with a given probability $p$, the degree distribution follows a binomial form \cite{Bollobas1985}
\begin{equation}
P(k)=\binom{N-1}{k} p^k(1-p)^{N-1-k},
\label{ER degree distribution 1}
\end{equation}
where $N=|V|$ is the number of nodes in the network. This distribution has a characterized scale represented by the average degree $\bar{k}=p(N-1)$. At the end of the last century, researchers turned to investigation of large-scale real networks where it turned out that their degree distributions often span several orders of magnitude and approximately follow a power-law form
\begin{equation}
P(k) \sim k^{-\gamma},
\end{equation}
with $\gamma$ being a positive exponent usually lying between 2 and 3 \cite{Albert2002}. Such networks are called \emph{scale-free networks} as they lack a characteristic scale of degree and the power-law function $P(k)$ is scale-invariant \cite{Caldarelli2007}. Note that detection of power-law distributions in empirical data requires solid statistical tools~\cite{Clauset2009,Goldstein2004TZ}. For a directed network, the out-degree of a node $x$, denoted by $k^{\mathrm{out}}$, is the number of edges starting at $x$, and the in-degree $k^{\mathrm{in}}$ is the number of edges ending at $x$. The in- and out-degree distribution of a directed network in general differ from each other.

\begin{figure}
\begin{center}
\centering
\figg{0.6}{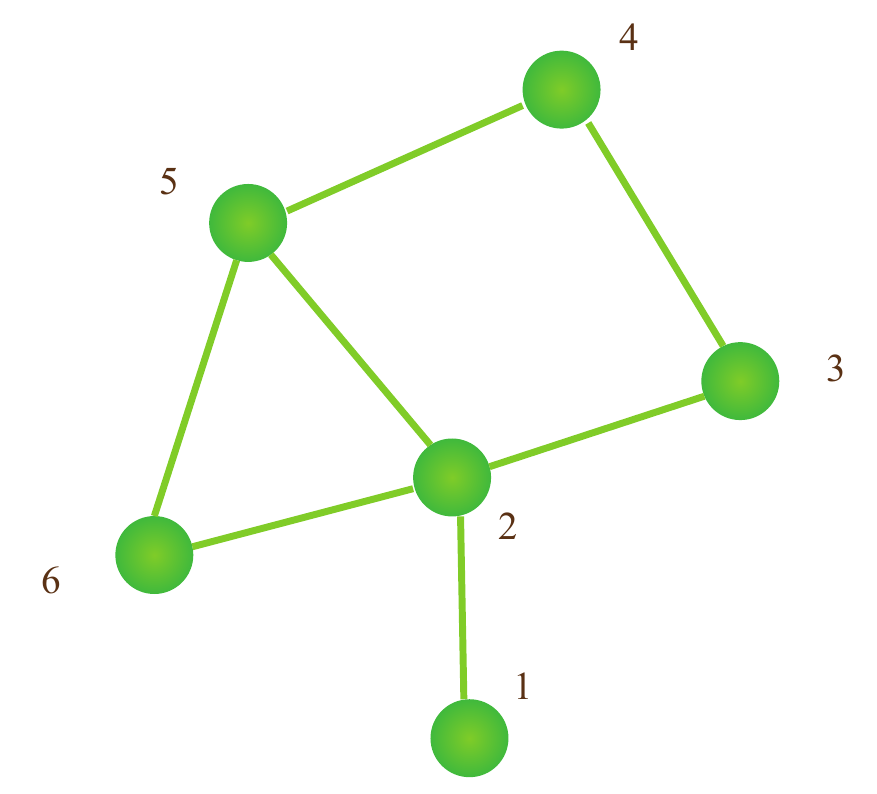}
\caption{A simple undirected network with 6 nodes and 7 edges. The node degrees are $k_1=1$, $k_2=k_5=3$ and $k_3=k_4=k_6=2$, corresponding to the distribution $P(1)=1/6$, $P(2)=1/2$ and $P(3)=1/3$. The diameter and average distance of this network are $d_{\text{max}}=3$ and $\bar{d}=1.6$, respectively. The clustering coefficients are $c_2=\frac{1}{6}$, $c_3=c_4=0$, $c_5=\frac{1}{3}$ and $c_6=1$, and the average clustering coefficient is $C=0.3$.}
\label{simple network illustration}
\end{center}
\end{figure}

Generally speaking, a network is said to be \emph{assortative} if its high-degree nodes tend to connect with high-degree nodes and the low-degree nodes tend to connect with low-degree nodes (it is said to be disassortative if the situation is opposite). This degree-degree correlation can be characterized by the average degree of the nearest neighbors \cite{Pastor-Satorras2001,Vazquez2002} or a variant of Pearson coefficient called \emph{assortativity coefficient} \cite{Newman2002,Newman2003b}. The assortativity coefficient $r$ lies in the range $-1\leq r \leq 1$. If $r>0$ the network is assortative; if $r<0$, the network is disassortative. Note that this coefficient is sensitive to degree heterogeneity. For example, $r$ will be negative in a network with very heterogeneous degree distribution (e.g., the Internet) regardless to the network's connecting patterns \cite{Zhou2007b}.

The number of edges in a path connecting two nodes is called length of the path, and \emph{distance} between two nodes is defined as the length of the shortest path that connects them. The \emph{diameter} of a network is the maximal distance among all node pairs and the \emph{average distance} is the mean distance averaged over all node pairs as
\begin{equation}
\bar{d}=\frac{1}{N(N-1)}\sum_{x\neq y}d_{xy},
\label{average distance}
\end{equation}
where $d_{xy}$ is the distance between $x$ and $y$.\footnote{When no path exists between two nodes, we say that their distance is infinite which makes the average distance automatically infinite too. This problem can be avoided either by excluding such node pairs from averaging or by using the harmonic mean \cite{Newman2003,Latora2001TZ}.}
Many real networks display a so-called \emph{small-world} phenomenon: their average distance does not grow faster than the logarithm of the network size \cite{Milgram1967,Watts1998}.

The importance of triadic clustering in social interaction systems has been realized for more than 100 years \cite{Simmel1908}. In social network analysis \cite{Wasserman1994}, this kind of clustering is called \emph{transitivity}, defined as three times the ratio of the total number of triangles in a network to the total number of connected node triples. In 1998, Watts and Strogatz \cite{Watts1998} proposed a similar index to quantify the triadic clustering, called \emph{clustering coefficient}. For a given node $x$, this coefficient is defined as the ratio of the number of existing edges between $x$'s neighbors to the number of neighbor pairs,
\begin{equation}
c_x=\frac{e_x}{\tfrac12 k_x(k_x-1)}
\label{node clustering coefficient}
\end{equation}
where $e_x$ denotes the number of edges between $k_x$ neighbors of node $x$ (this definition is meaningful only if $k_x>1$). The \emph{network clustering coefficient} is defined as the average of $c_x$ over all $x$ with $k_x>1$. It is also possible to define the clustering coefficient as the ratio of 3~$\times$~number of triangles in the network to the number of connected triples of vertices, which is sometimes referred to as ``fraction of transitive triples'' \cite{Newman2003}. Note that the two definitions can give substantially different results.

Figure \ref{simple network illustration} illustrates the above definitions for a simple undirected network. For more information about network measurements, readers are encouraged to refer an extensive review article \cite{Costa2007} on characterization of networks.

               % checked, done
\subsection{Bipartite Networks and Hypergraphs}
\label{sec:bipartite}
A network $G(V,E)$ is a \emph{bipartite network} if there exists a partition $(V_1,V_2)$ such that $V_1\cup V_2=V$, $V_1\cap V_2=\emptyset$, and every edge connects a node of $V_1$ and a node of $V_2$. Many real systems are naturally modeled as bipartite networks: the metabolic network \cite{Jeong2000} consists of chemical substances and chemical reactions, the collaboration network \cite{Newman2001} consists of acts and actors, the Internet telephone network consists of personal computers and phone numbers \cite{Xuan2009}, etc. We focus on a particular class of bipartite networks, called \emph{web-based user-object networks} \cite{Shang2010}, which represent interactions between users and objects in online service sites, such as collections of bookmarks in \emph{delicious.com} and purchases of books in \emph{amazon.com}. As we shall see later, these networks describe the fundamental structure of recommender systems. Web-based user-object networks are specific by their gradual evolution where both nodes and links are added gradually. By contrast, this cannot happen in, for example, act-actor networks (e.g., one can not add authors to a scientific paper after its publication).

\begin{figure}
\centering
\figg{0.6}{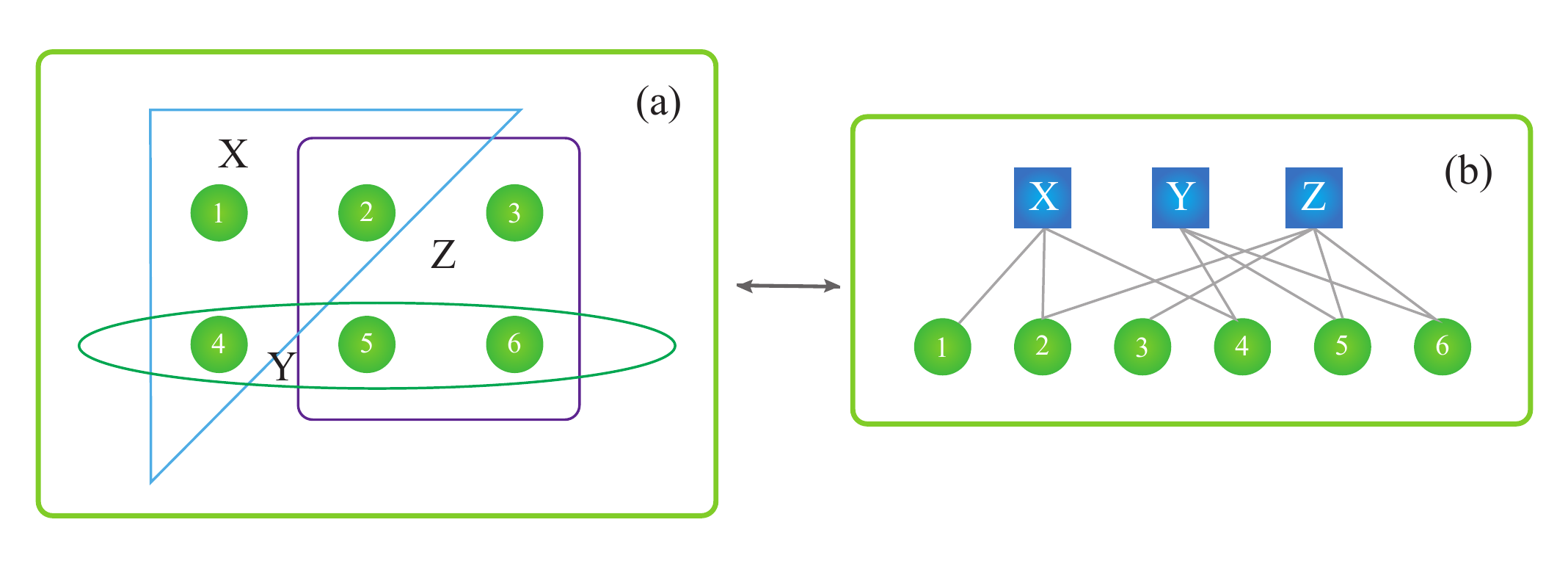}
\caption{An illustration of the one-to-one correspondence between a hypergraph (a) and a bipartite network (b). There are three hyperedges,
$X=\{1,2,4\}$, $Y=\{4,5,6\}$ and $Z=\{2,3,5,6\}$.}
\label{bipartite-hypergraph}
\end{figure}

Most web-based user-object networks share some structural properties. Their object-degree distributions obey a power-law-like form $P(k)\sim k^{-\gamma}$, with $\gamma\approx 1.6$ for the \emph{Internet Movie Database} (IMDb) \cite{Grujic2008}, $\gamma\approx 1.8$ for the music-sharing site \emph{audioscrobbler.com} \cite{Lambiotte2005}, $\gamma\approx 2.3$ for the e-commerce site \emph{amazon.com} \cite{Shang2010}, and $\gamma\approx 2.5$ for the bookmark-sharing site \emph{delicious.com} \cite{Shang2010}. The form of the user-degree distribution is usually between an exponential and a power law \cite{Shang2010}, and can be well fitted by the Weibull distribution
\cite{Laherrere1998}
\begin{equation}
P(k)\sim k^{\mu-1}\exp\big[-(k/k_0)^\mu\big]
\label{SED}
\end{equation}
where $k_0$ is a constant and $\mu$ is the stretching exponent. Connections between users and objects exhibit a disassortative mixing pattern \cite{Grujic2008,Shang2010}.

A straightforward extension of the definition of bipartite network
is the so-called \emph{multipartite network}. For an $r$-partite
network $G(V,E)$, there is an $r$-partition $V_1,V_2,\cdots,V_r$
such that $V=V_1\cup V_2\cup \cdots \cup V_r$, $V_i\cap
V_j=\emptyset$ whenever $i\neq j$, and no edge joins two nodes in the same set $V_i$ for all $1\leq i \leq r$. The tripartite network representation has found its application in \emph{collaborative tagging systems} (also called \emph{folksonomies} in the literature) \cite{Lambiotte2006,Cattuto2007,Palla2008,ZhaZK2011}, where users assign tags to online resources, such as photographs in \emph{flickr.com}, references in \emph{CiteULike.com} and bookmarks in \emph{delicious.com}.

\begin{figure}
\centering
\figg{0.8}{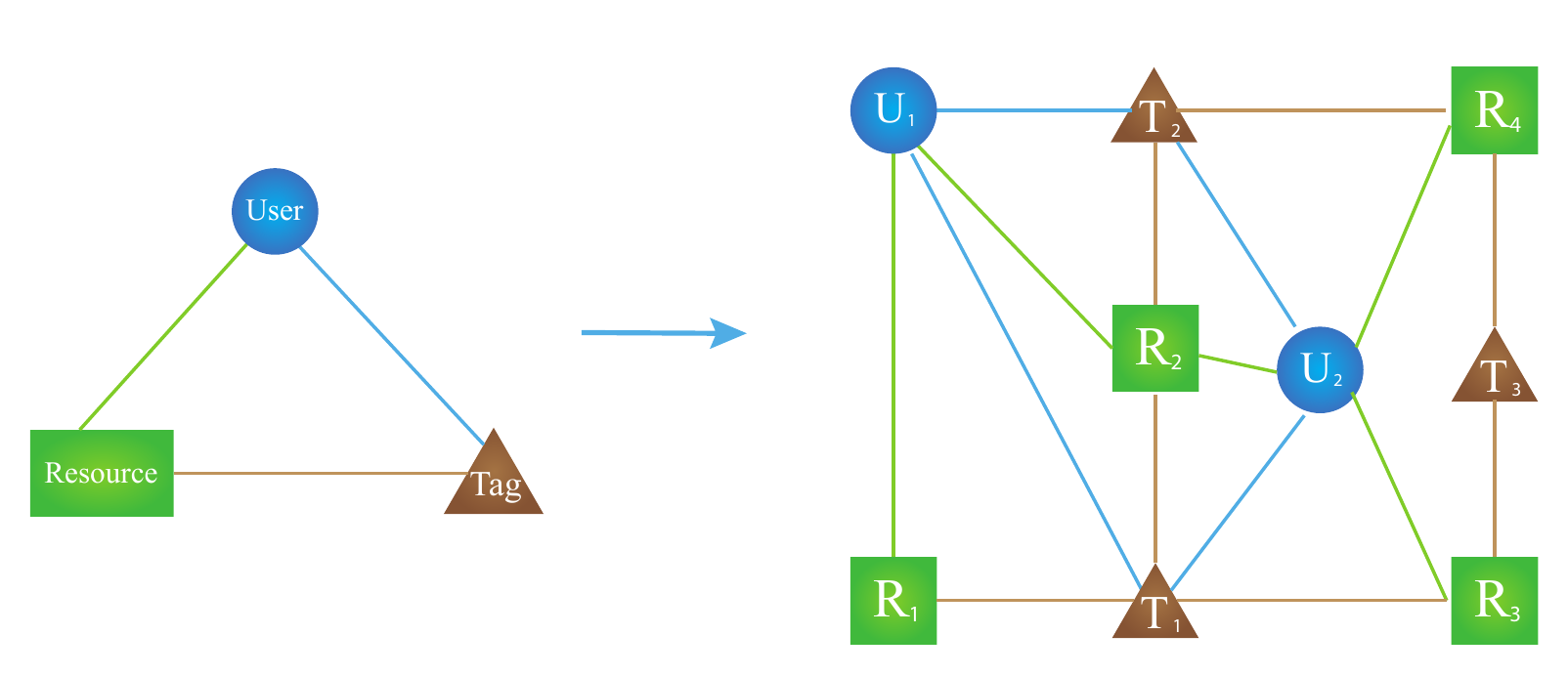}
\caption{ A hypergraph illustration of collaborative tagging networks. (left) A triangle-like hyperedge \cite{Ghoshal2009}, which contains three types of vertices, depicted by one red circle, one green rectangle and one blue triangle which respectively represent a user, a resource and a tag. (right) A descriptive hypergraph consists of two users, four resources and three tags. Take user $U_2$ and resource $R_1$ for example, the measurements are denoted as: (i) $U_2$ has participated in six hyperedges, which means its hyperdegree is 6; (ii) $U_2$ has directly connected to three resources and three tags. As defined by Eq.~(\ref{hyperedge_diversity}), it  suggests it possibly has 3$\times$3=9 hyperedges in maximal. Thus its clustering coefficient equals 6/9$\approx$0.667 , where 6 is its hyperdegree; Comparatively, as defined by Eq.~(\ref{hyperedge density}), its clustering coefficient $D_h(U_2)=\frac{12-6}{12-4}$=0.75; (iii) the shortest path from $U_2$ to $R_1$ is $U_2-T_1-R_1$, which indicates the distance between $U_2$ and $R_1$
is 2. }
\label{hypergraph}
\end{figure}

Note that some information is lost in the tripartite representation. For example, given an edge connecting a resource and a tag, we do not know which user (or users) contributed to this edge. To resolve this, \emph{hypergraph} \cite{Berge1973} can be used to give an exact representation of the full structure of a collaborative tagging system. In a hypergraph $H(V,E)$, the \emph{hyperedge} set $E$ is a subset of the power set of $V$, that is the set of all subsets of $V$. Link $e$ can therefore connect multiple nodes. Analogously to ordinary networks, node degree in a hypergraph is defined as the number of hyperedges adjacent to a node and the distance between two nodes is defined as the minimal number of hyperedges connecting these nodes. The clustering coefficient \cite{Cattuto2007,Zlatic2009} and community structure \cite{Zlatic2009,Vazquez2009} can also be defined and quantified following the definitions in ordinary networks. Notice that there is a one-to-one correspondence between a hypergraph and a bipartite network. Given a hypergraph $H(V,E)$, the corresponding bipartite network $G(V',E')$ contains two node sets, as $V'=V\cup E$, and $x\in V$ is connected with $Y\in E$ if and only if $x\in Y$ (see Figure \ref{bipartite-hypergraph} for an illustration).

Hypergraph representation has already found applications in ferromagnetic dynamics \cite{Bolle2006,Bolle2008}, population stratification \cite{Vazquez2008}, cellular networks \cite{Klamt2009}, academic team formation \cite{Taramasco2010}, and many other areas. Here we are concerned more about the hypergraph representation of collaborative tagging systems \cite{Zlatic2009,Ghoshal2009,Zhang2010} where each hyperedge joins three nodes (represented by a triangle-like in Figure~\ref{hypergraph}), user $u$, resource $r$ and tag $t$, indicating that $u$ has given $t$ to $r$. A resource can be collected by many users and given several tags by a user, and a tag can be associated with many resources, resulting in small-world hypergraphs \cite{Zlatic2009,Zhang2010} (Figure~\ref{hypergraph} shows both the basic unit and extensive description). Moreover, hypergraphs for collaborative tagging systems have been shown to be highly clustered, with heavy-tailed degree distribution\footnote{The degrees of users, resources and tags are usually investigated separately. For \emph{flickr.com} and \emph{CiteULike.com}, the user and tag degree distributions are power-law-like, while the resource degree distributions are much narrower because in \emph{flickr.com}, a photograph is only collected by a single user and in \emph{CiteULike.com}, a reference is rarely collected by many users \cite{Zlatic2009}. By contrast, in \emph{delicious.com}, a popular bookmark can be collected by thousands of users and thus the resource degree distribution is of a power-law kind \cite{Zhang2010}.} and of community structure \cite{Zlatic2009,Zhang2010}. A model for evolving hypergraphs can be found in \cite{Zhang2010}.

Generally, to evaluate a hypergraph from the perspective of complexity science, the following quantities (Figure~\ref{hypergraph}
gives a detailed description of these quantities) can be applied:

(i) \emph{hyperdegree}:  The degree of a node in a hypergraph can be naturally defined as the number of hyperedges adjacent to it.

(ii) \emph{hyperdegree distribution}: defined as the proportion that each hyperdegree occupies, where hyperdegree is defined as the number of hyperedges that a regular node participates in.

(iii) \emph{clustering coefficients}: defined as the proportion of real number of hyperedges to all the possible number of hyperedges that a
 regular node could have \cite{Cattuto2007}. e.g., the clustering coefficient for a user, $C_u$, is defined as
\begin{equation}
C_u = \frac{k_u}{R_u T_u},\label{hyperedge_diversity}
\end{equation}
where $k_u$ is the hyperdegree of user $u$, $R_u$ is the number of resources that $u$ collects and $T_u$ is the number of tags that $u$
possesses. The above definition measures the fraction of possible pairs present in the neighborhood of $u$. A larger $C_u$ indicates
that $u$ has more similar topic of resources, which might also show that $u$ has more concentrated on personalized or special topics,
while smaller $C_u$ might suggest that s/he has more diverse interests. Similar definitions can also be defined for measuring the clustering coefficient of resources and tags.

An alternative metric, named hyperedge density, is proposed by Zlati\'{c} \emph{et al} \cite{Zlatic2009}. Taking a user node $u$ again as an
example, they define the coordination number of $u$ as $z(u)=R_u+T_u$. Given $k(u)$, the maximal coordination number is $z_\texttt{max}(u)=2k(u)$, while the minimal coordination number is $z_\texttt{min}(u)=2n$ for $n(n-1)<k(u)\leq n^2$ and $z_\texttt{min}(u)=2n+1$ for $n^2<k(u)\leq n(n+1)$, with $n$ some integer. Obviously, a local tree structure leads to maximal coordination number, while the maximum overlap corresponds to the minimal coordination number. Therefore, they define the hyperedge density as \cite{Zlatic2009}:
\begin{equation}
D_h(u)=\frac{z_\texttt{max}(u)-z(u)}{z_\texttt{max}(u)-z_\texttt{min}(u)},\texttt{
} 0\leq D_h(u) \leq 1. \label{hyperedge density}
\end{equation}
The definition of hyperedge density for resources and tags is similar. Empirical analysis indicates a high clustering behavior under both metrics \cite{Cattuto2007,Zlatic2009}. The study of hypregraph for the collaborative tagging networks has just been unfolding, and how to properly quantify the clustering behavior, the correlations and similarities between nodes, and the community structure is still an open problem.

(iv) \emph{average distance}: defined as the average shortest path length between two random nodes in the whole network.

             % checked, done
\subsection{Recommender Systems}
\label{sec:rec_intro}

\begin{figure}
\centering
\figg{0.8}{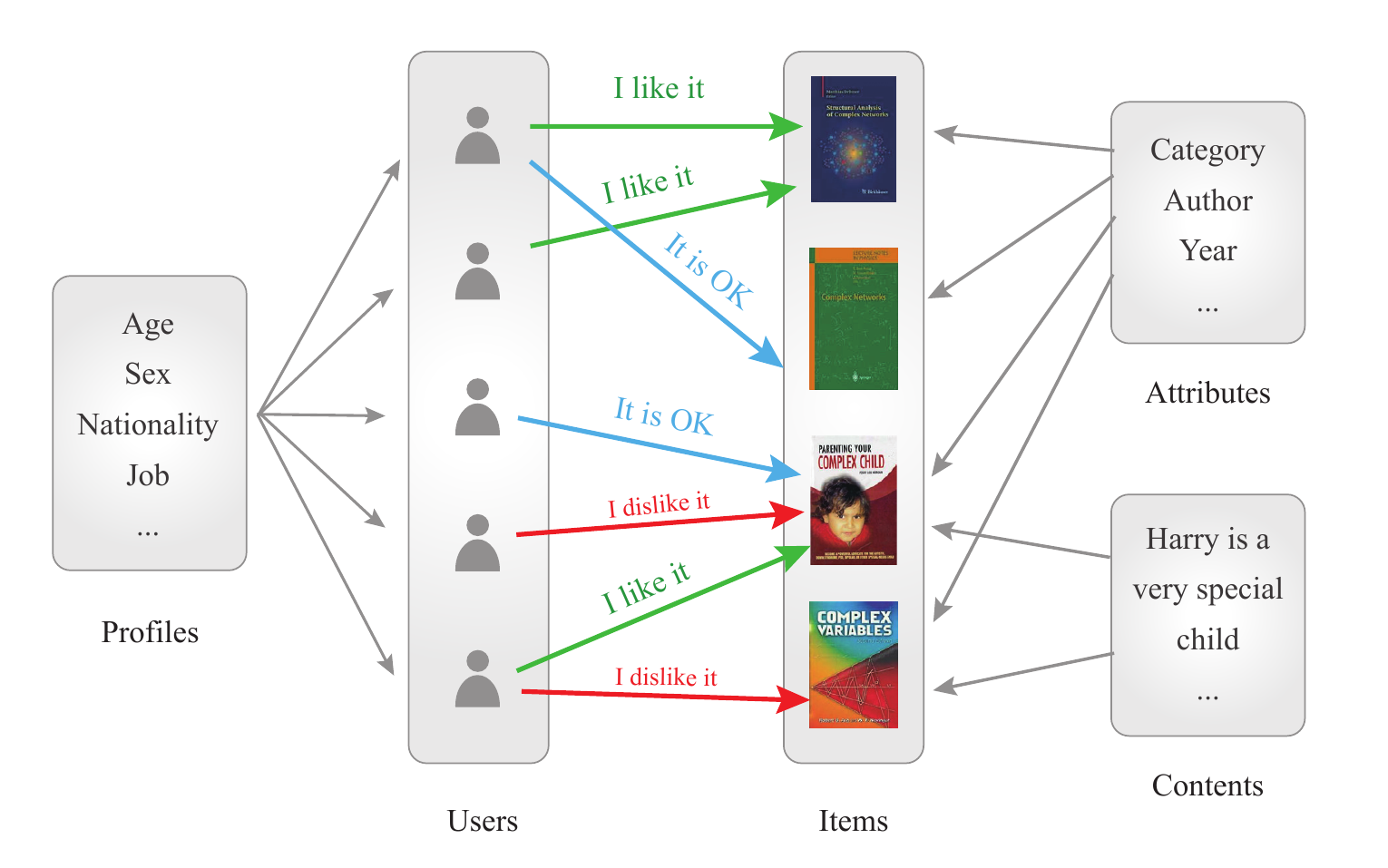}
\caption{Illustration of a recommender system consisted of five users and four books. The basic information contained by every recommender system is the relations between users and objects that can be represented by a bipartite graph. This illustration also exhibits some additional information frequently exploited in the design of recommendation algorithms, including user profiles, object attributes and object content.}
\label{fig:RS}
\end{figure}

A recommender system uses the input data to predict potential further likes and interests of its users. Users' past evaluations are typically an important part of the input data. Let $M$ be the number of users and let $N$ be the number of all objects that can be evaluated and recommended. Note that \emph{object} is simply as a generic term which can represent books, movies, or any other kind of consumed content. To stay in line with standard terminology, we sometimes use \emph{item} which has the same meaning. To make the notation more clear, we restrict to Latin indices $i$ and $j$ when enumerating the users and to Greek indices $\alpha$ and $\beta$ when enumerating the objects. Evaluation/rating of object $\alpha$ by user $i$ is denoted as $r_{i\alpha}$. This evaluation is often numerical in an integer rating scale (think of Amazon's five stars)---in this case we speak of explicit ratings. Note that the common case of binary ratings (like/dislike or good/bad) also belongs to this category. When objects are only collected (as in bookmark sharing systems) or simply consumed (as in online newspaper or magazine without rating systems) or when ``like'' is the only possible expression (as on Facebook), we are left with unary ratings. In this case, $r_{i\alpha}=1$ represents a collected/consumed/liked object and $r_{i\alpha}=0$ represents a non-existing evaluation (See Fig. \ref{fig:RS}). Inferring users' confidence levels of ratings is not a trival task, especially from the binary or unary ratings. Accessorial information about users' behavior may be helpful, for example, the users' confidence levels can be estimated by their watching time of television shows and with the help of this information, the quality of recommendation can be improved \cite{HuKoren2008}. Even if we have explict ratings, it does not mean we know how and why people vote with these ratings--Do they have standards of numerical ratings or they just use ratings to present orders? Recent evidence \cite{KorenSill2011} to some extent supports the latter ansatz.  

\begin{table}
\centering
\begin{tabular}{rccc}
                      & Alice & Bob & Carol\\
\hline
              Titanic &   5   &  1  &   5\\
2001: A Space Odyssey &   1   &  5  &   2\\
           Casablanca &   4   &  2  &   ?\\
\hline
\end{tabular}
\caption{Recommendation process in a nutshell: to estimate the potential favorable opinion of Carol about Casablanca, one can use the similarity of her with those of Alice. Alternatively, one can note that ratings of Titanic and Casablanca follow a similar pattern, suggesting that people who liked the former might also like the latter.}
\label{tab:rec_sys}
\end{table}

The goal of a recommender system is to deliver lists of personalized ``recommended'' objects to its users. To this end, evaluations can be predicted or, alternatively, recommendation scores can be assigned to objects yet unknown to a given user. Objects with the highest predicted ratings or the highest recommendation scores then constitue the recommendation list that is presented to the target user. There is an extensive set of performance metrics that can be used to evaluate the resulting recommendation lists (see Sec.~\ref{sec:metrics}). The usual classifications of recommender systems is as follows \cite{AdoTuz2005}:
\begin{enumerate}
\item\emph{Content-based recommendations:} Recommended objects are those with content similar to the content of previously preferred objects of a target user. We present them in Sec.~\ref{sec:ext_sim}.

\item\emph{Collaborative recommendations:} Recommended objects are selected on the basis of past evaluations of a large group of users. They can be divided into:

\begin{enumerate}
\item\emph{Memory-based collaborative filtering:} Recommended objects are those that were preferred by users who share similar preferences as the target user, or, those that are similar to the other objects preferred by the target user. We present them in Sec.~4 (Standard similarity-based methods) and Sec.~\ref{sec:social} (methods employing social filtering).

\item\emph{Model-based collaborative filtering:} Recommended objects are selected on models that are trained to identify patterns in the input data. We present them in Sections \ref{sec:dim_red} (dimensionality reduction methods) and \ref{sec:diffusion} (diffusion-based methods).
\end{enumerate}

\item\emph{Hybrid approaches:} These methods combine collaborative with content-based methods or with different variants of other collaborative methods. We present them in Sec.~\ref{sec:hybrid}.
\end{enumerate}

      % checked, done
\subsection{Evaluation Metrics for Recommendation}
\label{sec:metrics}
Given a target user $i$, a recommender system will sort all $i$'s uncollected objects and recommend the top-ranked objects. To evaluate recommendation algorithms, the data is usually divided into two parts: The training set $E^T$ and the probe set $E^P$. The training set is treated as known information, while no information from the probe set is allowed to be used for recommendation. In this section we briefly review basic metrics that are used to measure the quality of recommendations. How to choose a particular metric (or metrics) to evaluate recommendation performance depends on the goals that the system is supposed to fulfill. Of course, the ultimate evaluation of any recommender system is given by the judgement of its users.

\subsubsection{Accuracy Metrics}

\paragraph{Rating Accuracy Metrics}
The main purpose of recommender systems is to predict users' future likes and interests. A multitude of metrics exist to measure various aspects of recommendation performance. Two notable metrics, \emph{Mean Absolute Error} (MAE) and \emph{Root Mean Squared Error} (RMSE), are used to measure the closeness of predicted ratings to the true ratings. If $r_{u\alpha}$ is the true rating on object $\alpha$ by user $i$, $\tilde{r}_{i\alpha}$ is the predicted rating and $E^P$ is the set of hidden user-object ratings, MAE and RMSE are defined as
\begin{align}
\label{mae}
\mathrm{MAE}&=\frac1{\abs{E^P}}\sum_{(i,\alpha)\in{E^{P}}}
\abs{r_{i\alpha}-\tilde{r}_{i\alpha}},\\
\label{rmse}
\mathrm{RMSE}&=\Big(\frac1{\abs{E^P}}\sum_{(i,\alpha)\in{E^{P}}}
(r_{i\alpha}-\tilde{r}_{i\alpha})^2\Big)^{1/2}.
\end{align}
Lower MAE and RMSE correspond to higher prediction accuracy. Since RMSE squares the error before summing it, it tends to penalize large errors more heavily. As these metrics treat all ratings equally no matter what their positions are in the recommendation list, they are not optimal for some common tasks such as finding a small number of objects that are likely to be appreciated by a given user (\emph{Finding Good Objects}). Yet, due to their simplicity, RMSE and MAE are widely used in the evaluation of recommender systems.

\paragraph{Rating and Ranking Correlations}
Another way to evaluate the prediction accuracy is to calculate the correlation between the predicted and the true ratings. There are three well-known correlation measures, namely the Pearson product-moment correlation \cite{Rodgers1988}, the Spearman \cite{Spearman1904} correlation and Kendall's Tau \cite{Kendall1938}. The Pearson correlation measures the extent to which a linear relationship is present between the two sets of ratings. It is defined as
\begin{equation}
PCC=\frac{\sum_{\alpha} (\tilde{r}_{\alpha}-\bar{\tilde{r}})(r_{\alpha}-\bar{r})}
{\sqrt{\sum_{\alpha}(\tilde{r}_{\alpha}-\bar{\tilde{r}})^2}\sqrt{\sum_{\alpha}(r_{\alpha}-\bar{r})^2}},
\end{equation}
where $r_{\alpha}$ and $\tilde{r}_{\alpha}$ are the true and predicted ratings, respectively. The Spearman correlation coeffcient $\rho$ is defined in the same manner as the Pearson correlation, except that $r_{\alpha}$ and $\tilde{r}_{\alpha}$ are replaced by the ranks of the respective objects. Similarly to the Spearman correlation, Kendall's Tau also measures the extent to which the two rankings agree on the exact values of ratings. It is defined as $\tau=(C-D)/(C+D)$ where $C$ is the number of concordant pairs---pairs of objects that the system predicts in the correct ranked order and $D$ is the number of discordant pairs---pairs that the system predicts in the wrong order. $\tau=1$ when the true and predicted ranking are identical and $\tau=-1$ when they are exactly opposite. For the case when objects with equal true or predicted ratings exist, a variation of Kendall's Tau was proposed in \cite{Herlocker2004}
\begin{equation}
\tau\approx\frac{C-D}{\sqrt{(C+D+S_T)(C+D+S_P)}},
\end{equation}
where $S_T$ is the number of object pairs for which the true ratings are the same, and $S_P$ is the number of object pairs for which the predicted ratings are the same. Kendall's Tau metric applies equal weight to any interchange of successively ranked objects, no matter where it occurs. However, interchanges at different places, for example between top 1 and 2, and between 100 and 101, may have different impacts. Thus a possible improved metric could give more weight to object pairs at the top of the true ranking.

Similar to Kendall's Tau, the normalized distance-based performance measure (NDPM) was originally proposed by Yao \cite{Yao1995} to compare two different weakly ordered rankings. It is based on counting the number of contradictory pairs $C^-$ (for which the two rankings disagree) and compatible pairs $C^u$ (for which one ranking reports a tie while the other reports strict preference of one object over the other). Denoting the total number of strict preference relationships in the true ranking as $C$, NDPM is defined as
\begin{equation}
NDPM=\frac{2C^-+C^u}{2C}.
\end{equation}
Since this metric does not punish the situation where the true ranks are tied, it is more appropriate than correlation metrics for domains where users are interested in objects that are good-enough.

\paragraph{Classification Accuracy Metrics}
Classification metrics are appropriate for tasks such as ``Finding Good Objects", especially when only implicit ratings are available (\ie, we know which objects were favored by a user but not how much they were favored). When a ranked list of objects is given, the threshold for recommendations is ambiguous or variable. To evaluate this kind of systems, one popular metric is AUC (Area Under ROC Curve), where ROC stands for the receiver operating characteristic \cite{Hanely1982} (for how to draw a ROC curve see \cite{Herlocker2004}). AUC attempts to measure how a recommender system can successfully distinguish the relevant objects (those appreciated by a user) from the irrelevant objects (all the others). The simplest way to calculate AUC is by comparing the probability that the relevant objects will be recommended with that of the irrelevant objects. For $n$ independent comparisons (each comparison refers to choosing one relevant and one irrelevant object), if there are $n'$ times when the relevant object has higher score than the irrelevant and $n''$ times when the scores are equal, then according to \cite{Zhou2009}
\begin{equation}
\AUC=\frac{n'+0.5n''}{n}.
\end{equation}
Clearly, if all relevant objects have higher score than irrelevant objects, $\AUC=1$ which means a perfect recommendation list. For a randomly ranked recommendation list, $\AUC=0.5$. Therefore, the degree of which AUC exceeds 0.5 indicates the ability of a recommendation algorithm to identify relevant objects. Similar to AUC is a so-called \emph{Ranking Score} proposed in \cite{Zhou2007}. For a given user, we measure the relative ranking of a relevant object in this user's recommendation list: when there are $o$ objects to be recommended, a relevant object with ranking $r$ has the relative ranking $r/o$. By averaging over all users and their relevant objects, we obtain the mean ranking score $RS$---the smaller the ranking score, the higher the algorithm's accuracy, and vice versa.

Since real users are usually concerned only with the top part of the recommendation list, a more practical approach is to consider the number of a user's relevant objects ranked in the top-$L$ places. Precision and recall are the most popular metrics based on this. For a target user $i$, precision and recall of recommendation, $P_i(L)$ and $R_i(L)$, are defined as
\begin{equation}
P_i(L)=\frac{d_i(L)}{L},\quad
R_i(L)=\frac{d_i(L)}{D_i}
\end{equation}
where $d_i(L)$ indicates the number of relevant objects (objects collected by $i$ that are present in the probe set) in the top-$L$ places of the recommendation list, and $D_i$ is the total number of $i$'s relevant objects. Averaging the individual precision and recall over all users with at least one relevant object, we obtain the mean precision and recall, $P(L)$ and $R(L)$, respectively. These values can be compared with precision and recall resulting from random recommendation, leading to precision and recall enhancements as defined in \cite{Zhou_PNAS}
\begin{equation}
e_P(L)=P(L)\, \frac{MN}{D},\quad
e_R(L)=R(L)\, \frac{N}{L},
\end{equation}
where $M$ and $N$ are the number of users and objects, respectively, and $D$ is the total number of relevant objects. While precision usually decreases with $L$, recall always grows with $L$. One may combine them into a less $L$-dependent metric
\cite{Rijs79,Pazzani1997}
\begin{equation}
F_1(L)=\frac{2PR}{P+R}
\end{equation}
which is called $F_1$-score. Many other measurements which combine precision and recall are used to evaluate the effectiveness of information retrieval, but rarely applied to evaluate recommendation algorithms: \emph{Average Precision}, \emph{Precision-at-Depth}, \emph{R-Precision}, \emph{Reciprocal Rank} \cite{Buckley2005}, \emph{Binary Preference Measure} \cite{Buckley2004}. A detailed introduction and discussion of each combination index can be found in \cite{Moffat2008}.

\subsubsection{Rank-weighted Indexes} 
Since users have limited patience on inspecting individual objects in the recommended lists, user satisfaction is best measured by taking into account the position of each relevant object and assign weights to them accordingly. Here we introduce three representative indexes that follow this approach. For a detailed discussion of their strengths and weaknesses see \cite{Herlocker2004}.

\paragraph{Half-life Utility}
The \emph{half-life utility} metric attempts to evaluate the utility of a recommendation list to a user. It is based on the assumption that the likelihood that a user examines a recommended object decays exponentially with the object's ranking. The expected utility of recommendations given to user $i$ hence becomes \cite{Breese1998}
\begin{equation}
HL_i=\sum_{\alpha=1}^N \frac{\max(r_{i\alpha}-d,0)}{2^{(o_{i\alpha}-1)/(h-1)}},
\end{equation}
where objects are sorted by their recommendation score $\tilde r_{i\alpha}$ in descending order, $o_{i\alpha}$ represents the predicted ranking of object $\alpha$ in the recommendation list of user $i$, $d$ is the default rating (for example, the average rating), and the ``half-life'' $h$ is the rank of the object on the list for which there is a $50\%$ chance that the user will eventually examine it. This utility can be further normalized by the maximum utility (which is achieved when the user's all known ratings appear at the top of the recommendation list). When $HL_i$ is averaged over all users, we obtain an overall utility of the whole system.

\paragraph{Discounted Cumulative Gain}
For a recommendation list of length $L$, $DCG$ is defined as \cite{Jarvelin2002}
\begin{equation}
\label{DCG}
DCG(b)=\sum_{n=1}^b r_n+\sum_{n=b+1}^L \frac{r_n}{\log_b n},
\end{equation}
where $r_n$ indicates the relevance of the $n$-th ranked object ($r_n=1$ for a relevant object and zero otherwise) and $b$ is a persistence parameter which was suggested to be $2$. The intention of $DCG$ is that highly ranked relevant objects give more satisfaction and utility than badly ranked ones.

\paragraph{Rank-biased Precision}
This metric assumes that users always check the first object and progress from one object to the next one with certain (persistence) probability $p$ (with a complementary probability $1-p$, the examination of the recommendation list ends). For a list of length $L$, the rank-biased precision metric is defined as \cite{Moffat2008}
\begin{equation}
RBP=(1-p)\sum_{n=1}^L r_n p^{n-1},
\end{equation}
where $r_n$ is the same as in $DCG$. $RBP$ is similar to $DCG$, the difference is that $RBP$ discounts relevance via a geometric sequence, while $DCG$ does so using a log-harmonic form.

\subsubsection{Diversity and Novelty}
Even a successfully recommended relevant object has little value to a user when it is notorious. To complement the above accuracy-probing metrics, several diversity- and novelty-probing metrics have been proposed recently~\cite{McNee06,Zhou_PNAS,Castells2011} and we introduce them here.

\paragraph{Diversity}
Diversity in recommender systems refers to how different the recommended objects are with respect to each other. There are two levels to interpret diversity: one refers to the ability of an algorithm to return different results to different users---we call it \emph{Inter-user diversity} (\ie, the diversity between recommendation lists). The other one measures the extent to which an algorithm can provide diverse objects to each individual user---we call it \emph{Intra-user diversity} (\ie, the diversity within a recommendation list). Inter-user diversity \cite{Zhou2008} is defined by considering the variety of users' recommendation lists. Given users $i$ and $j$, the difference between the top $L$ places of their recommendation lists can be measured by the Hamming distance
\begin{equation}
H_{ij}(L) = 1 - \frac{Q_{ij}(L)}{L},
\end{equation}
where $Q_{ij}(L)$ is the number of common objects in the top-$L$ places of the lists of users $i$ and $j$. If the lists are identical, $H_{ij}(L)=0$, while if their lists are completely different, $H_{ij}(L)=1$. Averaging $H_{ij}(L)$ over all user pairs, we obtain the mean Hamming distance $H(L)$. The greater its value, the more diverse (more personalized) recommendation is given to the users.

Denoting the recommended objects for user $i$ as $\{o_1,o_2,\cdots,o_L\}$, similarity of these objects $s(o_{\alpha},o_{\beta})$ can be used to measure the intra-user diversity (this similarity can be obtained either directly from the input ratings or from object metadata) \cite{Zhou_NJP}. The average similarity of objects recommended to user $i$,
\begin{equation}
I_i(L)=\frac{1}{L(L-1)}\sum_{\alpha\neq\beta}s(o_{\alpha},o_{\beta}),
\end{equation}
can be further averaged over all users to obtain the mean intra-similarity of the recommendation lists, $I(L)$. The lower is this quantity, the more diverse objects are recommended to the users. Notably, intra-list diversity can be used to enhance improve recommendation lists by avoiding recommendation of excessively similar objects \cite{Ziegler05}. The rank-sensetive version can be obtained by introducing a discount function of the object's rank in recommendation list \cite{Castells2011}. 

\paragraph{Novelty and Surprisal}
The novelty in recommender systems refers to how different the recommended objects are with respect to what the users have already seen before. The simplest way to quantify the ability of an algorithm to generate novel and unexpected results is to measure the average popularity of the recommended objects
\begin{equation}
\label{novelty}
N(L) = \frac1{ML}\sum_{i=1}^M \sum_{\alpha\in O_R^i}{k_{\alpha}},
\end{equation}
where $O_R^i$ is the recommendation list of user $i$ and $k_\alpha$ denotes the degree of object $\alpha$ (\ie, the popularity of object $\alpha$). Lower popularity indicates higher novelty of the results. Another possibility to measure the unexpectedness is using the self-information (surprisal) \cite{Tribus1961} of recommended objects. Given an object $\alpha$, the chance that a randomly-selected user has collected it is $k_{\alpha}/M$ and thus its self-information is 
\begin{equation}
\label{Self-Information-EQ}
U_\alpha=\log_2(M/k_{\alpha}).
\end{equation}
A user-relative novelty variant can be defined by restricting the observations to the target user, namely caculating the mean self-information of target user's top-$L$ objects. Averaging over all users we obtain the mean top-$L$ surprisal $U(L)$. With a similar resulting formula, a discovery-based novelty was proposed in \cite{Castells2011} by considering the propability that an object is known or familiar to a random user.

\subsubsection{Coverage}
Coverage measures the percentage of objects that an algorithm is able to recommend to users in the system. Denoting the total number of distinct objects in top $L$ places of all recommendation lists as $N_d$, the $L$-dependent coverage is defined as 
\begin{equation}
\label{Coverage-EQ}
COV(L)=N_d/N.
\end{equation}
Low coverage indicates that the algorithm can access and recommend only a small number of distinct objects (usually the most popular ones) which often results in little diverse recommendations. On the contrary, algorithms with high coverage are more likely to provide diverse recommendations \cite{PRE066119}. From this viewpoint, coverage can be also considered as a diversity metric. In addition, coverage is helpful to better evaluate results of accuracy metrics \cite{Cacheda2011}: recommending popular objects is likely to be of high accuracy but of low coverage. A good recommendation method is expected to be of both high accuracy and coverage.

The choice of a particular metric (or metrics) to evaluate a recommender system depends on the goals that the system is supposed to fulfill. In practice, one may specify different goals for new and experienced users which further complicates the evaluation process. For a better overview, Table \ref{tab:metrics} summarizes the described metrics for evaluation of recommender systems.

\begin{table}
\small
\centering
\caption{Summary of the presented recommendation metrics. The third column represents the preference of the metric (\eg, smaller MAE means higher rating accuracy). The fourth
column describes the scope of the metric. The last two columns show whether the metric is obtained from a ranking and whether it depends of the length of the recommendation list $L$.}
\begin{tabular}{rlllll}
\toprule
                      Name & Symbol    & Preference & Scope                 & Rank& $L$  \\
\midrule
                       MAE &$MAE$      & small & rating accuracy            & No  & No \\
                      RMSE &$RMSE$     & small & rating accuracy            & No  & No \\
                   Pearson &$PCC$      & large & rating correlation         & No  & No\\
                  Spearman &$\rho$     & large & rating correlation         & Yes & No\\
             Kendall's Tau &$\tau$     & large & rating correlation         & Yes & No\\
                      NDPM &$NDPM$     & small & ranking correlation        & Yes & No\\
                 Precision &$P(L)$     & large & classification accuracy    & No  & Yes\\
                    Recall &$R(L)$     & large & classification accuracy    & No  & Yes\\
               $F_1$-score &$F_1(L)$   & large & classification accuracy    & No  & Yes\\
                       AUC &$AUC$      & large & classification accuracy    & No  & No\\
             Ranking score &$RS$       & small & ranking accuracy           & Yes & No\\
         Half-life utility &$HL(L)$    & large & satisfaction               & Yes & Yes\\
Discounted Cumulative Gain &$DCG(b,L)$ & large & satisfaction and precision & Yes & Yes\\
     Rank-biased Precision &$RBP(p,L)$ & large & satisfaction and precision & Yes & Yes\\
          Hamming distance &$H(L)$     & large & inter-diversity            & No  & Yes\\
          Intra-similarity &$I(L)$     & small & intra-diversity            & No  & Yes\\
                Popularity &$N(L)$     & small & surprisal and novelty      & No  & Yes\\
          Self-information &$U(L)$     & large & unexpectedness             & No  & Yes\\
                  Coverage &$COV(L)$   & large & coverage and diversity     & No  & Yes\\
\bottomrule
\end{tabular}
\label{tab:metrics}
\end{table}
           % checked, done
\section{Similarity-based methods}
\label{sec:sim-based}
\begin{figure}
\centering
\figg{1}{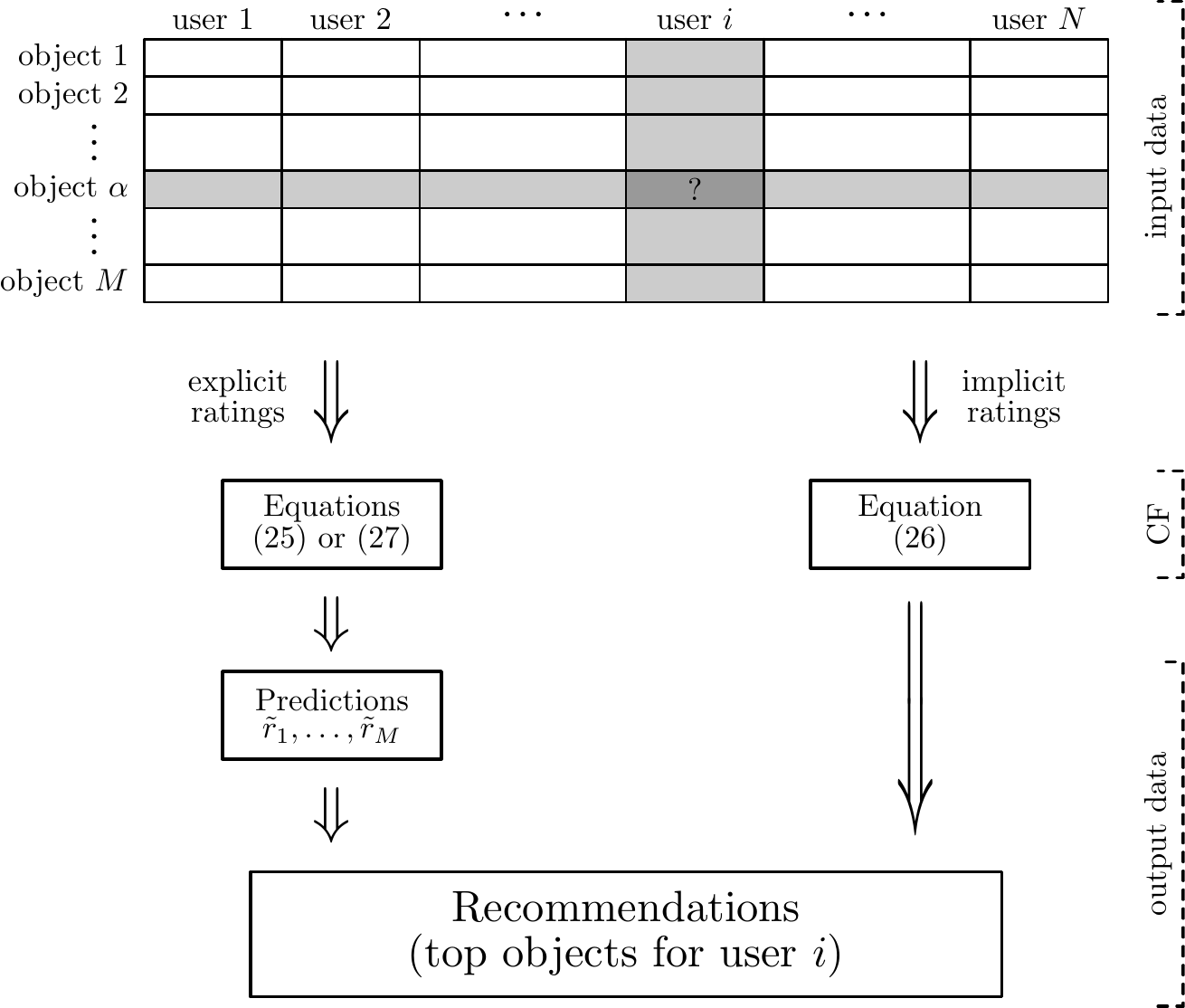}
\caption{A schematic representation of collaborative-filtering (CF) recommendation method: rating prediction for a given user-object pair is based on the user's and object's past ratings.}
\label{fig:CF}
\end{figure}

Similarity-based methods represent one of the most successful approaches to recommendation. They have been studied extensively and found various applications in e-commerce \cite{Linden2003,Sarwar2000}. This class of algorithms can be further divided into methods employing user and item similarity, respectively. The basic assumption of a method based on user similarity is that people who agree in their past evaluationes tend to agree again in their future evaluations. Thus, for a target user, the potential evaluation of an object is estimated according to the ratings from users (``taste mates'') who are similar to the target user (see \fig{fig:CF} for a schematic illustration). Different from user similarity, an algorithm based on item similarity recommends a user the objects that are similar to what this user has collected before. Note that, sometimes the opinions from dissimilar users \cite{Zeng2010} or the negative ratings \cite{Zeng2011,Kong2011} can play a significant (even positive) role in determining the recommendation, especially when the data set is very sparse and thus the information about relevance is more important than that about correlation \cite{Shang2009a}. For additional information see the recent review articles \cite{Su2009,Almazro2010}, and \cite{LPSurvey} is a nice survey that contains a number of similarity indices.  

\subsection{Algorithms}
Here we briefly introduce the conventional similarity-based algorithms which are often referred to as memory-based collaborative filtering techniques. The term ``collaborative filtering'' was introduced by creators of the first commercial recommender system, Tapestry \cite{Tapestry}, derives from the fact that it requires collaboration of multiple agents who share their data to obtain better recommendation. In the following sections, we describe basic algorithms as well as main approaches to the computation of similarity which is a critical component of the recommendation process.

\subsubsection{User similarity}
The goal is to make automated prediction of user preferences by collecting evaluation data from many other users, especially those whose evaluations are similar to evaluations from the target user. Denote the rating from user $u$ on object $\alpha$ as $r_{u\alpha}$ and let $\Gamma_u$ be the set of objects that user $u$ has evaluated. The average rating given by $u$ is $\bar{r}_u=\frac{1}{\abs{\Gamma_u}}\sum_{\alpha\in\Gamma_u}{r_{u\alpha}}$. According to the standard collaborative filtering, the predicted rating of user $u$ on object $\alpha$ is
\begin{equation}
\label{UCF} \tilde r_{u\alpha}=\bar{r}_{u}+
\kappa \sum_{v\in\hat{U}_u}{s_{uv}(r_{v\alpha}-\bar{r}_{v})}
\end{equation}
where $\hat{U}_u$ denotes the set of users that are most similar to user $u$, $s_{uv}$ denotes the similarity between user $u$ and user $v$ and $\kappa=\frac{1}{\sum_{v}{\abs{s_{uv}}}}$ is a normalization factor. If instead of explicit ratings, only the sets of objects collected by individual users are known (implicit ratings), we aim at predicting the objects which are most likely to be collected by a user in the future. According to \cite{Zeng2010}, \req{UCF} should be replaced with
\begin{equation}
\label{CF} p_{u\alpha} = \sum_{v\in\hat{U}_u} s_{uv}a_{v\alpha}
\end{equation}
where $p_{u\alpha}$ is the recommendation score of object $\alpha$ for user $u$ and $a_{v\alpha}$ is an element of the adjacency matrix of the user-object bipartite network ($a_{v\alpha}=1$ if user $v$ has collected object $\alpha$ and $a_{v\alpha}=0$ otherwise).

As has already been made explicit by \req{UCF} and \req{CF}, only users most similar to given user $u$ are usually considered. To obtain $\hat{U}_u$, two neighborhood selection strategies are usually applied: (i) correlation threshold \cite{Shardanand1995} is based on selecting all users $v$ whose similarity $s_{uv}$ surpasses a given threshold, (ii) maximum number of neighbors \cite{Resnick1994} consists of selecting those $k$ users that are most similar to $u$ (here $k$ is a parameter of the algorithm). Restricting computation to the most similar users is not only computationally advantageous but in general it leads to superior results \cite{Goldberg01}.

\subsubsection{Item similarity}
In this case, item-item similarity $s_{\alpha\beta}$ is employed instead of user-user similarity $s_{uv}$. The simplest way is to estimate unknown ratings using the weighted average \cite{Sarwar2001}
\begin{equation}
\label{ICF} \tilde r_{u\alpha}=
\frac{\sum_{\beta\in\Gamma_u} s_{\alpha\beta}r_{u\beta}}
{\sum_{\beta\in\Gamma_u}\abs{s_{\alpha\beta}}}
\end{equation}
where $\Gamma_u$ is the set of items evaluated by user $u$. Techniques limiting the computation of $\tilde r_{u\alpha}$ to items that are most similar to $\alpha$ can be applied similarly as described above for user similarity. One of the advantages of this approach is that similarity between items tends to be more static than similarity between users, allowing its values and neighborhoods to be computed offline (\emph{i.e.}, before recommendation for a particular user is requested---this allows to shorten the time needed to obtain the recommendation). Hybrid collaborative filtering algorithms combining user-, item- or attribute-based similarity were proposed \cite{Wang06,LiuZB2010}. Their results show that this approach not only improves the prediction accuracy but it is also more robust to data sparsity.

\subsubsection{Slope One predictor}
\begin{figure}
\centering
\figg{0.9}{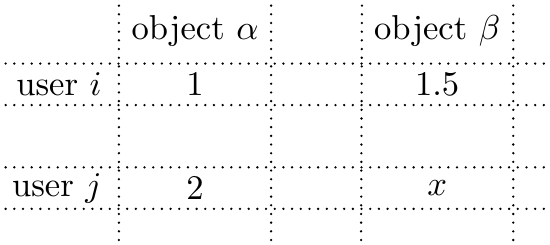}
\caption{In the depicted case, the Slope One prediction would be $x=2+(1.5-1)=2.5$.}
\label{fig:slope_one}
\end{figure}

Slope One predictor with the form $f(x)=x+b$, where $b$ is a constant and $x$ is a variable representing the rating values, is the simplest form of item-based collaborative filtering based on ratings \cite{Lemire2005}. It subtracts the average ratings of two items to measure how much more, on average, one item is liked than another. This difference is used to predict another user's rating of one of these two items, given his rating of the other. For example, consider a case where user $i$ gave score $1$ to item $\alpha$ and score $1.5$ to item $\beta$ while user $j$ gave score $2$ to item $\alpha$. Slope One then predicts that user $j$ will rate item $\beta$ with $2+(1.5-1)=2.5$ (see \fig{fig:slope_one} for an illustration).

The Slope One scheme takes into account both information from other users who rated the same item and from the other items rated by the same user. In particular, only ratings by users who have rated some common items with the target user and only ratings of items that the target user has also rated are involved in the prediction process. Denoting the set of users who rated items $\alpha$ and $\beta$ as $\mathrm{S}(\alpha,\beta)$, the average deviation of item $\beta$ with respect to item $\alpha$ is defined as
\begin{equation}
\label{dev}
\mathrm{dev}_{\beta\alpha}=\frac{\sum_{i\in\mathrm{S}(\alpha,\beta)}r_{i\beta}-r_{i\alpha}}{\abs{\mathrm{S}(\alpha,\beta)}}.
\end{equation}
Given a known rating $r_{u\alpha}$, Slope One predicts $u$'s rating on item $\beta$ as $r_{u\alpha}+\mathrm{dev}_{\beta\alpha}$. By varying $\alpha$ in Eq.~\ref{dev}, we obtain different predictions. A reasonable overall predictor is their average value
\begin{equation}
\label{Slope_ave}
\tilde r_{u\alpha}=\frac{1}{\abs{\mathrm{R}(u,\alpha)}}
\sum_{\alpha\in\mathrm{R}(u,\alpha)}(r_{u\beta}+\mathrm{dev}_{\alpha\beta}),
\end{equation}
where $\mathrm{R}(u,\alpha)$ is the set of items that have been both rated by $u$ and co-rated with item $\beta$. Note that predictions obtained from different items $\alpha$ have equal weight no matter how many users have co-rated $\alpha$ with $\beta$. To take into account the fact that the credibility of $\mathrm{dev}_{\alpha\beta}$ depends crucially on $\abs{\mathrm{S}(\alpha,\beta)}$ (the larger the overlap, the more trustful the value), one can introduce a Weighted Slope One prediction as
\begin{equation}
\tilde r_{u\alpha}^w=\frac{\sum_{\alpha}\abs{S(\alpha,\beta)}
(r_{u\beta}+\mathrm{dev}_{\alpha\beta})}{\sum_\alpha\abs{S(\alpha,\beta)}}.
\end{equation}
Another improvement of the basic Slope One algorithm is based on dividing the set of all items into items liked and disliked by a given user (a straightforward criterion to identify liked and disliked items is to check whether their rating were higher or lower than the average rating awarded by the given user). From these liked and disliked items, two separate predictions are then derived which are combined into one prediction at the very end. Denote by $S^{+1}(\alpha,\beta)$ and $S^{-1}(\alpha,\beta)$ the sets of users who like and dislike, respectively, both $\alpha$ and $\beta$. The deviations for liked and disliked items are
\begin{equation}
\mathrm{dev}^{+1}_{\beta\alpha}=\frac{1}{\abs{\mathrm{S}^{+1}(\beta,\alpha)}}
\sum_{i\in \mathrm{S}^{+1}(\beta,\alpha)}(r_{i\beta}-r_{i\alpha}),\quad
\textrm{dev}^{-1}_{\beta\alpha}=\frac{1}{\abs{\mathrm{S}^{-1}(\beta,\alpha)}}
\sum_{i\in \mathrm{S}^{-1}(\beta,\alpha)}(r_{i\beta}-r_{i\alpha}).
\end{equation}
The prediction for the rating of item $\beta$ based on the rating of item $\alpha$ is either $r_{j\alpha}+\textrm{dev}^{+1}_{\beta\alpha}$ or $r_{j\alpha}+\textrm{dev}^{-1}_{\beta\alpha}$ depending on whether the target user $j$ likes or dislikes item $\alpha$ respectively. The Bi-Polar Slope One is thus given by
\begin{equation}
p^{\textrm{bi}}_{j\beta}=
\frac{\sum_\alpha\abs{S^{+1}(\beta,\alpha)}(r_{j\alpha}+\textrm{dev}^{+1}_{\beta\alpha})+
\sum_\alpha\abs{S^{-1}(\beta,\alpha)}(r_{j\alpha}+\textrm{dev}^{-1}_{\beta\alpha})}
{\sum_\alpha\abs{S^{+1}(\beta,\alpha)}+\sum_\alpha\abs{S^{-1}(\beta,\alpha)}}
\end{equation}
where the weights are chosen similarly as for Weighted Slope One.

It was shown that Slope One can outperform linear regression (\ie, estimation by $f(x)=ax+b$) while having half the number of regressors \cite{Lemire2005,Cacheda2011}. This simple approach also reduces storage requirements and latency of the recommender system. Slope One has been used as a building block to improve other algorithms \cite{WangP2009,ZhangD2009,Gao2009}. For instance, it can be combined with user-based collaborative filtering to address the data sparsity problem via filling the vacant ratings of the user-item matrix by the Slope One scheme, and thus improving the prediction accuracy \cite{WangP2009}.

\subsection{How to define similarity}
The key problem of similarity-based algorithms is how to define similarity between users or objects. When explicit ratings are available, similarity is usually defined using a correlation metric such as Pearson, for example (two users are considered as similar when they tend to give similar ratings to the objects they rate). When there is no rating information available, similarity can be inferred from the structural properties of the input data (two users are considered as similar when they liked/bought many objects in common). Besides, external information such as users' attributes, tags and objects' content meta information can be utilized to estimate similarity better.

\subsubsection{Rating-based similarity}
In many online e-commerce services, users are allowed to evaluate the consumed objects by ratings. For example, in Yahoo Music, users vote each song with one to five stars representing "Never play again"($\star$), "It is ok"($\star\star$), "Like it" ($\star\star\star$), "Love it"($\star\star\star\star$) and "Can't get enough"($\star\star\star\star\star$). With explicit rating information we can measure the similarity between two users or between two objects by \emph{Cosine index} \cite{AdoTuz2005,2-ACM evaluating} which is defined as
\begin{equation}
s^{\cos}_{xy}=\frac{\vek{r}_x \cdot \vek{r}_y}{\abs{\vek{r}_x} \abs{\vek{r}_y}}.
\end{equation}
For quantifying the similarity between users, $\vek{r}_x,\vek{r}_y$ are rating vectors in the $N$-dimensional object space while for similarity between objects, $\vek{r}_x,\vek{r}_y$ are vectors in $M$-dimensional user space. Note that, in the calculation of rating-based similarity, it is necessary to eliminate the rating tendencies of users and/or on items, otherwise the similarity is less meaningful. Actually, according to a recently reported smart method, in some rating systems, via proper usage of rating tendencies, one could predict the unknown ratings with remarkably higher accuracy than the simply similarity-based methods \cite{Cacheda2011}.   

The rating correlation can also be measured by \emph{Pearson coefficient} (PC) \cite{AdoTuz2005,2-ACM evaluating}. To quantify similarity between users $u$ and $v$, it reads
\begin{equation}
\label{Pearson-user}
s^{PC}_{uv}=\frac{\sum_{\alpha\in O_{uv}}(r_{u\alpha}-\bar{r}_u)(r_{v\alpha}-\bar{r}_v)}
{\sqrt{\sum_{\alpha\in O_{uv}}(r_{u\alpha}-\bar{r}_u)^2}
\sqrt{\sum_{\alpha\in O_{uv}}(r_{v\alpha}-\bar{r}_v)^2}},
\end{equation}
where $O_{uv}=\Gamma_u\cap\Gamma_v$ indicates the set of objects rated by both $u$ and $v$. A \emph{constrained Pearson coefficient} proposed by Shardanand and Maes \cite{Shardanand1995} consists of substituting the user mean in Eq.~\ref{Pearson-user} with a ``central'' rating (for example, in the scale from 1 to 5, one can set the central rating to be 3). The idea is to take into account the difference between positive (above the central rating) and negative ratings (below the central rating). A \emph{weighted Pearson coefficient} is based on the idea of capturing the confidence that can be placed on similarity values (when two users evaluated only a few objects in common, their potentially high similarity should not be trusted as much as for a pair of users with many overlapping objects). It was proposed \cite{Herlocker1999} to weight the Pearson coefficient as
\begin{equation}
s_{uv}^{WPC}=
\begin{cases}
s_{uv}^{PC}\,\frac{\abs{O_{uv}}}{H} & \text{for $\abs{O_{uv}}\leq H$},\\
s_{uv}^{PC} & \text{otherwise}
\end{cases}
\label{weight}
\end{equation}
where $H$ is a threshold, determined experimentally, beyond which the correlation measure can be trusted.

Analogically, Pearson similarity between objects $\alpha$ and $\beta$ reads
\begin{equation}
s^{PC}_{\alpha\beta}=\frac{\sum_{u\in U_{\alpha\beta}}
(r_{u\alpha}-\bar{r}_\alpha)(r_{u\beta}-\bar{r}_\beta)}
{\sqrt{\sum_{u \in U_{\alpha\beta}}(r_{u\alpha}-\bar{r}_\alpha)^2}
\sqrt{\sum_{u \in U_{\alpha\beta}} (r_{u\beta}-\bar{r}_\beta)^2}},
\end{equation}
where $U_{\alpha\beta}$ is the set of users who rated both $\alpha$ and $\beta$, and $\bar{r}_\alpha$ is the average rating of object $\alpha$. Experiments have shown that the Pearson coefficient performs better than the cosine vector index \cite{Breese1998}. When only binary ratings are available (like or dislike, purchase or no purchase, click or no click, etc.), the cosine and Pearson coefficient can still be applied to quantify the similarity of vectors with binary elements. For example, Amazon's patented algorithm \cite{Linden2003} computes the cosine similarity between binary vectors representing users' purchases and use it in item-based collaborative filtering.

\subsubsection{Structural similarity} \label{structural_similarity}
As we have mentioned above, similarity can be defined using the external attributes such as tag and content information. However, the required data is usually very difficult to collect. Another simple and effective way to quantify the similarity, \emph{structural similarity} \cite{Liben-Nowell2007}, is based solely on the network structure of the data. Recent research shows that the structural-based similarity can produce better recommendations that the Pearson correlation coefficient, especially when the input data is very sparse \cite{Shang2009a}.

To calculate the structural similarity between users or objects, we generally project the user-object bipartite network which contains the complete information about the system into a monopartite user-user or object-object network (for more information on this aspect of similarity see \cite{Zhou2007}). In the simplest case, two users are considered similar if they have voted at least one common object (analogically, two objects are considered similar if they have been co-voted by at least one user). More refined similarity metrics that can be roughly categorized as node-dependent vs. path-dependent, local vs. global, parameter-free vs. parameter-dependent, and so on---here we review some of them.

(i) \emph{Node-dependent similarity.} The simplest weighted similarity index is Common Neighbors (CN) where the similarity of two nodes is directly given by the number of common neighbors (think of the number of users who bought both objects $\alpha$ and $\beta$ or the number of objects shared by users $u$ and $v$). By considering degrees of the two target nodes, six variations of CN were derived: Salton Index \cite{Salton1983}, Jaccard Index \cite{Jaccard1901}, S{\o}rensen Index \cite{Sorensen1948}, Hub Promoted Index (HPI) \cite{Ravasz2002}, Hub Depressed Index (HDI) and Leicht-Holme-Newman Index (LHN1\footnote{We use the abbreviation LHN1 to distinguish this index to another index named as LHN2 also proposed by Leicht, Holme and Newman.}) \cite{Leicht2006}. One can further take into account degrees of respective common neighbors to reward less-connected neighbors with a higher weight as in Adamic-Adar Index (AA) \cite{Adamic2003} and Resource Allocation Index (RA) \cite{Zhou2009}. Note that since AA uses a logarithmic weighting, it penalizes high-degree common neighbors less than RA. Finally, Preferential Attachment Index (PA) builds on the classical preferential attachment rule in network science \cite{Barabasi1999}. This index has been used to quantify the functional significance of links subject to various network-based dynamics, such as percolation \cite{Holme2002}, synchronization \cite{Yin2006} and transportation \cite{Zhang2007}. Note that these similarity can be computed also for a bipartite network where common neighbors are objects and users when considering user and object similarity, respectively. A summary of mathematical definitions of these similarity indices is shown in Table~\ref{node-d}.

\begin{table}
\caption{Mathematical definitions of the described node-dependent similarity indices. $\Gamma_x$ denotes the set of neighbors of node $x$ (which can be either a user or an object node) and $k_x$ is the degree of node $x$.}
\centering
\begin{tabular}{rl}
\toprule
\textbf{Index}    & Definition\\
\midrule
CN           & $s_{xy}=\abs{\Gamma_x\cap \Gamma_y}$\\
Salton       & $s_{xy}=\abs{\Gamma_x\cap \Gamma_y}/\sqrt{k_x k_y}$\\
Jaccard      & $s_{xy}=\abs{\Gamma_x\cap \Gamma_y}/\abs{\Gamma_x\cup\Gamma_y}$\\
S{\o}rensen  & $s_{xy}=2\abs{\Gamma_x\cap \Gamma_y}/(k_x+k_y)$\\
HPI          & $s_{xy}=\abs{\Gamma_x\cap \Gamma_y}/\min\{k_x,k_y\}$\\
HDI          & $s_{xy}=\abs{\Gamma_x\cap \Gamma_y}/\max\{k_x,k_y\}$\\
LHN1         & $s_{xy}=\abs{\Gamma_x\cap \Gamma_y}/(k_x k_y)$\\
AA           & $s_{xy}=\sum_{z\in \Gamma_x\cap \Gamma_y}1/\ln k_z$\\
RA           & $s_{xy}=\sum_{z\in \Gamma_x\cap \Gamma_y}1/k_z$\\
PA           & $s_{xy}=k_x k_y$\\
\bottomrule
\end{tabular}
\label{node-d}
\end{table}

(ii) \emph{Path-dependent similarity.} The basic assumption here is that two nodes are similar if they are connected by many paths. Since elements of an $n$-th power of the adjacency matrix, $\mathsf{A}^n$, are equal to the number of distinct paths between respective pairs of nodes, path-dependent similarity metrics can be usually written in a compact form such as
\begin{equation}
s^{LP}_{xy}=(\AM^2)_{xy}+\epsilon (\AM^3)_{xy}
\end{equation}
for the Local Path Index \cite{Lu2009} where only paths of length two and three count and $\epsilon$ is a damping parameter. (Note that in a bipartite network, only paths of an even length can exist between nodes of the same kind.) By including paths of all lengths, we obtain the classical Katz similarity \cite{Katz1953} which is defined as
\begin{equation}
s^{Katz}_{xy}=\beta A_{xy}+\beta^2(\AM^2)_{xy}+\beta^3(\AM^3)_{xy}+\dots,
\label{Sim11}
\end{equation}
where $\beta$ is a damping factor controlling the path weights. This can be written as $s^{Katz}=(I-\beta \AM)^{-1}-I$. A variant of the Katz index, \emph{Leicht-Holme-Newman Index} (LHN2) \cite{Leicht2006}, was proposed where the term $(\AM^l)_{xy}$ is replaced with
$(\AM^l)_{xy}/\EE{(\AM^l)_{xy}}$ where $\EE{X}$ is the expected value of $X$.

(iii) \emph{Random-walk-based similarity.} Another group of methods is based on random walks on networks.

\textbf{Average Commute Time}: The average commute time between nodes $x$ and $y$ is defined as the average number of steps required by a random walker starting from node $x$ to reach node $y$ plus that from $y$ to $x$. It can be obtained in terms of the pseudoinverse of the network's Laplacian matrix, $\LM^+$, as \cite{Klein1993,Fouss2007}
\begin{equation}
n(x,y)=\big((\LM^+)_{xx}+(\LM^+)_{yy}-2(\LM^+)_{xy}\big)E,
\end{equation}
where $E$ is the number of edges in the network. Assuming that two nodes are similar if they have small average commute time, similarity between nodes $x$ and $y$ can be defined as the reciprocal of their average commute time
\begin{equation}
s^{ACT}_{xy}=\frac{1}{(\LM^+)_{xx}+(\LM^+)_{yy}-2(\LM^+)_{xy}}.
\end{equation}
where the constant factor $E$ has been removed.

\textbf{Cosine-based on $L^+$}: This index is an inner-product-based measure. In the Euclidean space spanned by $\vek{v}_x=\Lambda^{\frac{1}{2}}\mathsf{U}^T\vek{e}_x$ where $\mathsf{U}$ is an orthonormal matrix composed of the eigenvectors of $\LM^+$ ordered in a decreasing order of their eigenvalues $\lambda_x$, $\Lambda=\mathrm{diag}(\lambda_x)$, $\vek{e}_x$ is a column base vector ($(\vek{e}_x)_y)=\delta_{xy}$) and $T$ is matrix transposition, elements of the pseudoinverse of the Laplacian matrix are the inner products of the node vectors, $(\LM^+)_{xy}=\vek{v}_x^T\vek{v}_y$. Consequently, cosine similarity is defined as \cite{Fouss2007}
\begin{equation}
s^{\cos+}_{xy}=\frac{\vek{v}_x^T\vek{v}_y}{\abs{\vek{v}_x}\abs{\vek{v}_y}}=
\frac{(\LM^+)_{xy}}{\sqrt{(\LM^+)_{xx}(\LM^+)_{yy}}}.
\end{equation}

\textbf{Random Walk with Restart}: This index is a direct application of the PageRank algorithm \cite{Brin1998}. Consider a random walker starting from node $x$ recursively moves to a random neighbor with probability $c$ and returns to node $x$ with probability $1-c$. Denoting by $q_{xy}$ the resulting stationary probability that the walker is located at node $y$, we can write
\begin{equation}
\vek{q}_x= c\PM^T\vek{q}_x+(1-c)\vek{e}_x \label{RWR1}
\end{equation}
where $\PM$ is the transition matrix with elements $P_{xy}=1/k_x$ if $x$ and $y$ are connected and $P_{xy}=0$ otherwise. The solution to this equation is
\begin{equation}
\vek{q}_x = (1-c) (I-c\PM^T)^{-1}\vec{e_x}. \label{RWR2}
\end{equation}
Finally, the similarity index is defined as
\begin{equation}
s^{RWR}_{xy}=q_{xy}+q_{yx}.
\end{equation}
A fast algorithm to calculate this index was proposed \cite{Tong2006} and the application to recommender systems was studied in \cite{Shang2009a} where it was found that this similarity performs better than the Pearson correlation coefficient.

\textbf{SimRank}: This index is defined based on the assumption that two nodes are similar if they are connected to similar nodes. This allows us to define SimRank in a self-consistent way \cite{Jeh2002} as
\begin{equation}
s^{SimRank}_{xy}=C\,
\frac{\sum_{z\in{\Gamma_x}}\sum_{z'\in{\Gamma_y}}s^{SimRank}_{zz'}}{k_x k_y},
\end{equation}
where $s_{xx}=1$ and $C\in{[0,1]}$ is a free parameter. SimRank can also be interpreted by the random-walk process: $s^{SimRank}_{xy}$ measures how fast are two random walkers, who respectively start at nodes $x$ and $y$, expected to meet at a certain node.

\textbf{Matrix Forest Index}: This index introduces similarity between $x$ and $y$ as the ratio of the number of spanning rooted forests such that nodes $x$ and $y$ belong to the same tree rooted at $x$ to all spanning rooted forests of the network (for details see \cite{Chebotarev1997}). Its mathematical definition
\begin{equation}
s^{MFI}=(I+\LM)^{-1}.
\end{equation}
can be further parametrized to obtain a variant of MFI
\begin{equation}
s^{PMFI}=(I+\alpha\LM)^{-1},\quad\alpha>0.
\end{equation}
According to the authors, $\alpha>0$ determines the proportion of accounting for long connections between vertices of the graph versus short ones.

\textbf{Local Random Walk}: To measure similarity between nodes $x$ and $y$, a random walker is introduced in node $x$ and thus the initial occupancy vector is $\vek{\pi}_{x}(0)=\vek{e}_x$. This vector evolves as $\vek{\pi}_x(t+1)=\PM^T\vek{\pi}_x(t)$ for $t\geq 0$. The LRW index at time step $t$ is defined \cite{Liu2010} as
\begin{equation}
s^{LRW}_{xy}(t)=q_x\pi_{xy}(t)+q_y\pi_{yx}(t)
\end{equation}
where $q$ is the initial configuration function and $t$ denotes the time step. In \cite{Liu2010} it was suggested to use a simple approach where $q$  is determined by node degree: $q_x=k_x/M$. Note that in bipartite networks, an even time step must be used to obtain similarity between nodes of the same kind.

\textbf{Superposed Random Walk}: Similar to the RWR index, in \cite{Liu2010} they proposed another index where the random walker is continuously released at the starting point, resulting in a higher similarity between the target node and its nearby nodes. The mathematical expression reads
\begin{equation}
s^{SRW}_{xy}(t)=\sum_{\tau=1}^{t}s_{xy}^{LRW}(\tau)=\sum_{\tau=1}^{t}{[q_x\pi_{xy}(\tau)+q_y\pi_{yx}(\tau)]}.
\end{equation}

In \cite{Fouss2007}, several random-walk-based similarity indices, such as ACT, $\cos+$ and MFI, were applied in collaborative filtering. Their experimental results show that in general, Laplacian-based similarities perform well.

\subsubsection{Similarity involving external information}
\label{sec:ext_sim}
Besides the fundamental user-object relations and the ratings, additional information can be exploited to define or improve the node similarity.

(i) \emph{Attributes.} The dimension and elements of the attribute vectors are defined in advance by some domain experts, and is identical to all users (objects) in the system. The similarity between two users (objects) is obtained by calculating the correlation of their corresponding attributes vectors. For example, user profiles, usually including age, sex, nationality, location, career, etc., can be simply applied to quantify the similarity between users based on the assumption that two users are similar when they have many common features. In \cite{Tso2006}, a hybrid method considering both attributes of objects and the ratings was shown to provide better recommendations than when these two sources of information are used independently. However, the application of attributes represents some risks to user privacy---as shown by a recent work on de-anonymization of large datasets \cite{NaraShm08}, collection and utilization of attribute data poses several sensitive issues. For more information on the issues of user privacy see \cite{Kobsa2006}.

(ii) \emph{Contents.}
Modern information retrieval techniques allow us to automatically extract content and meta information of the available objects. Object similarity can hence be calculated based on the content comparison of the given objects. This is usually referred as content-based recommendation in literature \cite{Pazzani2007}. Unlike collaborative filtering, in a content-based algorithm, recommendations are made based solely on the profile built up by analyzing the content of objects that the target user has rated in the past. The recommendation problem hence becomes a search for objects whose content is most similar to the content of objects already preferred by the target user. The classical method to weigh content is TF-IDF (term frequency - inverse document frequency) \cite{Salton1983}, which is a weighing metric often used in information retrieval and text mining. A term, $t$, in a given document, $d$, is weighted as,

\begin{equation}
    W_{t,d} = \mathrm{tf}(t,d) \times log\frac{|D|}{|{d:t\subseteq d}|},
\end{equation} 
where $\mathrm{tf}(t,d) $ is the frequency of $t$ in document $d$, $|D|$ is number of all observed documents. Then $W_{t,d}$ can be used to measure the similarly of objects defined in Sec.~\ref{structural_similarity}. In addition, if two users have collected objects with similar content, we may assume that these two users are similar.

Since both content-based method and collaborative filtering have their individual limitations, such as CF systems do not explicitly incorporate feature information and face the sparsity and cold-start problems, while content-based systems do not necessarily incorporate the information in preference similarity across individuals (see summaries and discussions in Refs. \cite{AdoTuz2005,Su2009}), many hybrid algorithms are proposed to avoid certain weaknesses in each approach and thereby improve the recommendation performance. The combination methods can be classified into four categories: (i) implement separate collaborative and content-based methods and then combine their predictions \cite{Claypool1999,Pazzani1999}; (ii) add content-based characteristics to collaborative models \cite{BaSho1997,Melville2002,Salter2006}; (iii) add collaborative characteristics to content-based models \cite{Soboroff1999} and (iv) develop a general unified model that integrate both content-based and collaborative characteristics \cite{Basu1998,Popescul2001,Schein02,Yu2003,Jin2005}. However, these methods are only effective if the objects contain rich content information that can be automatically extracted. This is the case for recommendation of books, articles and bookmarks, but not for videos, music tracks or pictures

(iii) \emph{Tags.} Collaborative tagging systems emerged with the advent of Web2.0 \cite{Zhang2010}. Different from traditional taxonomy with hierarchical structure, tagging systems allow users to freely assign keywords (which are usually referred to as \emph{tags}) to manage their own collections without the limitation of a preset vocabulary. Tags provide a rich source of information for recommendation purposes. With the tagging information, algorithms can be easily designed to calculate user similarity and object similarity by considering tag vectors in user and object space, respectively. To alleviate the effects of spam and magnify personalized user preferences, weighting techniques are often applied to measure the importance of each element in a given tag vector.

                        % checked, done
\section{Dimensionality Reduction Techniques}
\label{sec:dim_red}
\begin{figure}
\centering
\figg{0.8}{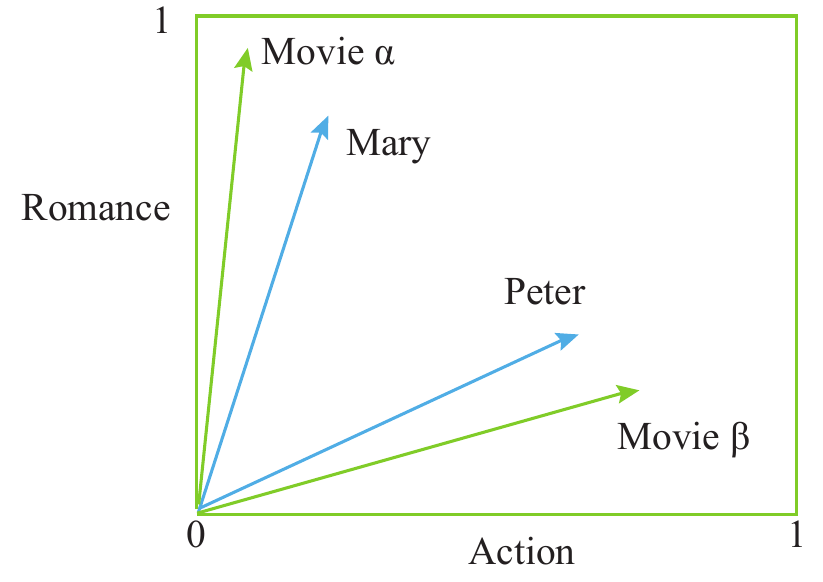}
\caption{An example of a user's movie selection process with hidden variables.}
\label{gr_latent}
\end{figure}

Dimensionality reduction aims at downsizing the amount of relevant data while preserving the major information content. It is often applied in areas such as data mining, machine learning and cluster analysis. Most techniques of dimensionality reduction involve feature extraction which makes use of hidden variables, or so called latent variables, to describe the underlying causes of co-occurrence data. In the context of movie selection, potential viewers may consider genres such as action, romance or comendy features in a movie, which consititute the latent variables. These latent variables are usually represented by multi-dimensional vectors. A simplified picture of two-dimensional vectors of action and romance is shown in \fig{gr_latent}, which shows that user \emph{Peter} has a preference in action movies, while user \emph{Mary} prefers romantic content. Given these vectors and the corresponding vectors of movies, we can define the expected rating of a user on a move as the scalar product of their vectors. For instance, we expect \emph{Peter} prefers movie $\beta$ rather than $\alpha$, while the opposite is true for \emph{Mary}. Recommendations can thus be made once the vectors are computed. If $\toth$ hidden variables are used, the latent vectors are $\toth$-dimensional and dimensionality reduction is achieved if $\toth(\totu+\toti)<\totu\toti$, since the number of relevant variables is reduced. In practise, these techniques are particularly suitable for large data sets which are costly to store and manipulate.

Instead of introducing latent variables which describe interests and genres, users and objects can also be assigned to individual classes which leads to reduction of data dimension. In this case, the co-ocurrence of a user-object pair is explained by the relation between the classes to which the user and the object belong to. Though the original intention for such classification is not to reduce the data dimension, the number of classes used is usually significantly smaller than the number of users and objects, which then results in reduction of dimensionality.

Dimensionality reduction is in particular well applicable in collaborative filtering (it is sometimes referred to as model-based collaborative filtering), as for most applications only a small fraction of user-object pairs are observed such that the number of relevant variables can be significantly reduced. Reductions in dimensionality effectively preserve the information content while drastically decreasing the computation complexity and memory requirements for making recommendations. In this section, several techniques of dimensionality reduction with implementation to recommender systems are discussed, including singular value decomposition (SVD) \cite{takacs2007}, Bayesian clustering \cite{ungar98}, probabilistic latent semantic analysis (pLSA) \cite{hofmann04} and latent dirichlet allocation (LDA) \cite{blei03}.

\subsection{Singular Value Decomposition (SVD)}
We start with a $\totu\times\toti$ matrix $\dataM$ whose element $\data_{\use\ite}$ corresponds to the rating of user $\use$ to object $\ite$ (if the rating has not yet been given, the corresponding element of $\dataM$ is zero). In the case without numeric ratings, $\dataM$ becomes the adjacency matrix as $\data_{\use\ite}=0,1$ for connected and unconnected user-object pair, respectively. Recommendation process then aims to determine which presently zero entries of $\dataM$ have high chance to be non-zero in the future. Note that $\dataM$ is a sparse matrix for most applications because only a small fraction of all its elements are different from zero.

Dimensionality reduction is achieved by introducing $\toth$ hidden variables which categorize tastes of users and attributes of objects. The original $\dataM$ is approximated as the product of two matrices
\begin{equation}
\label{eq_mf}
\dataM \approx \svdU\svdV
\end{equation}
where $\svdU$ and $\svdV$ are respectively matrices with dimension $\totu\times \toth$ and $\toth\times\toti$. They contain the taste information for users and content information for objects, respectively, expressing them in terms of $\toth$ hidden variables. Because of these hidden variables, SVD belongs to a broad class of latent semantic analysis (LSA) techniques. From the product of $\svdU$ and $\svdV$ in \req{eq_mf}, we see that objects are selected by users based on the overlap between a user's tastes and a movie's attributes. When the number of hidden variables is smaller than $\totu$ and $\toti$, the number of parameters needed to describe the system reduces from  $\totu\toti$ for the original $\dataM$ to $\totu \toth+\toth\toti$ for the product $\svdU\svdV$. This approach is also known as matrix factorization (MF) as $\dataM$ is factorized into a product of matrices.

To obtain $\svdU$ and $\svdV$, singular value decompostion (SVD) is a common algebraic tool in LSA which results in downsizing of relevant variables and, at the same time, finding a good approximation of $\dataM$. In SVD, $\dataM$ is factorized as
\begin{equation}
\label{eq_svd}
\dataM = \svdU \Sigma \svdV
\end{equation}
where $\Sigma$ is a $\toth\times\toth$ diagonal matrix, and equality in the above factorization holds with $\toth=\min(\totu, \toti)$. The matrix $\Sigma$ contains the so-called \emph{singular values} of $\dataM$, which are indeed the square root of the eigenvalues of $\dataM\dataM^*$ (or $\dataM^*\dataM$). To benefit from dimensionality reduction, we put $\toth<\min(\totu,\toti)$ which corresponds to the $\toth$-rank approximation in SVD and include only the $\toth$ largest singular values in $\Sigma$ and replace the others by zero. Equality in \req{eq_svd} no longer holds and $\dataM$ is approximated with $\tilde\dataM$ given by
\begin{equation}
\label{eq_svdApprox}
\dataM\approx \tilde\dataM = \svdU \tilde\Sigma \svdV
\end{equation}
where $\tilde\Sigma$ is the $\toth$-rank approximation of $\Sigma$.

It can be shown that $\tilde\dataM$ can be found by minimizing the Frobenius norm of the matrix $\dataM-\tilde\dataM$, that is
\begin{equation}
\tilde\dataM = \underset{\{\tilde\dataM'\}}{\operatorname{argmin}}
\norm{\dataM - \tilde\dataM'}
\end{equation}
with the rank of $\tilde\dataM'$ restricted to be $\toth$. The Frobenius norm is given by
\begin{equation}
\label{eq_norm}
\norm{\dataM-\tilde\dataM}=\sqrt{\sum_{\use\ite}(\data_{\use\ite}-\tilde\data_{\use\ite})^2},
\end{equation}
which corresponds to the root square error of $\tilde\dataM$ with respect to $\dataM$, with $\tilde\data_{\use\ite}$ denoting the element $(\use, \ite)$ in $\tilde\dataM$. SVD thus provides a simple cost function for measuring the agreement between $\dataM$ and $\tilde\dataM$. To obtain $\tilde\dataM$ explicitly for a particular $\dataM$, an simple iterative approach \cite{takacs2007} based on gradient descent can be employed. $\Sigma$ in \req{eq_svdApprox} is first absorbed into either $\svdU$ or $\svdV$ to obtain $\tilde\dataM=\svdU\svdV$ as in \req{eq_mf}. Our task is to obtain the optimal $\svdU$ and $\svdV$ for which $\tilde\dataM=\svdU\svdV$ minimizes the norm in \req{eq_norm}. We now express $\tilde\data_{\use\ite}$ as
\begin{equation}
\tilde\data_{\use\ite}=\sum_{\hid=1}^{\toth}\svdu_{\use \hid}\svdv_{\hid\ite}.
\end{equation}
After substitution of the above expression, we minimize the difference between $\data_{\use\ite}$ and $\tilde\data_{\use\ite}$ by differientiation of $(\data_{\use\ite}-\tilde\data_{\use\ite})^2$, namely
\begin{align}
\frac{\partial}{\partial \svdu_{\use\hid}}(\data_{\use\ite}-\tilde\data_{\use\ite})^2&=
-2\svdu_{\use\hid}\bigg(\data_{\use\ite}-
\sum_{\hid=1}^{\toth}\svdu_{\use\hid}\svdv_{\hid\ite}\bigg),\\
\frac{\partial}{\partial \svdv_{\hid\ite}}(\data_{\use\ite}-\tilde\data_{\use\ite})^2&=
-2\svdv_{\hid\ite}\bigg(\data_{\use\ite}-
\sum_{\hid=1}^{\toth}\svdu_{\use\hid}\svdv_{\hid\ite}\bigg).
\end{align}
Since the norm is non-negative by definition, we minimize the norm square which leads to the same result while encountering simpler expressions in the process. The obtained gradients can be used to write a gradient descent-based updating procedure for $\svdu_{\use \hid}$ and $\svdv_{\hid\ite}$ in the form
\begin{align}
\svdu_{\use\hid}(t+1)&=\svdu_{\use\hid}(t)+2\eta\svdu_{\use\hid}(t)e_{\use\ite}(t),\\
\svdv_{\hid\ite}(t+1)&=v_{\hid\ite}(t)+2\eta\svdv_{\hid\ite}(t)e_{\use\ite}(t),
\end{align}
where $t$ denotes the iteration step and
\begin{eqnarray}
e_{\use \ite}(t) = \data_{\use \ite}- \sum_{\hid=1}^{\toth}\svdu_{\use \hid}(t)\svdv_{\hid\ite}(t).
\end{eqnarray}
The learning rate $\eta>0$ should be small to avoid big jumps in the solution space. With random initial conditions on $\svdu_{\use\hid}$ and $\svdv_{\hid\ite}$, these equations are iterated until the squared norm shows no further decrease. Other procedures such as the variants of stochastic gradient descent may be applied to improve computational efficiency \cite{germulla11}. As suggested by \cite{takacs2007}, substracting a small term from the gradient can prevent large resulting weights of $\svdu_{\use \hid}$ and $\svdv_{\hid\ite}$ (this is equivalent to minimizing $\norm{\dataM-\tilde\dataM}^2+\frac{\lambda}{2}\norm{\svdU}^2+\frac{\lambda}{2}\norm{\svdV}^2$, or similar to using the Tikhonov regularization which gives preference to solutions with small norms in ill-posed problems), which leads to the following update rule
\begin{align}
\svdu_{\use \hid}(t+1)&=\svdu_{\use\hid}(t)+
\eta\big(2\svdu_{\use\hid}(t)e_{\use\ite}(t)-\lambda\svdu_{\use\hid}(t)\big),\\
\svdv_{\hid\ite}(t+1)&=\svdv_{\hid\ite}(t)+
\eta\big(2\svdv_{\hid\ite}(t)e_{\use\ite}(t)-\lambda\svdv_{\use\hid}(t)\big).
\end{align}
where a new parameter $\lambda\geq0$ is introduced to this end; $\lambda>0$ usually leads to better accuracy of the results. Using the resulting $\svdu_{\use \hid}$ and $\svdv_{\hid\ite}$, we can compute $\tilde\dataM=\svdU\svdV$ which has non-zero values on entries where the input matrix $\dataM$ are zero (representing unexpressed evaluations)---element $\tilde \data_{\use\ite}$ then predicts the possible rating given by user $\use$ to object $\ite$.

Note that while $\norm{\dataM - \tilde\dataM}$ measures the ``error'' of $\tilde\dataM$ with respect to $\dataM$, it is usually very different from (greater than) the error of the ultimate rating estimates. The reason for this, of course, lies in the fact that $\norm{\dataM - \tilde\dataM}$ is minimized while knowing $\dataM$, and the $\toth(\totu+\toti)$ free parameters usually allow us to achieve very low value of $\norm{\dataM - \tilde\dataM}$. To measure performance of this method correctly, the available data must be divided into a training set (which is used to ``learn'' $\svdU$ and $\svdV$) and a test set (see Sec.~\ref{sec:rec_intro} for details). Increasing $\toth$ does not automatically improve the results: over-fitting the data (providing too many free parameters) can lead to inferior accuracy. The use of this and other machine-learning methods hence requires a certain amount of tests and/or experience.

In addition to a simple iteration procedure, SVD also enjoys a flexibility in dealing with additional data. For instance, one can easily include the influence of individual rating bias in the framework. Suppose user $\use$ tends to give an average of $b_\use$ more score for all his items when compared to other users, while object $\ite$ tends to receive $b_\ite$ more scores when compared to other items, the predicted scores can be expressed as \cite{koren09}
\begin{equation}
\tilde\data_{\use\ite}=\mu+b_\use+b_\ite+\sum_{\hid=1}^{\toth}\svdu_{\use \hid}\svdv_{\hid\ite},
\end{equation}
where $\mu$ is the average value among all user-object pairs. In this case, $\norm{\dataM - \tilde\dataM}$ is minimized with respect to $b_\use$, $b_\ite$, $\svdu_{\use \hid}$ and $\svdv_{\hid\ite}$ for all $\use, \hid$ and $\ite$. Other than individual bias, there are other variants of SVD which utilizes the relations between users in social network, for instance the similarity in taste between friends, to improve the recommendation accuracy by the factorized matrices \cite{ma11}.

Apart from the case with user-object ratings as the only input data, the above-described procedure can be generalized to incorporate additional information \cite{takacs2007}. Suppose $d_{\use\ite}$ is the additional information (\eg date) associated with a given user-object pair. Transforming $d_{\use\ite}$ into positive integers $l$ with $1\le l\le L$, the elements $y_{kl}$ in the $\toth\times L$ matrix $Y$ contain the relation between each of these additional data and the hidden variables. For example, a large number of movies with romantic content are reviewed on the Valentine day, leading to a large value in the corresponding entries of romance and Valentine day in $Y$. With the additional information, the matrix $\tilde\dataM$ can be expressed as
\begin{eqnarray}
\tilde\data_{\use\ite}=\sum_{\hid=1}^{\toth}
\svdu_{\use \hid}\svdv_{\hid\ite}y_{\hid d_{\use\ite}}.
\end{eqnarray}
Similar derivations based on gradient descent provide updating procedures for
$\svdu_{\use \hid}$, $\svdv_{\hid\ite}$ and $y_{\hid, d_{\use\ite}}$ which can then be used to obtain $\tilde\dataM$ with the smallest error with respect to $\dataM$.

\subsection{Bayesian Clustering}
Before describing a probabilistic version of LSA, we first introduce the Bayesian clustering method  which is also probabilistic but simpler in formulation. In Bayesian networks, the value of a variable depends only on the value of its parent variables. For example, the probability of a variable $x$ is described by the conditional probability $P(x\vert\text{pa}_x)$ where $\text{pa}_x$ is the parent variable of $x$. The joint probability of several independent variables can be factorized as $P(x_1,\dots,x_N)=\prod_{i=1}^{N}P(x_i\vert\text{pa}_{x_i})$ which represents the dependency structure of the variables. However, obtaining the most relevant dependency structure, \ie the dependency relation between different nodes in the Bayesian network, is not a trivial task.

For the purpose of personalized recommendation, we describe a two-sided clustering \cite{ungar98, getoor99} which is easy to implement. To obtain the rating for an unobserved user-object pair, one classifies users and objects respectively into $\toth_{\rm user}$ and $\toth_{\rm object}$ classes. The values of $\toth_{\rm user}$ and $\toth_{\rm object}$ are parameters on the algorithm, similarly to $\toth$ which was a parameter of SVD. We assume that there is a simple Bayesian network that underlies the input data---the simplest assumption is that $\score_{\use\ite}$ depends only on the user's class $c_\use$ and the object's class $c_\ite$. The probability of $\score_{\use\ite}$ can then be written as
$$
P(\score_{\use\ite}) =
\sum_{c_\use=1}^{\toth_{\rm user}}\sum_{c_\ite=1}^{\toth_{\rm object}}
P(\score_{\use\ite}\vert c_\use,c_\ite)P(c_\use)P(c_\ite),
$$
To obtain an estimate of $\score_{\use\ite}$, we need to find $P(c_\use)$,  $P(c_\ite)$ and $P(\score_{\use\ite}\vert c_\use,c_\ite)$, which is effectively $P(\score|x,y)$ as the rating $\score$ is dependent merely on the user class $x$ and the object class $y$. This can be done by applying the inference methods including the marginal estimation by belief propagation \cite{pearl82, yedidia05} and likelihood maximization by expectation maximization \cite{gupta10}. Here we describe another simple scheme, known as the Gibbs sampling method \cite{geman84}, which is similar to the heat bath algorithm in statistical physics \cite{griffiths04, newman99b}. Gibbs sampling is useful when the joint distribution of all variables is difficult to sample (in our case, $P(\{\score\vert(x,y)\},\{c_\use\}, \{c_\ite\})$), while sampling the conditional probability of individual variables given all other variable is comparatively easy (\eg, $P(c_{\use'}\vert\{\score\vert(x, y)\}, \{c_\use\}_{-\use'}, \{c_\ite\})$). It is similar to the heat bath algorithm as it models the state of a system moving in a phase space and samples at certain time intervals the required (physical) quantities.

Here we describe the Gibbs sampling scheme suggested in \cite{ungar98} to sample the state $(\{\score\vert(x,y)\}, \{c_\use\}, \{c_\ite\})$ of the system which allows us to evaluate the predicted ratings for any unobserved user-object pair. This scheme was developed to sample binary ratings so we assume that $\score_{\use\ite}=1$ when user $\use$ has collected $\ite$ or rated object by a score above a certain threshold; $\score_{\use\ite}=0$ otherwise. In this case, one can represent $P(\score\vert x,y)$ by a single value variable $P_{xy}$ corresponding to the probability that a user from category $x$ likes an object from category $y$.

We start the algorithm with a random latent class assignment for all users and objects and evaluate $N_x$, $N_y$ and $N_{xy}$, respectively correspond to the number of users in class $x$, the number of objects in class $y$, and the number of observed user-object pairs (\ie, $\score_{\use\ite}=1$) for all pairs of classes. We then draw values of $P_{xy}$ from the beta distribution with parameters $(N_{xy}+1, N_x N_y-N_{xy}+1)$, or simply approximate $P_{xy}$ by its mean $P_{xy}=N_{xy}/N_xN_y$. Similary, the variables $P_x$ and $P_y$, respectively defined as the probability that a random user or object is classified in class $x$ or $y$, are drawn from Dirichlet distributions or simply approximated by $P_x=N_x/\totu$ and $P_y=N_y/\toti$. All these values of $P_{xy}$, $P_x$ and $P_y$ are used to evalulate the transition probability of the system from the present state to another state in the phase space. We then successively pick either a random user or a random object and update its latent class as \cite{ungar98}
\begin{eqnarray}
P(c_\use=x) \propto P_x\prod_{y=1}^{\toth_{\rm object}}
P_{xy}^{\sum_\ite \score_{\use\ite}\delta_{c_\ite, y}}
(1-P_{xy})^{\sum_\ite (1-\score_{\use\ite})\delta_{c_\ite, y}},
\end{eqnarray}
for users, and
\begin{eqnarray}
P(c_\ite=y) \propto P_y\prod_{x=1}^{\toth_{\rm user}}
P_{xy}^{\sum_\use \score_{\use\ite}\delta_{c_\use, x}}
(1-P_{xy})^{\sum_\use (1-\score_{\use\ite})\delta_{c_\use, x}}.
\end{eqnarray}
for objects. These equations involve large powers of probability values and may lead to inaccurate numerics during the computation. Instead of computating the probability direcly, one can first compute the powers and convert them to probability during the update of latent classes. The values of $N_x$, $N_y$ and $N_{xy}$, and thus of $P_x$, $P_y$ and $P_{xy}$, are updated after each update of user class or object class. After a sufficient number of iterations, we can start sampling estimated ratings of unobserved user-object pairs at regular time intervals which should be long enough to ensure low correlation between consecutive sampled states. For instance, the predicted rating for user $\use$ on object $\ite$ can be obtained by
\begin{eqnarray}
\tilde\score_{\use\ite} = \sum_{t} P_{xy}(t_{c}+tT)\delta_{x, c_\use(t_{c}+tT)}\delta_{u, c_\ite(t_{c}+tT)},
\end{eqnarray}
where $t_c$ and $T$ are respectively the convergence time and the sampling time interval. We can also store the state $(\{P_{xy}\}, \{c_\use\}, \{c_\ite\})$ at each sampling time and use the above equation to obtain the predictions afterwards.

We remark that the above two-sided Bayesian network corresponds to the simplest dependency structure which relates ratings to merely user and object classes. More comprehensive Bayesian relation can be derived, for instance, to include the individual rating preference for objects \cite{zhang07} or to model the possibility of mixed membership \cite{shan08}. Another class of extensions is to build a probabilistic relational model (PRM) \cite{getoor99} to predict ratings by utilizing other meta-data including age, occupation and gender of users, or category, price and origin of objects. Although this comprehensive information generally improves recommendation results, to determine a valid dependency structure of meta-data is a non-trivial task in PRM.

\subsection{Probabilistic Latent Semantic Analysis (pLSA)}
\label{sec:pLSA}
Probabilistic latent semantic analysis (pLSA) is similar to LSA in the sense that hidden variables are introduced in to explain the co-occurrence pairs of data. Unlike the algebraic SVD employed in LSA, pLSA is a statistical technique based on a probabilistic model. Well developed inference methods including likehood maximization \cite{gupta10} and Gibbs sampling \cite{geman84} can thus be employed in pLSA. pLSA models the relations between users and objects through the implicit overlap of genres, as compared to the two-sided Bayesian clustering where each user and object belong to a single specific category. In pLSA, the co-occurrence probability $P(\use,\ite)$ of user $\use$ and object $\ite$ is expressed using the conditional probability given a hidden variable $\hid$
\begin{equation}
\label{eq_phid}
P(\use,\ite) = \sum_{\hid=1}^{\toth}P(\use\vert\hid)P(\ite\vert\hid)P(\hid).
\end{equation}
Since $P(\use\vert\hid)P(\hid) = P(\hid\vert\use)P(\use)$, this can be written as
\begin{equation}
\label{eq_per}
P(\use,\ite) = P(\use)\sum_{\hid=1}^{\toth}P(\ite\vert\hid)P(\hid\vert\use)
\end{equation}
which leads to the condititonal probability $P(\ite\vert\use)$ of an object $\ite$ to be collected
given the user $\use$,
\begin{equation}
\label{6eq_person}
P(\ite\vert\use) = \sum_{\hid=1}^{\toth}P(\ite\vert\hid)P(\hid\vert\use),
\end{equation}
which is already a quantity useful for personalized recommendation. Unlike the Bayesian clustering approach where the co-ocurrences of users and objects are characterized by the coupled probability $P_{xy}$ between the classes, users and objects are rendered independent in pLSA given the hidden variables---the co-ocurrence probabilities are factorized. Our task is to obtain suitable forms of $P(\ite\vert\hid)$ and $P(\hid\vert\use)$ which provide accurate predictions of collected and recommended objects through $P(\ite\vert\use)$. We note that instead of $P(\ite\vert\use)$, $P(\use,\ite)$ can also be expressed as
\begin{equation}
P(\use,\ite) = P(\ite)\sum_{\hid=1}^{\toth}P(\use\vert\hid)P(\hid\vert\ite)
\end{equation}
to obtain $P(\use\vert\ite)$ which can be of interest for some purposes.

To obtain $P(\ite\vert\hid)$ and $P(\hid\vert\use)$, one can adopt a variational approch described in \cite{hofmann04} to maximize the per-link log-likelihood of the observed dataset which is given by
\begin{equation}
\label{6eq_L}
L(\bphi,\btheta) = \frac{1}{E}\sum_{(\use,\ite)}\log P(\ite\vert\use) =
\frac1E\sum_{(\use,\ite)}\log\bigg(\sum_{\hid=1}^{\toth}P(\ite\vert\hid)P(\hid\vert\use)\bigg)
\end{equation}
with respect to vectors $\bphi$ and $\btheta$ which parametrize $P(\ite\vert\hid)$ and $P(\hid\vert\use)$ by $P(\ite\vert\hid) = \phiki$ and $P(\hid\vert\use)=\thetauk$. Here $E$ is the total number of user-object links. We remark that the sum over $(\use,\ite)$ includes only the observed user-object pairs. We then employ the expectation maximization (EM) algorithms \cite{gupta10} to find the value of $\theta$ that maximize $L(\bphi,\btheta)$. To achieve the goal, one can introduce the variational probability distribution $Q(\hid\vert\use,\ite)$ in \req{6eq_L} for each observed pair $(\use,\ite)$, with the constraint $\sum_{\hid=1}^{\toth}Q(\hid\vert\use,\ite)=1$, which allows us to rewrite $L(\bphi, \btheta)$ as
\begin{align}
L(\bphi,\btheta) &= \frac{1}{E}\sum_{(\use,\ite)}\log\bigg(
\sum_{\hid=1}^{\toth} Q(\hid\vert\use,\ite)
\frac{P(\ite\vert\hid)P(\hid\vert\use)}{Q(\hid\vert\use,\ite)}\bigg)\\
&\geq \frac{1}{E}\sum_{(\use,\ite)}\sum_{\hid=1}^{\toth} Q(\hid\vert\use,\ite)
\log\frac{P(\ite\vert\hid)P(\hid\vert\use)}{Q(\hid\vert\use,\ite)}:=\cF(Q, \bphi, \btheta)
\end{align}
with the inequality is justified by Jensens' inequality. $\cF(Q,\bphi,\btheta)$ can be written as
\begin{equation}
\cF(Q,\bphi,\btheta) = \frac{1}{E}\sum_{(\use,\ite)}\bigg\{\sum_{\hid=1}^{\toth} Q(\hid\vert\use, \ite) \log[P(\ite\vert\hid)P(\hid\vert\use)]+S_{\use\ite}(Q)\bigg\}
\end{equation}
with $S_{\use\ite}(Q)$ being the entropy of the probability distribution $Q$ for the pair $(\use,\ite)$ which is
\begin{equation}
\label{6eq_S}
S_{\use\ite}(Q)=-\sum_{\hid=1}^{\toth}Q(\hid\vert\use,\ite)\log Q(\hid\vert\use,\ite).
\end{equation}
Since $\cF$ serves as the lower bound of likelihood function $L(\bphi,\btheta)$, we maximize $\cF$ with respect to $Q$, $\bphi$ and $\btheta$. The expectation maximization algorithm thus maximizes $\cF$ by finding the optimal $Q$, $\bphi$, $\btheta$ alternatively. We first obtain the distribution $Q$ which maximizes $\cF$ by assuming a particular form of $P(\ite\vert\hid)$ and $P(\hid\vert\use)$, \ie holding $\bphi$ and $\btheta$ constant. Maximization of $\cF$ in this step is subject to the normalization of $Q(k\vert\use,\ite)$ for every observed user-object pair which leads us to the Lagrangian
\begin{equation}
\cL(Q,\bphi_t,\btheta_t) = \cF(Q,\bphi_t,\btheta_t) +
\sum_{(\use,\ite)}\lambda_{\use\ite}\bigg(\sum_{\hid=1}^{\toth}Q(\hid\vert\use,\ite)-1\bigg)
\end{equation}
where $\bphi_t$ and $\btheta_t$ denotes respectively $\bphi$ and $\btheta$ after $t$ iteration steps, or equivalently, $P_t(\ite\vert\hid)$ and $P_t(\hid\vert\use)$. $\cL(Q, \bphi_t, \btheta_t)$ can then be differentiated to obtain the optimal $Q$ for every observed user-object pair in the form
\begin{equation}
\label{6eq_Q}
Q_t(\hid\vert\use,\ite)=\frac{P_t(\ite\vert\hid)P_t(\hid\vert\use)}
{\sum_{\hid'=1}^{\toth}P_t(\ite\vert\hid')P_t(\hid'\vert\use)}=
\frac{\phi^{(\hid)}_{\ite, t}\theta^{(\use)}_{\hid,t}}
{\sum_{\hid'=1}^{\toth}\phi^{(\hid')}_{\ite, t}\theta^{(\use)}_{\hid',t}}
\end{equation}
This optimal $Q$ is obtained from $\btheta_t$ and $\bphi_t$, thus we label it as $Q_t$.

We then proceed to obtain the distributions $P(\ite\vert\hid)$ and $P(\hid\vert\use)$, \ie the value of $\phi^{(\hid)}_\ite$ and $\theta^{(\use)}_\hid$, by assuming $Q$ is held fixed at $Q=Q_t$ obtained from \req{6eq_Q}. Due to the normalization of $P(\ite\vert\hid)$ for all $\ite$ and $P(\hid\vert\use)=1$ for all $\use$, the corresponding Lagrangian has the form
\begin{equation}
\cL(Q_t,\theta) = \cF(Q_t,\theta) +
\sum_{\hid}\lambda_{\hid}\bigg(\sum_{\ite=1}^{\toti}P(\ite\vert\hid)-1\bigg)+
\sum_{\use}\lambda_{\use}\bigg(\sum_{\hid=1}^{\toth}P(\hid\vert\use)-1\bigg).
\end{equation}
After differentiation, the optimal $P(\ite\vert\hid)$ and $P(\hid\vert\use)$ can be found
\begin{align}
\label{6eq_P}
P_{t+1}(\ite\vert\hid)&=\phi^{(\hid)}_{\use, t+1}=\frac{\sum_{\use}Q_t(\hid\vert\use,\ite)}{\sum_{\ite'}\sum_{\use}Q_t(\hid\vert\use,\ite')},\\
P_{t+1}(\hid\vert\use)&=\theta^{(\use)}_{\hid, t+1}=\frac{\sum_{\ite}Q_t(\hid\vert\use,\ite)}{\sum_{\hid'}\sum_{\ite}Q_t(\hid'\vert\use,\ite)}
\end{align}
where the summation involving $\use$ and $\ite$ runs only over the observed user-object pairs. The optimal $P(\ite\vert\hid)$ and $P(\hid\vert\use)$ are obtained from $Q=Q_t$. We label them as $P_{t+1}(\ite\vert\hid)$ and $P_{t+1}(\hid\vert\use)$, respectively, because they constitute the basis for the next iteration step where $Q_{t+1}$ is found. After stationary values of $Q_t$, $\bphi_t$ and $\btheta_t$ are found, \req{6eq_person} is used to obtain personalized recommendations.

As the above pLSA model considers a multinomial distribution $\phiki$ which can be used to model only binary preferences, one may consider the generlized pLSA \cite{hofmann03} which allows for numeric ratings. The fast increasing number of independent variables used by pLSA (there are $\toth(\totu+\toti)$ of them) and the cold-start problem for new objects can be alleviated by asssuming prior distributions on $\phik$ and $\thetau$, such as Dirichlet priors discussed in the following section \cite{blei03}.

\subsection{Latent Dirichlet Allocation (LDA)}
\label{sec:LDA}
Latent Dirichlet Allocation (LDA) \cite{blei03} is similar to pLSA in the sense that hidden variables are present in a probabilistic way. While pLSA does not assume a specific prior distribution over $P(\hid\vert\use)$, LDA assumes that priors that have the form of the Dirichlet distribution. LDA was applied to predict review scores based on the content of reviews \cite{blei07} and to uncover implicit community structures in a social network \cite{chen09}. It can be also extended to include general meta-data of users and objects \cite{agarwal10}.

\begin{figure}
\centering
\figg{0.9}{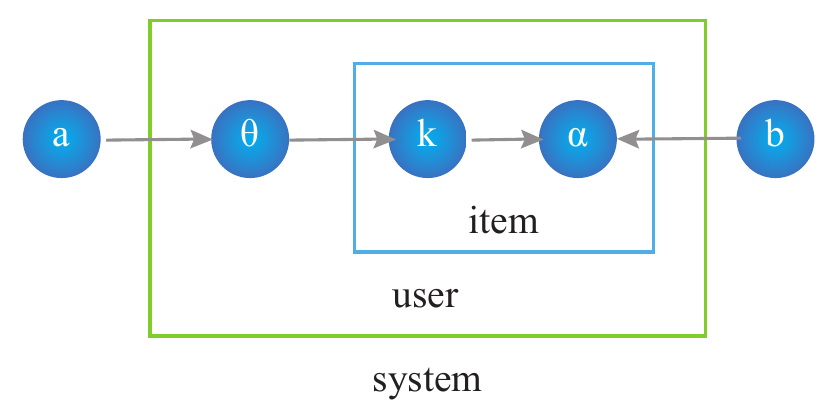}
\caption{A so-called plate notation representing the LDA recommendation model: $\dira$ and $\dirb$ represent the model's parameters at the system level, which characterize respectively the Dirichlet distribution of $\theta^{(\use)}_\hid$ and $\phi^{(\hid)}_\ite$, where $P(\hid\vert\use)=\theta^{(\use)}_\hid$ and $P(\ite\vert\hid)=\phi^{(\hid)}_\ite$.}
\label{gr_lda}
\end{figure}

For each user $\use$ there is a distribution $P(\hid\vert\thetau)$ where $\thetau$ is a $\toth$-dimensional multinomial distribution with $\thetauk$ is the probability of user $\use$ belonging to the latent class $\hid$, \ie $P(\hid\vert\use)=\thetauk$. Unlike pLSA, the variable $\thetau$ in LDA has a Dirichlet prior distribution with a $\toth$-dimensional parameter $\dira$. The probability of observing user $\use$ with the collected object set $\{\ite\}_\use$ is
\begin{eqnarray}
\label{6eq_lda}
P(\{\ite\}_\use\vert\dira,\bphi) = \int\dd\thetau P(\thetau\vert\dira)\bigg[
\prod_{\co=1}^{k_\use}\sum_{\hid=1}^\toth P(\ite_{\use,\co}\vert\hid)P(\hid\vert\thetau)\bigg].
\end{eqnarray}
where $k_\use$ is the number of objects collected by user $\use$ and $\ite_{\use,\co}$ is the $\co$-th object collected by $\use$. The Dirichlet prior distribution $P(\thetau\vert\dira)$ is
\begin{equation}
P(\thetau\vert\dira)=\frac{\Gamma\big(\sum_{\hid=1}^\toth\diraa_\hid\big)}{\prod_{\hid=1}^\toth\Gamma(\diraa_\hid)}
\prod_{\hid=1}^\toth\big(\thetauk\big)^{\diraa_\hid-1},
\end{equation}
where $\Gamma(x)$ is the Gamma function, the constraint $\sum_{\hid=1}^\toth\thetauk=1$ holds and $\theta_{\use,\hid}>0$ for all $\use$ and $\hid$.  Prior distributions for all $\theta^{(\use)}$ share the same parameter $\dira$. The LDA model can be represented by a so-called plate notation which is shown \fig{gr_lda}. The probability of the observed data $\{(\use,\ite)\}$,
\begin{eqnarray}
P(\{(\use,\ite)\}\vert\dira,\bphi) = \prod_{\use}\int\dd\thetau P(\thetau\vert\dira)\bigg[
\prod_{\co=1}^{k_\use}\sum_{\hid=1}^\toth P(\ite_{\use,\co}\vert\hid)P(\hid\vert\thetau)\bigg],
\end{eqnarray}
depends on the parameter vectors $\dira$ and $\dirb$.

In the original formulation of LDA \cite{blei03}, $P(\ite\vert\hid)$ is given by a multinomial distribution parametrized by $\bphi$ such that $P(\ite\vert\hid)=\phi^{(\hid)}_\ite$ as in the case of pLSA. Some variants of LDA consider $P(\ite\vert\hid)$ following a Dirchlet prior distribution, which is known as a smoothed version of LDA \cite{wei06}. In this case, $\dirb$ which characterizes the prior distribution of $P(\ite\vert\hid)=\phi^{(\hid)}_\ite$ in \req{6eq_lda} corresponds to a $\toti$-dimensional parameter of the Dirichlet prior distribution
\begin{equation}
P(\phik\vert\dirb)=\frac{\Gamma\big(\sum_{\ite=1}^\toti\dirbb_\ite\big)}{\prod_{\ite=1}^\toti\Gamma(\dirbb_\ite)}\prod_{\ite=1}^\toti\big(\phiki\big)^{\dirbb_\ite-1}.
\end{equation}
Prior distributions for all $\phi^{(\hid)}$ share the same parameter $\dirb$.

In order to obtain rating predictions for unobserved user-object pairs, one has to find $P(\{\phik\}, \{\thetau\}\vert\{(\use\ite)\}, \dira, \dirb)$ and use $\{\phik\}$ and $\{\thetau\}$ to make personalized predictions for user $\use$. For instance, the predicted score for an unobserved pair $(\use,\ite)$ is given by $\tilde\score_{\use\ite}=\sum_{\hid=1}^\toth \phiki\thetauk$. Since the distribution of $\{\phik\},\{\thetau\}$ is in general intractable, one can follow \cite{blei03} and adopt a variational approach to maximize likelihood---similarly as we did in the case of pLSA. Here we describe the Gibbs sampling for the smoothed LDA as an alternative inference method \cite{griffiths04}. The procedures are similar to the Gibbs sampling in Bayesian networks, except that one assigns a latent class for each user-object pair, instead of classes for individual users and objects.

To derive an equation for Gibbs sampling in LDA, we first assign an index $\co$ for each observed user-object pair $(\use,\ite)$ so that $\hid_\co$ is a latent variable drawn from the multinomial distribution $\theta^{(\use_\co)}$ and $\ite_\co$ is an object drawn from the multinomial distribution $\phi^{(\hid_\co)}$. As shown in \cite{griffiths04,griffiths02}, the inference method is much simplified by assuming a symmetry Dirichlet prior with homogeneous $\dira$ and $\dirb$, \ie $\diraa_1=\cdots=\diraa_\toth:=\diraa$ and $\dirbb_1=\cdots=\dirbb_\toti:=\dirbb$. Then one can show that the conditional probability for an observed user-object pair $\co'$ characterized by $\hid_{\co'}$ is given by
\begin{align}
\label{6eq_ldagibbs}
P\left(\hid_{\co'}=\hid\vert\{\hid_\co\}_{-\co'}, \{(\use_\co,\ite_\co)\}\right)
&\propto P\left(\ite_{\co'}\vert\hid_{\co'}=\hid, \{\hid_\co\}_{-\co'}, \{(\use_\co,\ite_\co)\}_{-\co'}\right)
P\left(\hid_{\co'}=\hid\vert \{\hid_\co\}_{-\co'}\right)\nonumber\\
&\propto \frac{n^{(\hid)}_{-\co', \ite_{\co'}}+\dirbb}{n^{(\hid)}_{-\co'}+\toti\dirbb}\,
\frac{n^{(\use_{\co'})}_{-\co', \hid}+\diraa}{n^{(\use_{\co'})}_{-\co'}+\toth\diraa}
\end{align}
where all $n_{-\co'}$'s are evaluated in the absence of pair $\co'$: $n^{(\hid)}_{-\co'}$ is the number of observed pairs charaterized by latent class $\hid$, $n^{(\hid)}_{-\co', \ite}$ is the number of observed pairs of object $\ite$ charaterized by latent class $\hid$, $n^{(\use)}_{-\co'}$ is the number of observed pairs of user $\use$ (degree of $\use$) and $n^{(\use)}_{-\co', \hid}$ is the number of observed pairs with user $\use$ characterized by latent class $\hid$.

The Gibbs sampling process runs as follows. We first start with a random assignment of latent class to each observed user-object pair and successively pick a random user-object pair to update its latent class according to \req{6eq_ldagibbs}. This corresponds to a shift of the system state from one to another; $n^{(\hid)}$, $n^{(\hid)}_{\ite_{\co'}}$, $n^{(\use_{\co'})}$ and $n^{(\use_{\co'})}_{\hid}$ are updated after each new assignment of the latent class. After a sufficient number of iterations, one can sample $\phik$ and $\thetau$ at a regular time interval
\begin{equation}
\phiki = \frac{n^{(\hid)}_{\ite}+\dirbb}{n^{(\hid)}+\toti\dirbb},\quad
\thetauk = \frac{n^{(\use)}_{\hid}+\diraa}{n^{(\use)}+\toth\diraa}.
\end{equation}
With these samples of $\phik$ and $\thetau$, the predicted score for an unobserved user-object pair can be computed as
\begin{eqnarray}
\tilde\score_{\use\ite}=\sum_t \sum_{\hid=1}^\toth \phiki(t_c+tT)\thetauk(t_c+tT),
\end{eqnarray}
where $t_c$ and $T$ are respectively the convergence time and the sampling time interval. As in the Gibbs sampling of Bayesian clustering, states of $\phik$ and $\thetau$ can be stored and use to compute the predicted scores later. When the input data is large, one can distribute the Gibbs sampling to several processors to shorten the computation time \cite{dnewman07}.

                  % checked, done
\section{Diffusion-based methods}
\label{sec:diffusion}
Similarly as the classical PageRank algorithm~\cite{Brin1998} brought the order to the Internet by analyzing the directed network of links among web pages, one could aim to obtain recommendations using a network representation of the input data with user preferences. The algorithms presented in this section are all based on specific transformations (projections) of the input data to object-object networks. Personalized recommendations for an individual user are then obtained by using this user's past preferences as ``sources'' in a given network and propagating them to yet unevaluated objects.

\subsection{Heat diffusion algorithm (HDiff)}
\label{sec:heat-diff}
This algorithm is based on projecting the input data on a simple object network characterized by a symmetric adjacency matrix $\mathsf{A}$ with elements either one (for similar objects) or zero (for dissimilar objects). It recommends objects to an individual user by a process motivated by heat diffusion: objects liked and disliked by this user are represented as hot and cold spots respectively, and recommendation is made according to the equilibrium ``temperature'' of the nodes in the networks~\cite{ZhBlYu07}. The discrete Laplace operator of the network has the form $\mathsf{L}=\mathsf{1}_N-\mathsf{D}^{-1}\mathsf{A}$ where $\mathsf{D}$ is the network's diagonal degree matrix with elements $D_{\alpha\beta}=k_{\alpha}\delta_{\alpha\beta}$. This operator is a discrete analog of the heat diffusion operator $-\nabla^2$ which is well-known in physics. The resulting temperature vector for user $i$, $\vek{h}_i$, is the solution of the heat diffusion equation
\begin{equation}
\label{heat-diff-eq}
\mathsf{L}\vek{h}_i=\vek{f}_i
\end{equation}
and has both variable part (which we seek) and fixed part. Fixed elements of $\vek{h}_i$ correspond to objects already evaluated by user $i$; they are set to $1$ (objects liked by the user---they act as heat sources) or $0$ (objects disliked by the user---they act as heat sinks). Mathematically this corresponds to the Dirichlet boundary condition. The external flux vector $\vek{f}_i$ is non-zero only for objects evaluated by user $i$ and allows for fixed values attributed to sources and sinks. \req{heat-diff-eq} can be solved using the Green's function method and the involved computational cost can be lowered by utilizing various algebraical properties of $\mathsf{L}$~\cite{ZhBlYu07}. At the same time, it is straightforward to find the equilibrium $\vek{h}_i$ iteratively by setting the initial temperature vector $\vek{h}_i^{(0)}$ to contain only the fixed heat sources and sinks and iterate
\begin{equation}
\label{heat-diff-iter}
\vek{h}_i^{(n+1)}=\mathsf{L}'\vek{h}_i^{(n)}
\end{equation}
where $\mathsf{L}'_i$ is the same as the Laplace operator above except that it keeps elements in $\vek{h}_i$ corresponding to $i$'s evaluated objects unchanged.

Given this mathematical framework, it is still an open question how exactly to apply it to a given rating matrix $\mathsf{R}$. The procedure adopted in~\cite{ZhBlYu07} is that the Pearson's correlation coefficient for ratings of objects $\alpha$ and $\beta$, $C_{\alpha\beta}$, is compared with a specified threshold $C_t$ and $A_{\alpha\beta}=1$ when $C_{\alpha\beta}\geq C_t$ and it is zero otherwise. The threshold $C_t$ is set so that the resulting number of links is the same as the number of non-zero entries in $\mathsf{R}^T\mathsf{R}$ (which is equivalent to the number of object pairs co-evaluated by at least one user). The boundary condition for user $i$ is composed of the ratings given by this user to different objects (note that the heat diffusion equations above are not constrained to the binary case like/dislike-hot/cold and can be used to arbitrary-valued ratings).

\subsection{Multilevel spreading algorithm (MultiS)}
\begin{figure}
\centering
\figg{0.7}{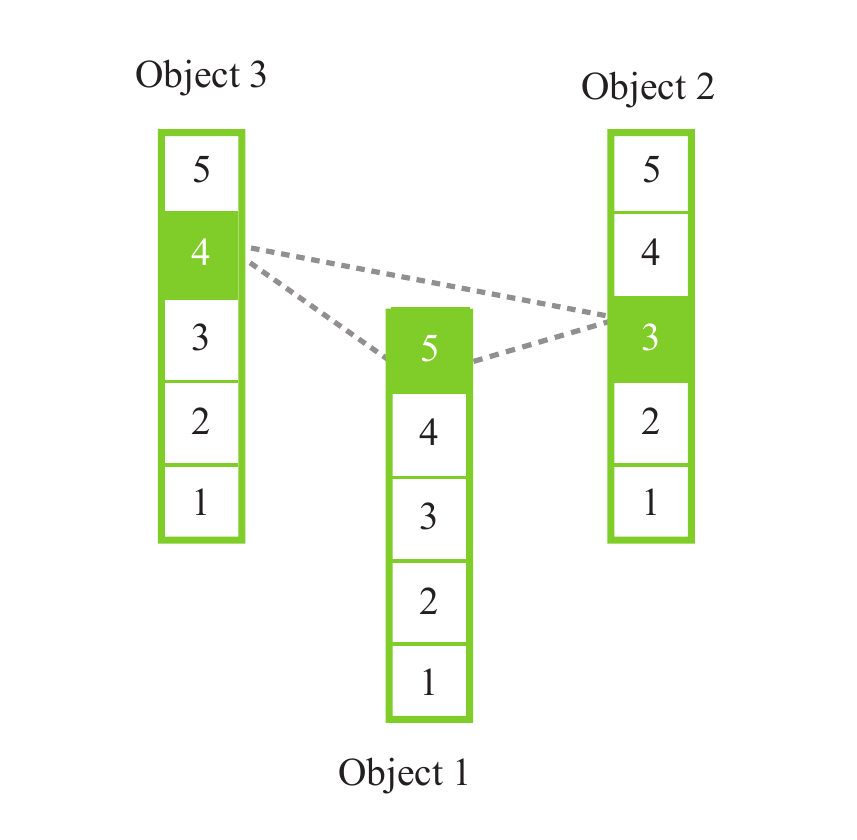}
\caption{Graphical representation of the links created by a user who has rated only objects 1 (rating 5), 2 (rating 3) and 3 (rating 4).}
\label{fig:QD-illust}
\end{figure}

This algorithm can be applied when ratings $r_{i\alpha}$ are given in a discrete scale.\footnote{It is also possible to apply a binning procedure to continuous-valued ratings and hence transform them into an integer scale but that has not been used in practice yet.}
For example, Amazon.com employs a five-star scale where one and five stars correspond to the worst and best rating, respectively. For the sake of simplicity, we assume a five-level rating scale in the rest of this subsection (generalization to a different number of levels is straightforward). As in other diffusion-based methods, the recommendation process starts with the preparation of a particular object-object projection of the rating data. To eliminate the loss of information in the projection, instead of merely creating a link between two objects, in this multilevel spreading algorithms, links are created between ratings given to a pair of objects~\cite{ZMRZLY07}. As a result we obtain $5^2=25$ separate connections (channels) for each object pair. This is illustrated in fig.~\ref{fig:QD-illust} on an example of a user who has rated three movies; as a result, three links are created between the given movies. When all data are processed, contributions from all users accumulate and a weighted object-object network is created. Note that splitting the connection between two objects into multiple separate channels aggravates the data sparsity problem and can lead to inferior performance of this algorithm in some cases.

Between a given pair of objects we create multiple links which are conveniently stored in a $5\times5$ matrix. By representing integer ratings $r_{i\alpha}$ with column vectors in 5 dimensional space (unknown rating with $\vek{v}_{i\alpha}=(0,0,0,0,0)^T$, rating $r_{i\alpha}=1$ with $\vek{v}_{i\alpha}=(1,0,0,0,0)^T$, rating $r_{i\alpha}=2$ with $\vek{v}_{i\alpha}=(0,1,0,0,0)^T$, etc.), connection matrix for objects $\alpha$ and $\beta$ has the form
\begin{equation}
\label{Wmatrix-QD}
\mathsf{W}_{\alpha\beta}=\sum_{i=1}^M
\frac{\vek{v}_{i\alpha}\vek{v}_{i\beta}^T}{k_i-1}.
\end{equation}
Weights of individual users are inversely proportional to their number of evaluations---this aims to compensate the quadratically growing number of links created by an individual user, $k_i(k_i-1)/2$, hence in total we get a linear relation between user's number of evaluations and the cumulative weight of user's ratings.\footnote{Since the users who have evaluated only one object add no links to the object-object network, the divergence of the weight $1/(k_i-1)$ at $k_i=1$ is not an obstacle.}

Matrices $\mathsf{W}_{\alpha\beta}$ form a symmetric matrix $\mathsf{W}$ with dimensions $5N\times5N$. By the column normalization of $\mathsf{W}$ we obtain an asymmetric matrix $\Omega$ which describes a diffusion process on the underlying network with the outgoing weights from any node in the graph normalized to unity (if one chooses the row normalization instead, the resulting process is equivalent to heat conduction in the network; for a mathematically-oriented review of flows in networks see~\cite{StYu07}). Elements with large weights in $\Omega$ represent strong patterns in user ratings (\eg, most of those who rated movie X with 5 gave 3 to movie Y). Similarly as for other diffusion-based methods, we obtain personal recommendations for user $i$ by combining the aggregate matrix $\Omega$ with opinions already expressed by $i$. This opinions are stored in a $5N$-dimensional vector $\vek{h}_i^{(0)}$ (the first 5 elements correspond to object 1, next 5 elements to object 2, etc.). As in Sec.~\ref{sec:heat-diff}, we seek the stationary solution of the equation
\begin{equation}
\label{QD-eq}
\Omega_i\vek{h}_i=\vek{h}_i,
\end{equation}
where $\Omega_i$ is the same as $\Omega$ except it keeps the elements corresponding to the objects evaluated by user $i$ unchanged. Resulting vectors $\vek{h}_i^{(n)}$ contain information about objects unrated by user $i$. This information can be used to obtain rating predictions by the standard weighted average. For example, if for a given object in $\vek{h}_i$ we obtain the 5-tuple $(0.0,0.2,0.4,0.4,0.0)^T$, the rating prediction is $0.2\times 2+0.4\times 3+0.4\times 4=3.2$. Numeric tests presented in~\cite{ZMRZLY07} suggest that $\vek{h}_i^{(1)}$ is a good enough predictor. Sophisticated techniques to avoid multiple iterations in \req{QD-eq} \cite{ZMRZLY07} are hence not fundamental for practical applications of this algorithm. An alterative way is to map each object to several channels with the number of channels being equal to the number of different ratings. So that if a user $i$ has collected an object $\alpha$ with rating 2, her will only connect to $\alpha^{(2)}$. After that, one can directly apply the probabilistic spreading process (see the next subsection) to obtain the similarity and then integrate it into the collaborative filtering framework to obtain better recommendation \cite{Shang2009}.

\subsection{Probabilistic spreading algorithm (ProbS)}
\label{sec:spreading_algo}
This algorithm is suitable for data without explicit ratings, i.e. only the sets of object collected/visited by each user is known. Elements of the rating matrix $\mathsf{R}$ are hence $r_{i\alpha}=1$ (when user $i$ has collected/visited object $\alpha$) or $r_{i\alpha}=0$ (otherwise). More explicit preference indicators can be easily mapped to this form, albeit losing information in the process, whereas the converse is not so.

\begin{figure}
\centering
\figg{0.6}{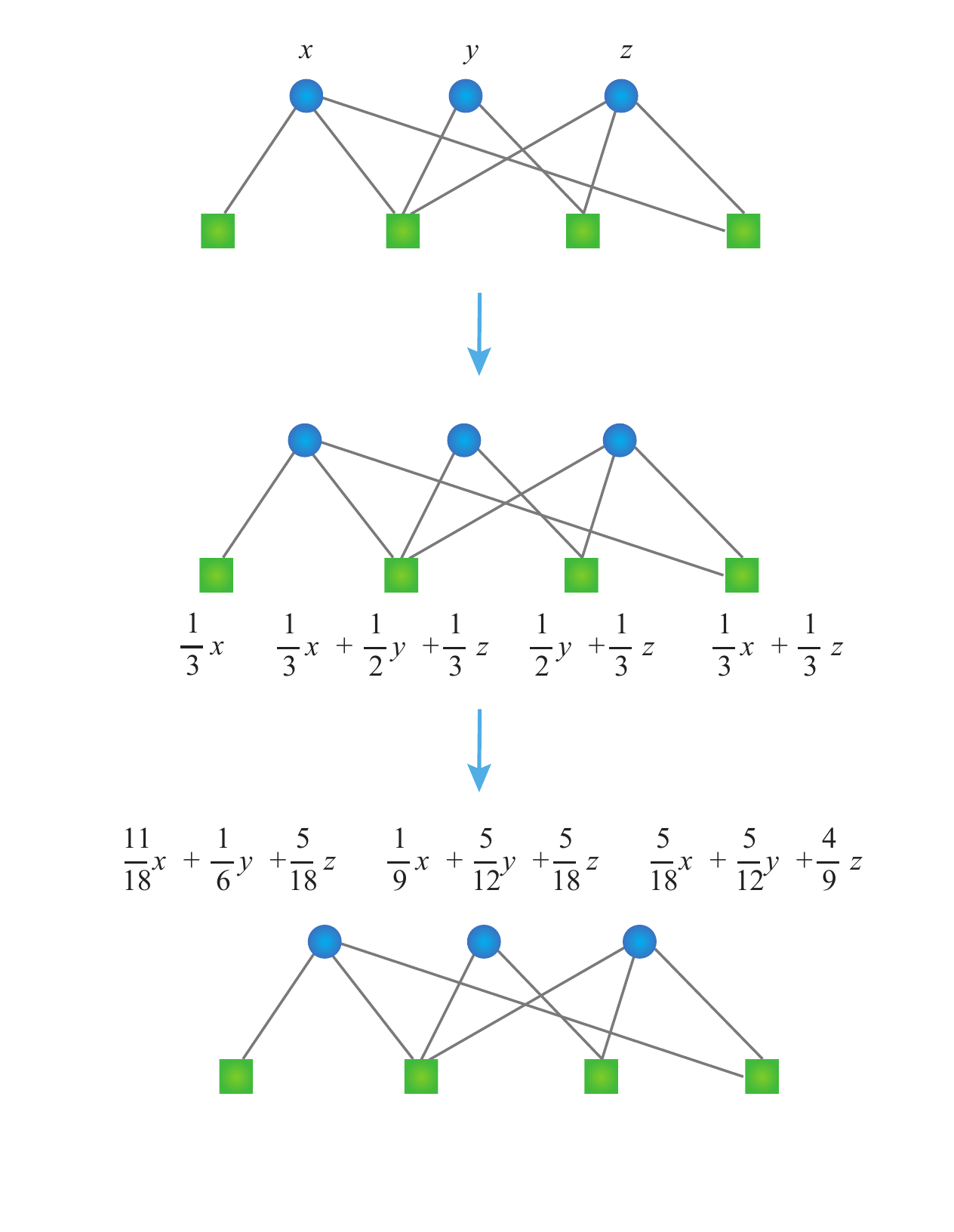}
\caption{Illustration of the ProbS's resource-allocation process in a simple bipartite network. The assigned resources first flow from object-nodes (circles) to user-nodes (squares) and then return back to object-nodes.}
\label{fig:diff_steps}
\end{figure}

The spreading recommendation algorithm proposed in~\cite{Zhou2007} is based on a projection of the input data (which can be represented by an unweighted user-object network) to an object-object network. In this projection, the weight $W_{\alpha\beta}$ can be considered as the importance of node $\alpha$ with respect to node $\beta$ and in general it differs from $W_{\beta\alpha}$. A suitable form of $W_{\alpha\beta}$ can be obtained by studying the original bipartite network where a certain amount of a resource (a scalar quantity which reflects, for example, social influence in a recommender system) is assigned to each object node. Since the network is unweighted, the unbiased allocation of the initial resource is split equally among all its neighboring user-nodes. Consequently, resources collected by user-nodes are equally redistributed back to their neighboring object-nodes. This is equivalent to random walk from the initial source nodes to a distance of two in the user-object bipartite graph. An illustration of this resource-allocation process for a simple bipartite network is shown in Fig.~\ref{fig:diff_steps}.

Denoting the initial object resource values as $x_{\alpha}$, the two resource-distribution steps can be merged to one and the final resource values read $\tilde{x}_{\alpha}=\sum_{\beta=1}^N W_{\alpha\beta}^P x_{\beta}$ where
\begin{equation}
\label{Wmatrix-ProbS}
W_{\alpha\beta}^P=\frac{1}{k_{\beta}}\sum_{i=1}^M
\frac{r_{i\alpha}r_{i\beta}}{k_i}.
\end{equation}
The superscript $P$ stands for ``probabilistic'' and serves to distinguish the current spreading process from its modifications that we shall discuss later. Note that the resulting $N \times N$ transition matrix is column normalized with $W_{\alpha\beta}^P$ representing the fraction of the initial $\beta$'s resource transferred to $\alpha$. Recommendations for a given user $i$ are obtained by setting the initial resource vector $\vek{h}^i$ in accordance with the objects the user has already collected, that is, by setting $h_{\beta}^i=r_{i\beta}$. Recommendation scores of objects are then obtained by
$$
\tilde{h}_{\alpha}^i=\sum_{\beta=1}^N
\mathsf{W}_{\alpha\beta}^P h_{\beta}^i.
$$
Objects recommended to user $i$ are then selected according to $\tilde{h}_{\alpha}^i$ (the higher the value, the better).

The original ProbS algorithm has been later improved in various directions. In \cite{Zhou2008}, the authors suggested a heterogeneous distribution of the initial resources among the nodes, $h_\alpha=a_{i\alpha} k_{\alpha}^\theta$, and showed that when the parameter $\theta$ is tuned appropriately, it can help increase the accuracy of recommendations and it also makes the recommendations more personalized. The optimal value of $\theta$ is close to -1 (e.g., -0.8 to -1.0 for MovieLens, depending on the size of the selected data set \cite{Zhou2008,Jia2008}), indicating that each item should be assigned more or less the same amount of total initial resource. In \cite{Zhou_NJP}, it was proposed to construct the transition matrix as $\mathsf{W}+\eta\mathsf{W}^2$ where $\mathsf{W}$ is defined by \req{Wmatrix-ProbS} and $\eta$ is a free parameter. By effectively removing redundant correlations (the optimal value of $\eta$ is usually negative), this method succeeded in outperforming ProbS and other derived methods in terms of accuracy and diversity of recommendations. Similar method can also be applied in designing more accurate similarity index for collaborative filtering \cite{LiuJG2010} and link prediction \cite{Lu2009}. In \cite{LiuJG2009a}, they proposed to increase the method's accuracy by giving preference to objects with degree similar to the average degree of objects collected by a given user. In addition, the degree correlation \cite{LiuJG2009b}, users' tastes \cite{LiuJG2009c}, user behavior patterns \cite{ZhangCJ2012} can also be accounted to improve the recommendation accuracy.

\begin{figure}
\centering
\figg{0.6}{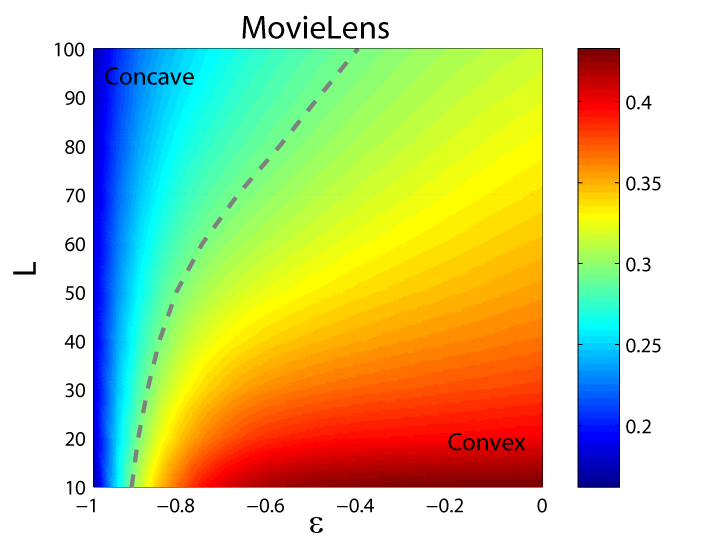}
\figg{0.6}{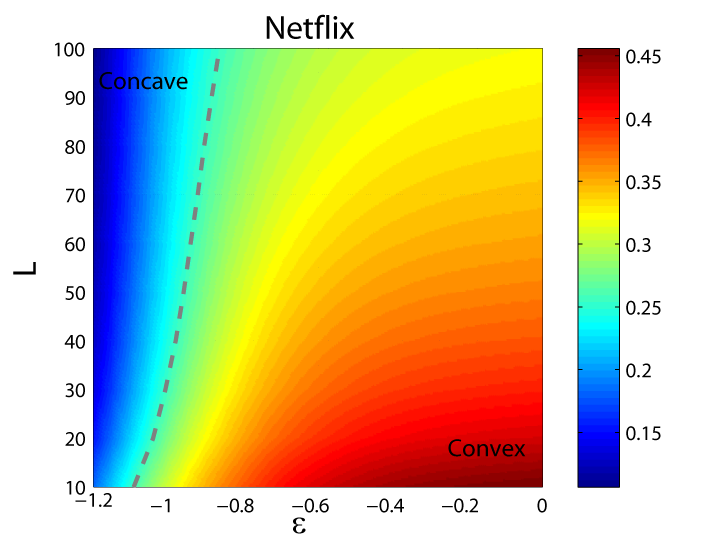}
\caption{(Color online) Intra-similarity $I(L)$ (see Eq.~21) for MovieLens and Netflix. With the parameter combination $(\varepsilon,L)$ along this line, the intra-similarity equals the value of the system.}
\label{fig:ly}
\end{figure}

Finally, we introduce a preferential diffusion method, which was proposed to enhance the algorithm's ability to find unpopular and niche objects \cite{PRE066119}. The basic idea is that at the last step (\ie, diffusion from users to objects), the amount of resource that object $\alpha$ receives is proportional to $k_{\alpha}^\varepsilon$ where $\varepsilon\leq 0$ is a free parameter. When $\varepsilon=0$, this method is identical to ProbS. It was shown that PD not only provides more accurate recommendations than ProbS but it also generates more diverse and novel recommendations by recommending relevant unpopular items. The authors further compared the intra-similarity of recommended items with that of the whole system. As shown in Fig.~\ref{fig:ly}, they draw a line that divides the parameter space into two phases: In the left region, especially the area corresponding to smaller $\varepsilon$ and larger $L$, PD is like a concave lens that broadens the user's vision, while in the right region, corresponding to larger $\varepsilon$ and smaller $L$, PD likes a convex lens that narrows the user's vision. Of course, we prefer the former case since it embodies the merit of personalization. 

Note that the resource-allocation process can also be applied in unipartite networks. Considering a pair of nodes $i$ and $j$, $i$ can send some resource to $j$ with their common neighbors playing the role of transmitters. In the simplest case, we assume that each transmitter has a unit of resource, and will equally distribute it to all neighbors. This defines a similarity index (called \emph{resource allocation index} \cite{Zhou2009}) between nodes:
\begin{equation}
s_{ij}=\sum_{l \in \Gamma_i\cap \Gamma_j}\frac{1}{k_l},
\end{equation}
where $\Gamma_i$ denotes the set of $i$'s neighboring nodes. Recent works showed that despite its simplicity, this index performs better than many known local similarity indices in link prediction \cite{Zhou2009}, community detection \cite{PanY2010}, and the characterization of weighted transportation networks \cite{WangYL2009}.

\subsection{Hybrid spreading-relevant algorithms}
\label{sec:hybrid_spreading}

\begin{figure}
\centering
\figg{0.8}{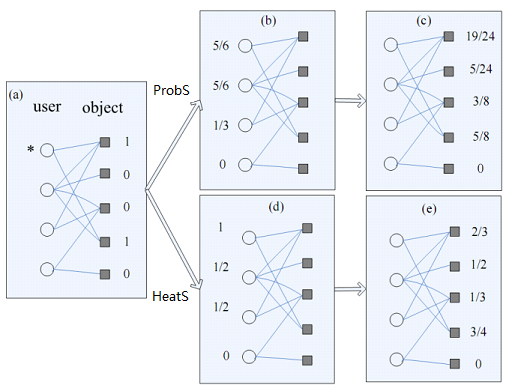}
\caption{(Color online) Comparison of ProbS and HeatS, where the target user is marked by a star and the collected objects are of initial resource 1. The final scores after ProbS and HeatS are listed in the right sides of plots (c) and (e).}
\label{fig:ph}
\end{figure}

To answer the need of diversity in algorithm-based recommendation (see Section \ref{sec:challenges} for a discussion of this problem and possible solutions), a hybrid algorithm was proposed in~\cite{Zhou_PNAS} which combines accuracy-focused ProbS with diversity-favoring heat spreading. As in probabilistic spreading, heat spreading works by assigning objects an initial level of ``resource'' denoted by the vector $\boldsymbol{h}$, and then redistributing it via the transformation $\tilde{\boldsymbol{h}}=\mathsf{W}^H\boldsymbol{h}$. The transition matrix of heat spreading reads
\begin{equation}
\label{Wmatrix-HeatS}
W_{\alpha\beta}^H=\frac1{k_{\alpha}}\sum_{i=1}^M
\frac{r_{i\alpha}r_{i\beta}}{k_i}
\end{equation}
which, in contrast to $\mathsf{W}^P$ obtained with \req{Wmatrix-ProbS}, is row-normalized and corresponds to a heat diffusion process (thus named HeatS) on the given user-object network.

Figure~\ref{fig:ph} illustrates the procedures of ProbS and HeatS. According to the final scores, ProbS will recommend the third object to the target user, while HeatS will recommend the second. Generally speaking, HeatS is able to find out unpopular (i.e., of low degree) objects, yet ProbS tends to recommend popular objects and thus lacks diversity and novelty. On the other hand, recommendations obtained by HeatS are too peculiar to be useful\footnote{Compared with the similarity-based methods and ProbS, the AUC value and precision of HeatS are considerably lower. Therefore, using HeatS alone seems not proper. Recent works \cite{LiuJG2011a,LiuJG2011b} indicate that some weighted version of HeatS could also give highly accurate recommendation.}. To integrate the advantages from both two algorithms, in~\cite{Zhou_PNAS} they proposed an elegant hybrid of $\mathsf{W}^H$ and $\mathsf{W}^P$ (named HybridS) in the form
\begin{equation}
\label{Whybrid1}
W_{\alpha\beta}^{H+P} =
\frac1{k_{\alpha}^{1-\lambda}k_{\beta}^{\lambda}}
\sum_{i=1}^M \frac{r_{i\alpha}r_{i\beta}}{k_i},
\end{equation}
where $\lambda=0$ gives pure heat spreading and $\lambda=1$ gives pure probabilistic spreading. As before, the resulting recommendation scores for user $i$ are computed as $\tilde{h}_{\alpha}^i=\sum_{\beta=1}^N W_{\alpha\beta}^{H+P} h_{\beta}^i$ where the initial resource values are set as $h_{\beta}^i=r_{i\beta}$. Results shown in~\cite{Zhou_PNAS} show that this combination of two different algorithms allows us not merely to compromise between diversity and accuracy but to simultaneously improve both aspects. By tuning the degree of hybridization, represented by the parameter $\lambda$, the algorithms can be tailored to many custom situations and requirements. Note that, the parameter $\lambda$ is not necessarily to be the same for different users and objects, namely each user $i$ can have her own parameter $\lambda_i$ or each object $\alpha$ can have its own parameter $\lambda_\alpha$, in this way, the algorithmic performance can be further improved \cite{Qiu2011}. And the initial resource distribution is not necessarily to be homogeneous, and by introducing the heterogeneity, the algorithmic performance can be improved \cite{LiuC2010}.

Similar to the Hybrid spreading algorithm, \emph{B-Rank} combines a random walk process with heat diffusion but it does so for data containing explicit ratings~\cite{Blattner09}. The transition matrix is introduced in the form
\begin{equation}
P_{\alpha\beta}=\frac{1-\delta_{\alpha\beta}}{n_{\alpha}}
\sum_{i=1}^M w_i x_{i\alpha}x_{i\beta}
\end{equation}
where $x_{i\alpha}=1$ if user $i$ has rated object $\alpha$ and $x_{i\alpha}=0$ otherwise, $w_i$ is the weight of user $i$ and $n_{\alpha}$ is a term which makes the matrix $\mathsf{P}$ row-normalized. User weights $w_i$ are set all identical but can be potentially set heterogeneous, for example, to give more weight to reliable users or suppress spammers (these possibilities have not been studied yet). Due to the term $1-\delta_{\alpha\beta}$, $P_{\alpha\alpha}=0$ and the corresponding random walk on the weighted object-object network is thus non-lazy (there is no possibility to return to the initial node after one step). Using the vector with ratings of user $i$, $h_{\alpha}^i=r_{i\alpha}$, object scores corresponding to the forward and backward propagation of $\vek{h}^i$ are computed as $\vek{F}^i=\mathsf{P}^T\vek{h}^i$ and $\vek{B}^i=\mathsf{P}\vek{h}^i$ (forward and backward propagation correspond to random walk and heat diffusion, respectively---for details see~\cite{StYu07}). The final score of object $\alpha$ is obtained as $f_{\alpha}^i=F_{\alpha}^iB_{\alpha}^i$ (the higher, the better). Note that this algorithm does not aim at predicting missing scores (hence the traditional measures of recommender systems, MAE and RMSE cannot be applied to evaluate it). Instead, it provides a personalized ranking (hence the name, 'B-Rank') of objects for each user.

The above-mentioned diffusion processes can be applied in computing the similarity between users or items, and then integrated into the similarity-based methods. Liu \emph{et al.} \cite{LiuJG2009a} defined the similarity between users according to ProbS\footnote{Similar to Eq.~\ref{Wmatrix-ProbS}, but the spreading process starts from user side and ends at user sides.}, and showed that based on the routine collaborative filtering algorithm, the proposed similarity index improved the algorithmic performance compared with the Pearson similarity index. Pan \emph{et al.} \cite{Pan2010} applied the HybridS process to define similarity, which outperforms the cosine similarity under the framework of collaborative filtering.

 % checked, done
\section{Social filtering}
\label{sec:social}
Recommendations made by a recommender system are often less appreciated than those coming from our friends \cite{Sinha2001} and social influences may play a more important role than similarity of past activities \cite{Salganik2006,Bonhard2006}. In addition, accuracy of recommendation can be improved by analyzing social relationships, such as coauthorships in academic literature recommendation \cite{Hwang2010} and friendships and memberships in product recommendation \cite{Symeonidis2011}. Many real systems, such as \emph{Delicious.com} and \emph{Facebook.com}, allow users to recommend objects to their friends. Similarly, users can subscribe to articles from selected bloggers in blogging sites (\emph{Twitter.com} and others) as well as to news alerts from information dissemination systems (\emph{Elesvier.com} and others). In this chapter, we will first present empirical evidences that demonstrate the presence of social influence on information filtering. Then we will introduce two basic ways social filtering are employed in recommendation: by quantifying and utilizing trust relationships between users and by using the opinion ``taste mates'' to select the content to be recommended.

\subsection{Social Influences on Recommendations}
Social influences, also called the word-of-mouth effect in the literatures, are known to be crucial to many sociometric processes, such as decisions making, opinions spreading and the propagation of innovation and fashion \cite{Herr1991,Bone1995,Fortunato2007,Ellero2009}. Scientists have been aware of the commercial values of social influences for a long time \cite{Arndt1967}, yet large-scale applications for commercial purpose only emerged when the Internet era began. Besides, the availability and the great variety of data provide us good opportunities to quantitatively understand social influences \cite{Dellarocas2003}. This section focuses on the social recommendations, whose effects can be roughly divided into two classes: one is on users' prior expectations, leading to the increase of sales; another is on users' posterior evaluations, resulting in the enhancement of the user loyalty.

\begin{figure}
\centering
\figg{0.5}{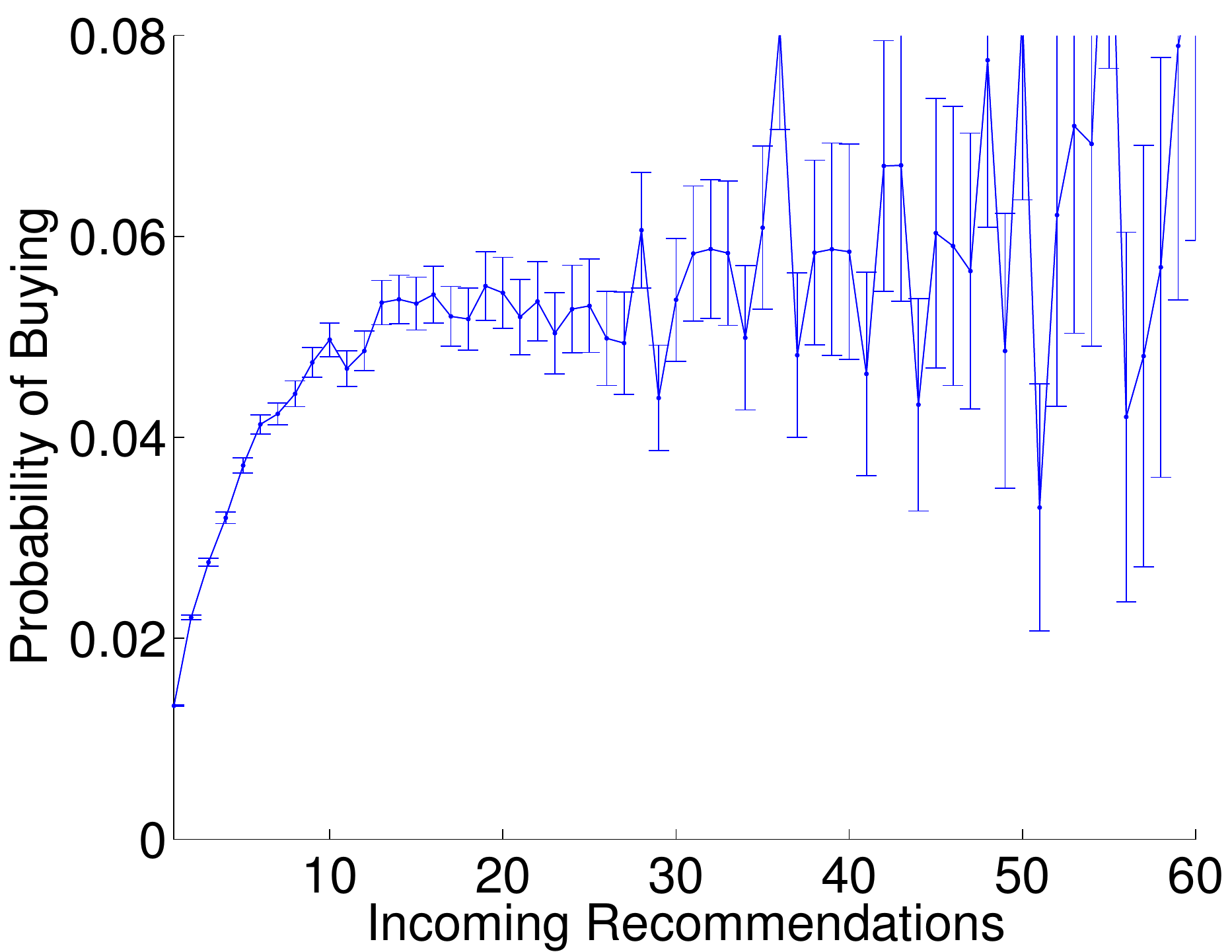}
\caption{Probability of buying a DVD given a number of incoming recommendations (taken from \cite{Leskovec2007}).}
\label{fig:DVD-recommendation}
\end{figure}

Positive effects of social recommendations on prior expectation have already been demonstrated in a number of real examples. They are found in a wide range of systems, including product reviews \cite{Chevalier2006,Yeung2011}, e-mails \cite{Phelps2004}, blogs \cite{Agarwal2008} and microblogs \cite{Jansen2009}. In \cite{Leskovec2007}, the authors studied the effects of social influences on purchase preference: users of an e-commerce system were given the option to recommend an item to their friends through e-mails after purchase. The first person to purchase the same item through a referral link from e-mails got a 10\% discount, and when this happens, the recommender will receive a 10\% credit. As shown in Fig.~\ref{fig:DVD-recommendation}, the purchase probability for a DVD grows remarkably with the increasing number of received recommendations from friends on this DVD. There is a saturation at about 10 recommendations, after which the purchase probability does not increase any more. In other examples, social influences can be much more complicated. In \cite{Leskovec2007}, they reported a similar experiment with book sales where in contrast to the common sense, recommendations had little or even negative effect on the purchase probability. Social influences may also vary across topics and items: \cite{Romero2011} showed that different tags and topics spread on Twitter differently and \cite{Tang2009} found that strength and direction of social influence are topic-dependent. A recent work shows that the mutual interaction between opinions of past viewers and potential future viewers leads to a complex dynamics that agrees qualitatively with movie popularity behavior seen in real systems \cite{Yeung2011b}.

\begin{figure}
\centering
\figg{1}{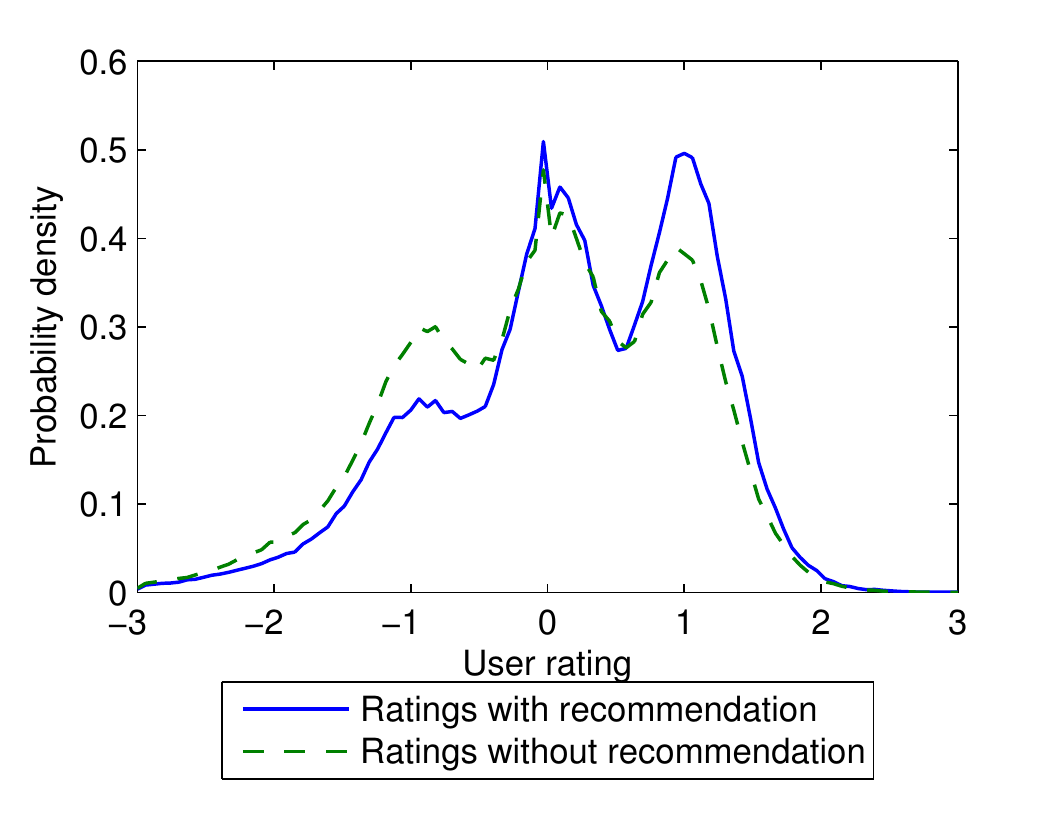}
\caption{The probability distributions of ratings, posted with (solid line) or without (dashed line) a word-of-mouth recommendation in Douban (taken from \cite{Huang2011}).}
\label{fig:Douban}
\end{figure}

In comparison, the issue about how social recommendations affect users' posterior evaluation received less attention. \cite{Huang2011} empirically analyzed two web sites, \emph{Douban.com} and \emph{Goodreads.com}, where millions of users rate books, music and movies, and share their ratings and reviews with friends and followers. On these social network sites, the phenomenon that users recommend favorites to friends and followers plays an important role in shaping users' behaviors and collections. Fig.~\ref{fig:Douban} compares the probability distribution of ratings on items in Douban with and without recommendations (the result is very similar in Goodreads). This demonstrates that an individual is more likely to give a high rating to an item with word-of-mouth recommendations, compared with items without recommendations.

There are also indirect evidences about the positive social influences, for example, in Twitter, statistically a tweet will spread to about $10^3$ users if it gets retweeted \cite{Kwak2010}, and in Taobao\footnote{Taobao.com is a Chinese consumer marketplace that is the world's largest e-commerce website.}, the communication between buyers is a fundamental driving force for purchasing activity \cite{Guo2011}.

Many ingredients could result in positive social influences in online recommendation. Firstly, the word-of-mouth influences and role-model effects from social mates are very strong in offline society, for example, an experiment in Nepal \cite{Oster2009} shows that on average the probability of a woman to use a novel menstrual cup \emph{Take-Up} will increase by 18.6\% if one more of her friends has used Take-Up. Secondly, the friendship network and interest-based network are strongly correlated with each other \cite{Yang2011}, and friends tend to visit the same items and vote them with similar ratings \cite{HeChu2010}. 

\subsection{Trust-Aware Recommender Algorithms}
It is the basic paradigm of recommender systems that when computing recommendations for an individual user, evaluations of the others are not weighed equally and preference is given to those who have similar rating patterns as the given user. This approach neglects an important facet of the evaluation process: it is not only personal tastes but also social relationships and the quality of evaluations that differs from one user to another. To make a better use of social relationships, various recommendation algorithms relying explicitly on trust or user reputation~\cite{SaSi05,JoIsBo07} have been developed and applied by many commercial web sites such as eBay.com~\cite{Dellarocas03}. Trust can be used instead of user similarity~\cite{Golbeck06}, in combination with collaborative filtering to help deal with data sparsity and the cold start problem~\cite{MaBh04,Massa2007,Jamali2009}, or it can help to further filter recommendations by prioritizing those approved by trusted sources (see~\cite{DoSm05} for a review). The use of reputation in recommendation is further supported by the evidence that trust correlates with user similarity~\cite{ZiLa04}, meaning that by introducing trust we are unlikely to conflict with users' interests and preferences. As noted in~\cite{ReZe02}, even an imperfect reputation system may be beneficial as (i) it provides an incentive for good behaviors, (ii) imposes costs on participants to get established, and (iii) swiftly reacts to bad behavior. The use of trust and reputation has also its drawbacks, which includes: (i) time consuming computation, (ii) low incentives for users to provide the required feedback, (iii) privacy concern for data of trust relationship, and (iv) low availability of trust datasets for tests of algorithms. However, without algorithms for trust and reputation, online transactions would be dramatically affected, if not halted.

The difference between reputation and trust is that while the former characterizes general beliefs about a user's trustworthiness, the latter concept is personal and relates to a user's willingness to rely on the actions of a given user. The terms local and global trust metric are sometimes used instead of trust and reputation, respectively. To establish trust or reputation, one may let the involved participants to rate each other and use these evaluations to derive trust or reputation scores~\cite{JoIsBo07}. When mutual user evaluations are given, an initial seed of trusted users can be used to find all the trustworthy users as in the classical Advogato trust metric (see \url{http://www.advogato.org/trust-metric.html}). Other popular trust metrics, such as PageRank~\cite{Brin1998} and Eigentrust~\cite{KaScGM03}, use the spreading activation approach by which nodes are intially loaded and then propagate their load within the network~\cite{Quil68} (diffusion-based recommendation methods are presented in Section~\ref{sec:diffusion} also use this approach). When trust relations between the users are weighted (with trust values $1$ and $0$ representing total trust and distrust, respectively), the plain trust propagation can be shown to be insufficient---the Appleseed algorithm solves this problem by creating virtual trust edges for backward propagation~\cite{ZiLa04b}.

It is a great disadvantage of reputation-aware recommender systems that they usually require substantial input on the the user side to evaluate trust and reputation. Some trust-aware reputation algorithms hence tried not to rely on explicit evaluations of other users. In~\cite{MaHuSi06}, they propose to detect noisy ratings by comparing the actual ratings with the predicted ratings obtained by a recommendation algorithm, with data only from a set of implicitly trusted users. This ``reputation of ratings'' can in turn be used to build reputation of users~\cite{DoSm05}. A different approach was proposed by~\cite{WaBaSch08} where the authors use the information contained in social relationships between the users (which may stem from users' family or friendship relations), the transitivity of trust (as in~\cite{GuKuRaTo04}), and the propagation of users' queries in social networks. When computing recommendations for a specific user $i$, the greatest weight is hence given to the users who can be connected with user $i$ in the social network by a short path with high trust values along its edges. The proposed system is shown to assign correctly trust values and self-organizes to a state producing highly accurate recommendations (when compared to a simple benchmark strategy when one of the recommendations from peers is chosen at random).

\subsection{Adaptive Social Recommendation Models}
While the above described trust-aware systems make use of existing social relationships, adaptive social recommendation models build a network of users based on their evaluations. In \cite{Medo2009}, the authors proposed a model where the recommended items spread over the network similarly as an epidemics \cite{Pastor-Satorras2001,Zhou2006} or rumor \cite{Moreno2004,Castellano2009} spreads in a society. Simultaneously with this spreading, the network of users evolves and adapts to best capture users' similarities. This epidemic-like spreading of a successful item is of particular importance in the case when individual items swiftly loose their relevance \cite{MedoM2011}---as it is the case for news stories, for example---because it combines personalization with the speed of access. It is very different from the currently popular services such as \emph{digg.com} and \emph{reddit.com} which still rely on centralized distribution of items where only those of very general interest can become popular and be accessed by many. For a recent review of approaches to news recommendation see \cite{Billsus2007}.

\begin{figure}
\centering
\figg{1}{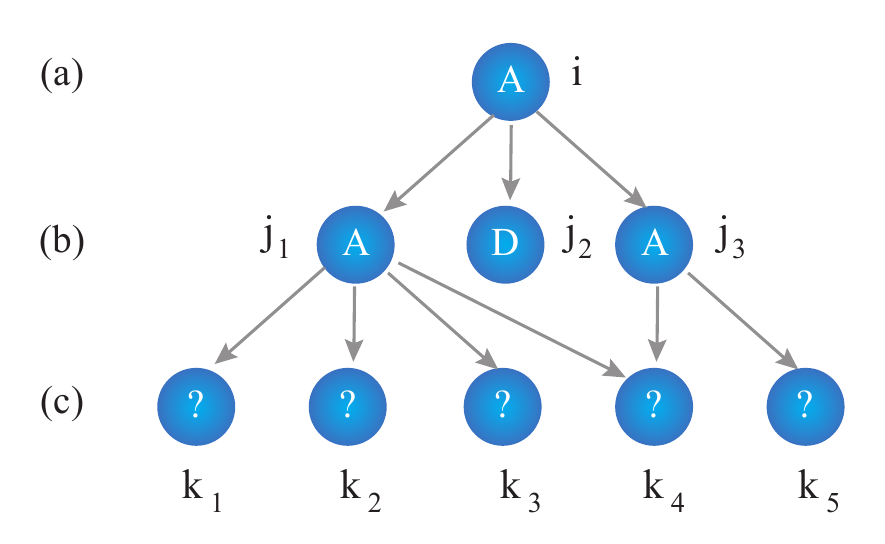}
\caption{Illustration of the news propagation in an adaptive network model \cite{Medo2009}. User $i$ added a~new news, which is automatically considered as approved (A) and sent to users $j_1,j_2,j_3$ who are $i$'s followers. While user $j_2$ dislikes (D) the news, users $j_1$ and $j_3$ approve it and pass it further to their followers $k_1,\dots,k_5$ who have not evaluated the news yet (which is denoted with question marks). User $k_4$ receives the news from the authorities $j_1$ and $j_3$, yielding the news's recommendation score $s_{j_1k_4}+s_{j_3k_4}$. At the same time, user $k_5$ receives the news only from the authority $j_3$ and hence for this user, the recommendation score is only $s_{j_3k_5}$.}
\label{fig:newsbag}
\end{figure}

Here we describe briefly the model introduced in \cite{Medo2009}.
In this model, users either ``approve'' or ``disapprove'' the consumed items. Each user $i$ has $S$ sources (\ie, $S$ other users from whom $i$ receives news) and thus the system can be described by a directed network with a constrained node in-degree $S$. When a news is approved by user $i$, it is added to the recommendation lists to all $i$'s followers (\ie, all users who have $i$ as one of their sources). This spreading process is illustrated in Fig.~\ref{fig:newsbag}. Similarity between users $i$ and $j$ is defined according to the agreement of their past evaluations,
\begin{equation}
s_{ij}=\frac{n_A}{n_A+n_D}\left(1-\frac{1}{\sqrt{n_A+n_D}}\right)
\end{equation}
where $n_A$ and $n_D$ denote the number of news for which evaluations of $i$ and $j$ agree and disagee evaluations, respectively. The term $1/\sqrt{n_A+n_D}$ aims at penalizing user pairs with little overlap of evaluated news as they may seem to be a great match for each other simply because agreeing in evaluations of a few news. To summarize, items at the top of a user's recommendation list are probably recommended by multiple sources of this user or, at least, by a source whose similarity with this user is high.

Apart from using user similarity in the recommendation process (the recommendation score of item $\alpha$ for user $i$ is given by the similarity between $i$ and the sources of $i$ who approved this news), it is also crucial for updating the source-follower network. This updating aims at maximizing the similarity of each user with their sources. As shown in \cite{Medo2009} by agent-based simulations, the system can evolve from a random initial state to a highly organized state where taste mates are connected and news spread effectively. The original network can be improved by rewiring, through ineffective random replacements or computationally demanding global optimization. One can also combine simple greedy optimization with random assignment \cite{Cimini2011}, repeated trials \cite{Wei2011} or exploration of the directed user network both in the direction of followers and sources \cite{Chen2011}. Robustness of this recommendation approach can be enhanced by introducing the user reputation \cite{Cimini2011}. This adaptive evolution of the network of user-user interactions can be used to explain the widely observed scale-free leadership structure in social recommender systems \cite{Zhou2011}, and recent analysis suggested that users could get better information by selecting proper leaders in social sharing websites \cite{Yeung2010}. 
        % checked, done
\section{Meta approaches}
\label{sec:meta}
Input data for recommendation can be extended far behind the traditional user-item-rating scheme. In this section, we briefly review the so-called meta approaches where additional information of various kind (tags assigned to items by users or time stamps of past evaluations) enters the recommendation process or simply several recommendation methods are combined together (either in an iterative self-evaluating way or by forming a hybrid algorithm). As possibilities for extensions are relatively easy to find, this direction has seen high activity over the past years.

\subsection{Tag-aware methods}
\label{sec:tag}
In the past decade, the advent of \emph{Web 2.0} and its affiliated applications brought us a new paradigm to facilitate the user-generated creation on the Internet. One such example are user-driven platforms that allow users to store resources (bookmarks, images, documents, and others) and to associate them with personalized words, so-called \emph{tags}. The resulting ternary user-object-tags structure, so-called \emph{folksonomy}, represents a rich data structure that is of interest for computer scientists, sociologists, linguists, and also physicists~\cite{Zhang2010}. When viewed from the perspective of an individual user, these tags constitute a personalized folksonomy~\cite{Lipczak2008} where tags, although only simple words, contain highly abstracted yet personalized information. Different from other kinds of metadata (such as profile, attributes, and content), tags are not predefined by domain experts or administrators. This approach has the advantage of being scalable and requiring no specific skills, hence allowing every individual to participate and contribute. Despite the lack of imposed organization, shared vocabularies were shown to emerge in folksonomies~\cite{Halpin07}, making them increasingly accessibile to advanced information filtering techniques. Tags therefore represent a simple yet promising tool to provide reasonable recommendations and solve some outstanding problems in recommender systems, \eg the cold-start problem~\cite{ZhaZK20102}. The social impact~\cite{Turner1991} and dynamical properties of folksonomies~\cite{Cattuto20071,LiuC2011} are expected to be applied to obtain trustworthy and real-time recommendations in tagging systems. In addition, the hypergraph~\cite{Zhang2010} theory is considered to fully utilize the complete network structure of tagging platforms without using any hybrid methods and losing any information, which gives promises for generally more reliable recommendations. At the same time, there are also drawbacks caused by the freedom of tags: \eg polysemy, synonymy, and ambiguity~\cite{WuH2006}. To alleviate these problems, advanced methods such as tag hierarchical clustering \cite{Shepitsen2008}, introduction of ontologies \cite{Mika2007} and recommendation of tags~\cite{SongY2011} were proposed. 

\begin{figure}
\centering
\figg{0.58}{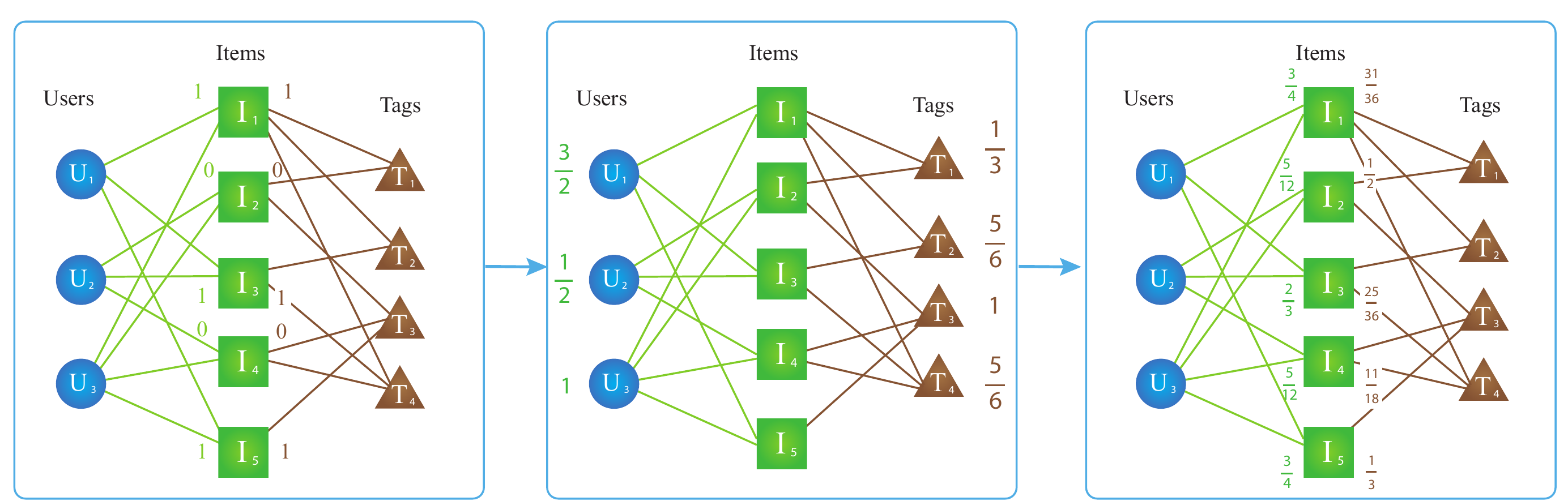}
\caption{(Color online) Illustration of a user-item-tag tripartite graph consists of three users, five items and four tags, as well as the recommendation process described in \cite{ZhaZK20101}. The tripartite graph is decomposed to user-item (black links) and item-tag (red links) bipartite graphs connected by items. This is how the scoring process works for a given target user $U_1$. (a) Firstly, highlight the items, $I_1$, $I_3$, $I_5$, collected by the target user $U_1$ and mark them with unit resource. In the depicted case we have: $f_{I_1}=f_{I_3}=f_{I_5}=1$, and $f_{I_2}=f_{I_4}=0$. (b) Secondly, distribute the resources from items to their corresponding users and tags, respectively: $f_{U_3}=f_{I_1}\times\tfrac{1}{2}+
f_{I_2}\times\tfrac{1}{2}+f_{I_5}\times\tfrac{1}{2}=1\times\tfrac{1}{2}+0+1\times\tfrac{1}{2}=1$ and $f_{T_4}=f_{I_1}\times\tfrac{1}{3}+f_{I_3}\times\tfrac{1}{2}+f_{I_4}\times\tfrac{1}{2}=
1\times\tfrac{1}{3}+1\times\tfrac{1}{2}+0=\tfrac{5}{6}$; (c) Finally, redistribute the resources from users and tags to their neighboring items: $f^p_{I_4}=f_{U_2}\times\tfrac{1}{3}+
f_{U_3}\times\tfrac{1}{4}=\tfrac{1}{2}\times\tfrac{1}{3}+1\times\tfrac{1}{4}=\tfrac{5}{12}$ and $f^{pt}_{I_4}=f_{T_3}\times\tfrac{1}{3}+f_{T_4}\times\tfrac{1}{3}=1\times\tfrac{1}{3}+
\tfrac{5}{6}\times\tfrac{1}{3}=\tfrac{11}{18}$.}
\label{fig:tag}
\end{figure}

Unlike being used as a traditional filter, researches tend to apply more sophisticated theories and methods (\eg, social impact) in designing tag-aware recommendation algorithms. The \emph{FolkRank}~\cite{Hotho2006}, a modified PageRank algorithm, was proposed to rank tags in folksonomies by assuming that important tags are given by important users. Due to the success of collaborative filtering, many works were devoted to using tags to measure similarity among users or objects, and then fuse with the standard memory-based collaborative filtering framework \cite{Xu2006, Tso2008, Shang20101}. In \cite{ZhaZK2011}, the present tag-aware algorithms are classified as follows:
\begin{enumerate}
\item Topic-based models. They implement probability-based methods such as pLSA and LDA (see Sections \ref{sec:pLSA} and \ref{sec:LDA}) to extract latent topics from the available tags in the user or object space and then produce recommendations using classical probability-based models \cite{SaidA2009, SiX2009, KrestelR2009}.
\item Network-based models. They implement graph theory-based methods such as ProbS (see Sec.~\ref{sec:spreading_algo}) to represent tags as nodes in a tripartite user-object-tag network and apply a diffusion process to generate recommendations~\cite{ZhaZK20101} (see Fig. \ref{fig:tag}).
\item Tensor-based models. They implement tensor factorization \cite{Kolda2009} to reduce the ternary relation into low-rank feature matrices, alleviate the sparsity problem in large-scale datasets and ultimately provide personalized recommendations~\cite{Rendle2009,Symeonidis2009}.
\end{enumerate}

\subsection{Time-aware methods}
\label{sec:time}

\begin{figure}
\centering
\figg{0.9}{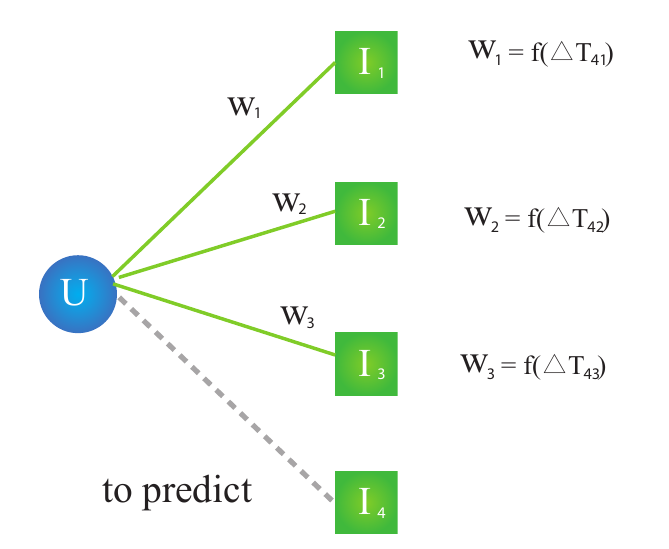}
\caption{(Color online) Illustration of time based recommendation where the target user has collected three items, $I_1$, $I_2$ and $I_3$, which are then used to predict the newest item $I_4$. $f$ is the decay function to weight the time difference between $I_4$ and other items that were collected before. }
\label{fig:time}
\end{figure}

\begin{figure}
\centering
\figg{0.4}{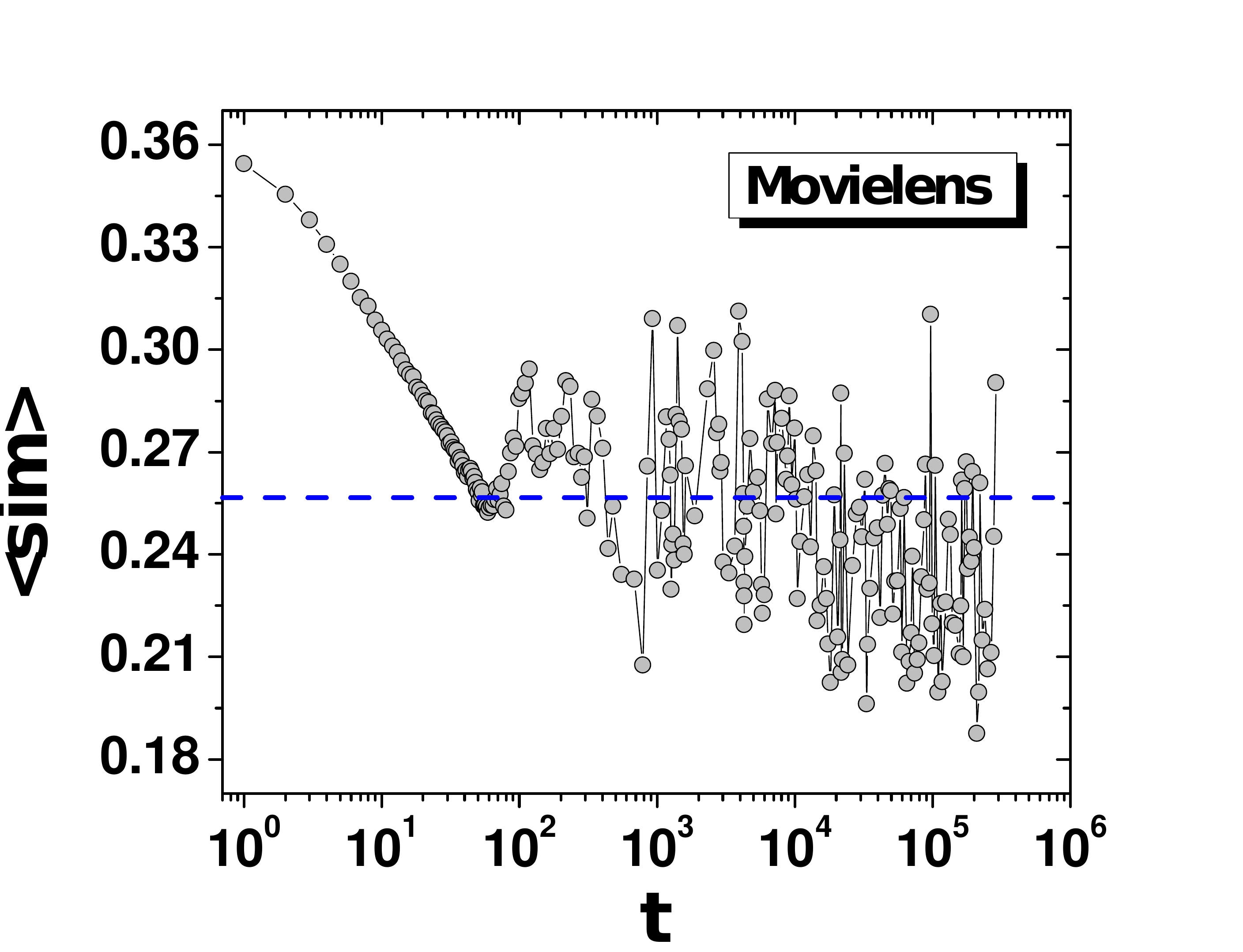}
\caption{(Color online) Illustration of users' interests changing with time in the \emph{MovieLens} data. Shifts of user interests are represented by the average similarity among item pairs of the target user within an observed time $t$ (shown with gray circles). The dashed line represents the average similarity of all users.}
\label{fig:interest-decay}
\end{figure}

Nowadays, huge quantities of information emerge every second. We receive news from various media, such as newspapers, TV programs, websites, etc. Due to its timeliness and in particular its convenience, more and more people prefer to read news online (\eg, using RSS feeds) instead of from traditional media like newspapers. However, given the enormous amount of online news, one challenging issue is the novelty of news recommendation, \ie, how to appropriately and efficiently recommend news to readers, matching their reading preferences as much as possible. The analysis of data from a popular platform for sharing news stories, Digg.com, by Wu and Huberman \cite{WuF2007} shows that the novelty of news half there decays in a very short period. Another typical instance is the communication system of e-commerce websites, which require real-time feedback among various agents \cite{ChouPH2010}. Such information updates so frequently that it is impossible for individuals to evaluate each, or even read them in time. Consequently, an urgent question emerges: how to automatically filter out the irrelevant information, receive timely news and perform immediate yet appropriate response? One promising solution lies in time-aware recommender systems, which can hopefully help to address aforementioned issues. Collaborative filtering (see Sec. \ref{sec:sim-based}), as the most widely adopted method in recommender systems, is the first one to be considered. Following the classical collaborative filtering framework, most of the related work focuses on designing time factors to suppress old evaluations or objects. Generally, such kind of weight is expressed by various decay functions (see Fig. \ref{fig:time}), which suggests that user interests in a single topic would decay with time (see Fig. \ref{fig:interest-decay}). Ding and Li \cite{DingY2005} weighed different items by putting smaller weights on older ones. Similar methods used the time factor to adaptively choose temporal neighborhoods and then obtain recommendations via the refined neighbors \cite{LathiaN2009, CamposPG2010, WuP2010}. Another kind of attempts consider decaying time to weigh user-item binary edges in bipartite networks. Liu and Deng \cite{LiuJ2009} hypothesized the time effect decayed in a exponential manner, which could also been found in other empirical studies \cite{WuF2007} and models \cite{KorenY2009}. A broad picture of collaborative filtering with time can be found in a recent Ph.D. thesis \cite{LathiaNK2010}.

Another important issue in recommender systems is that of \emph{novelty}. Although an item with the highest recommendation score is the most possible candidate for a target user, it may fail to be picked due to the diversity of human tastes. In such cases, these items should not always occupy the recommendation list and be recommended over and over again. Therefore, temporal diversity \cite{LathiaN2010} becomes crucial in designing time-based algorithms. Xiang \emph{et al.} divided user interests into two general categories: longer-term and short-term \cite{XiangL2010}. Longer-term interests govern the essential preferences of users and would not easily change over time. By contrast, short-term interests are more likely to be effected by social environment (Fig. \ref{fig:interest-decay} shows such difference). By identifying and making use of the differences between them using a time factor, in \cite{XiangL2010} they successfully provided more reliable yet interesting recommendations, which can also be found in relative works \cite{LiuNN2010}.

Besides recommendation, time also plays an important role in various fields, such as network growth \cite{PastorR2001, MedoM2011}, identification of original scientific contributions \cite{NewmanM20091, GualdiS20111}, selecting the backbone structure of citation networks \cite{GualdiS20112}, and aging effects in synchronization \cite{HwangDU2005}. However, how to appropriately use the time information to discover the underlying dynamics of user preferences and help us master the current information era still remains an important research challenge.

\subsection{Iterative refinement}
Iteratively solved self-consistent equations are widely applied in characterizing structural and functional features of nodes and/or edges in networked systems. Their solution is usually used to describe a stable distribution of a certain quantity in a system consisting of many interacting individuals, where the amount of this quantity assigned to individual $i$ is affected by both the interacting rules and the amounts of this quantity assigned to other individuals interacting with $i$. In a directed network, the significance of a node is not only determined by its attributes (if applicable), but also contributed by the significance of its downstream nodes. For example, the classical set of self-consistent equations for PageRank value $G(i)$ of the web page $i$ has the form \cite{Brin1998}
\begin{equation}
G(i)=c+(1-c)\sum_{j\in \Gamma_i}\frac{G(j)}{k_j^{out}},
\end{equation}
where $j$ runs over all the web pages that point to $i$, $c$ is the return probability accounting for the random browsing, and $k_j^{out}$ is $j$'s out-degree. This set of equations represents a particular Markovian process on a network and can be successfully solved using an iterative approach thanks to the fast convergence to a stationary solution observed in most cases \cite{Franceschet}. Besides web pages, similar self-consistent equations are also successfully applied in ranking people \cite{Lu2011}, genes \cite{Zhao2011}, and so on. If instead of referring to nodes, the individuals refer to pairs of nodes, this approach can be used to quantify node similarity \cite{Leicht2006,Sun2009}. In more complicated situations, the individuals can be users and items, or scientists and publications, and similar iterative equations embodied in bipartite networks can be applied in building quantitative reputation systems, namely simultaneously estimate people's reputation and objects' quality \cite{Yu2006,Laureti2006,Jiang2010,ZhouYB2010,ZhouYB2011,MedoWakeling2010}.

Besides the above-mentioned iterative equations, a closely related framework is that of the so-called \emph{self-consistent refinement} \cite{Maslov2001,Ren2008}. In the link prediction and personalized recommendation, the known information is the adjacency matrix representing a unipartite or a bipartite network and the task is to estimate the likelihoods of link existence for currently zero elements of the adjacency matrix. For recommender systems with ratings, the algorithms need to predict unknown ratings according to the rating matrix. Denoting $\mR$ the known matrix and $\tilde\mR$ the predicted matrix (\ie, output), the procedure of many algorithms can be written in a generic form
\begin{equation}
\tilde\mR=\mathfrak{D}(\mR), \label{non-iterative}
\end{equation}
where $\mathfrak{D}$ is a matrix operator.\footnote{See \cite{Ren2008} on how to use this generic form to represent the well-known similarity-based and spectrum-based algorithms for rating prediction.}
Denoting the initial configuration as $\mR^{(0)}$ and the initial time step $k=0$, a generic framework of self-consistent refinement reads \cite{Ren2008}: (i) Implement the operation $\mathfrak{D}(\mR^{(k)})$; (ii) Set the elements of $\mR^{(k+1)}$ as
\begin{equation}
R_{i\alpha}^{(k+1)}=\left\{
  \begin{array}{lr}
  \mathfrak{D}(\mR^{(k)})_{i\alpha} & \text{when }R_{i\alpha}^{(0)}=0, \\
  R_{i\alpha}                       & \text{when }R_{i\alpha}^{(0)}\neq 0.
  \end{array}
\right.
\end{equation}
Then, set $k=k+1$. (iii) Repeat (i) and (ii) until the difference between $\mR^{(k)}$ and $\mR^{(k-1)}$ is smaller than a given terminating threshold.

Considering the matrix series $\mR^{(0)},\mR^{(1)},\cdots,\mR^{(T)}$ ($T$ denotes the last time step) as a certain dynamics driven by the operator $\mathfrak{D}$, all the elements corresponding to the known items (\ie, $R_{i\alpha}^{(0)}\neq 0$) can be treated as the \emph{boundary conditions} expressing to the known and fixed information.\footnote{This is the essential difference between the self-consistent refinement and the above-mentioned iterative equations like PageRank, since in the latter case, every matrix element is free to be changed.}
If $\tilde{\mR}$ is an ideal prediction, it should satisfy the self-consistent condition $\tilde{\mR}= \mathfrak{D}(\tilde{\mR})$. However, this equation is not hold for most known algorithms. In contrast, the convergent matrix $\mR^{(T)}$ is self-consistent.

\begin{figure}
\centering
\figg{0.5}{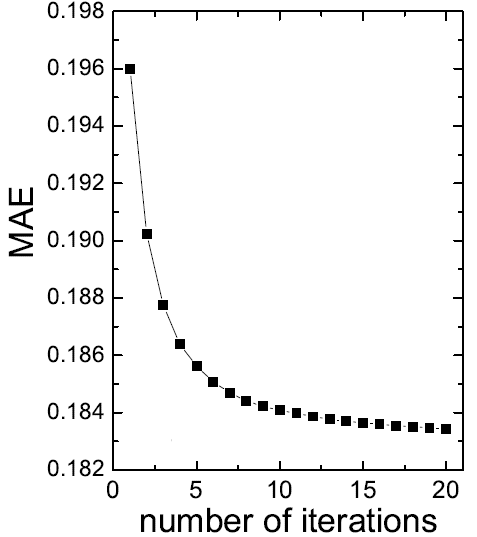}
\caption{Prediction error (MAE) versus iterative step for MovieLens data. We use
the \emph{MovieLens} data that consists of $N=3020$ users, $M=1809$ movies, and $2.24\times 10^5$
discrete ratings from 1 to 5. All the ratings are sorted according to their
time stamps with 90\% earlier ratings as the training set, and the remaining ratings (with later time stamps) as the testing set.}
\label{fig:SCR}
\end{figure}

As shown in \cite{Ren2008}, applying the self-consistent framework can lead to great improvements compared with non-iterative methods employing Eq.~(\ref{non-iterative}). We next show a simple example for similarity-based recommendation. Taking into account the different evaluation scales of different users \cite{Blattner2007}, we subtract the corresponding user average from each evaluated entry in the rating matrix $R$ and get a new matrix $R'$. The
predicted rating is calculated by using a weighted average, as:
\begin{equation}
\tilde{R}_{i\alpha}=\frac{\sum_{\beta}\Omega_{\alpha\beta}\cdot
R'_{i\beta}}{\sum_{\beta}|\Omega_{\alpha\beta}|},
\end{equation}
where the similarity between items $\alpha$ and $\beta$ is defined by Eq. (34). As shown in Fig.~\ref{fig:SCR}, experiment on MovieLens verifies the advantages of the iterative refinement, where the original similarity-based algorithm corresponds to the first iteration step.

\subsection{Hybrid algorithms}
\label{sec:hybrid}
Even for a good recommender algorithms it is difficult to address diverse needs of its heterogenous users. Hybrid methods overcome this problem by aptly combining recommendations obtained by different methods~\cite{Burke2002,Burke2007}. Hybridization is hence often used in practical implementations of recommender systems, even in very early ones~\cite{BaSho1997}. One of the most important applications of hybrid recommendation algorithms is to solve the cold-start problem, by combining collaborative and content data in such a way that even a new object that has never been rated before can be recommended~\cite{SPUP2002} (similarly, a new user who has never rated anything can receive some recommendations). Since hybrid algorithms combine different approaches to recommendation, they also have the potential to improve the diversity of recommendations~\cite{Zhou_PNAS}. The following classification of hybridization methods is taken from~\cite{AdoTuz2005}:
\begin{enumerate}
\item implement collaborative and content-based methods separately and combine their predictions,
\item incorporate some content-based characteristics into a collaborative approach,
\item incorporate some collaborative characteristics into a content-based approach, and
\item construct a general unifying model that incorporates both content-based and collaborative characteristics.
\end{enumerate}
Now we can extend class 4 to include also unifying models that incorporate two or more collaborative methods (see Sec.~\ref{sec:hybrid_spreading}).

Hybrid methods range from simple, such as using a linear combination of ratings obtained by different methods~\cite{Claypool1999}, to very complex, such as employing Markov chain Monte Carlo methods~\cite{CaLo1996} to model combined collaborative and content data. Recent Netflix prize (see Sec.~\ref{sec:netflix_prize}) has provoked interest in sophisticated methods for combined predictions (also called \emph{ensemble learning} or \emph{blending}) which were shown to be very successful in lowering the error of predictions~\cite{JaToLe2010}. The main idea of blending is that the prediction vectors of $F$ distinct recommendation methods, denoted as $\boldsymbol{x}_1,\dots,\boldsymbol{x}_F$, are combined by a function $\Omega: \mathbb{R}^F\to\mathbb{R}$ so that the prediction error on a test set (evaluated by RMSE, for example) is minimized. The optimal weighting function $\Omega$ is obtained by linear regression, neural networks, or bagging predictors~\cite{Breiman1996,Witten2011}. For details and evaluation of different blending schemes on the extensive dataset provided by Netflix for the competition see~\cite{JaToLe2010}. Some challenges can also be partially solved by hybridization, for example, the link prediction algorithm can be used to generate artifical links that could eventually improve the recommendation in very sparse data \cite{LPSurvey,Esslimani2011}.
                    % checked, done
\section{Performance evaluation}
\label{sec:performance}
In this section we briefly compare performance of individual algorithms presented in this review. To test the algorithms, we use two standard data sets: Movielens 1M\footnote{This data set can be obtained at \url{http://www.grouplens.org/node/73}.} and Netflix\footnote{This data set can be obtained at \url{www.netflixprize.com}.}. 
The Movielens data set, which contains ratings from 6,040 users to 3706 movies (corresponding to the rating density of $4.5\cdot 10^{-2}$), is used in its original form. Our subset of the original Netflix data set was created by randomly selecting 8,000 users from the original data set released by Netflix for the Netflix Prize and keeping all their evaluations (see \ref{sec:netflix_prize} for details on the competition). In this way, a data set with 17,148 objects (DVDs rented by the company to its users) and 1,632,172 ratings in the integer scale from one to five (corresponding to the rating density of $1.2\cdot 10^{-2}$) was created. Both data sets use the integer rating scale from one to five. For methods requiring no explicit ratings (binary data), we assume that all ratings greater or equal than three represent objects \emph{liked} by the users and hence constitute a corresponding user-object link (if the rating is less than three, no link is formed). As a result, there are 1,387,039 and 836,478 links in the Netflix and Movielens data set, respectively. Table~\ref{tab:datasets} summarizes basic properties of the data sets.

\begin{figure}
\centering
\figg{0.3}{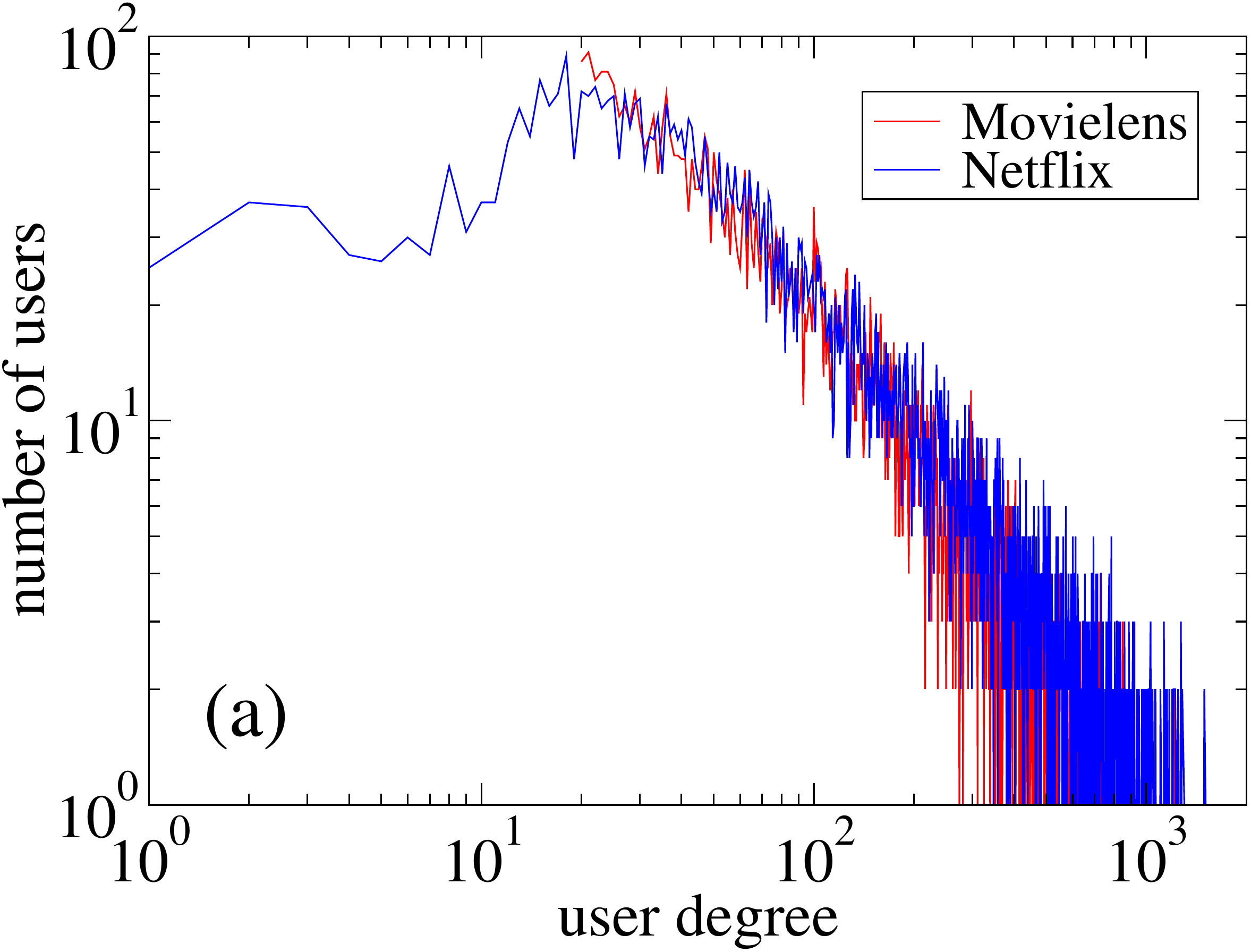}\qquad\figg{0.3}{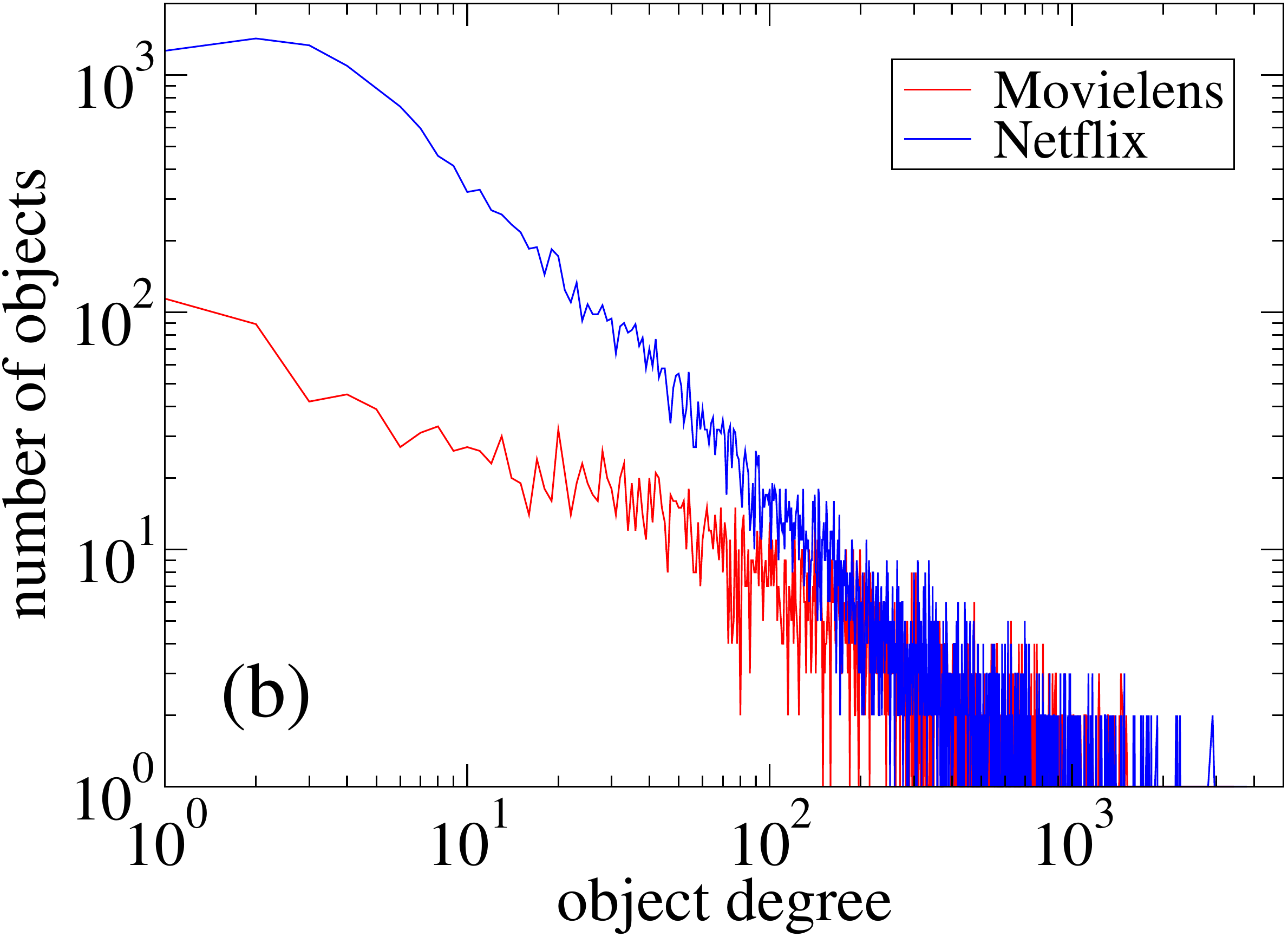}
\caption{User and object degree distribution for the two data sets used to evaluate recommendation methods.}
\label{fig:histograms}
\end{figure}

Our two data sets differ not only by their density and user/object ratio---histograms of user and object degrees in Fig.~\ref{fig:histograms} reveal further differences. Firstly, Movielens data were originally prepared in such a way that all users rated at least twenty movies. This constraint has not been applied to the Netflix data set, resulting in a considerable number of users with only little data on their past preferences available. Secondly, Netflix data also contains a large portion of movies that have been rated only a few times (this probably reflects the fact that the company rents a wide variety of DVDs, many of which are of interest for only a small part of the customers). Unsurprisingly, all degree distributions are broad and right-skewed, as similar to many other social systems~\cite{Newman05}.

\begin{table}
\centering
\begin{tabular}{rrrrrrr}
\toprule
 data set & users & objects &   ratings & density & $k_U$ & $k_O$\\
\midrule
Movielens & 6,040 &   3,706 & 1,000,209 & $4.5\%$ &   166 &   270\\
  Netflix & 8,000 &  17,148 & 1,632,172 & $1.2\%$ &   204 &    95\\
\bottomrule
\end{tabular}
\caption{Data sets used for evaluation of recommendation methods. Data density is defined as the ratio of the available ratings to the maximum possible number of ratings; $k_U$ and $k_O$ denote the average user and object degree, respectively.}
\label{tab:datasets}
\end{table}

To test a recommendation method, we employ the standard approach. First, a randomly selected small part of the input data is moved into a so-called probe. In our case, the probe contains $10\%$ of ratings present in the input data set. The remaining $90\%$ of the data is then given to the recommendation method and are used to estimate the missing ratings. The estimated missing ratings are then compared with the true ratings present in the probe set. This comparison is done by means of root mean square error (RMSE) and mean absolute error (MAE) in the case of data with explicit ratings and by means of precision, recall and the average relative rank in the case of data without explicit ratings (see Section~\ref{sec:metrics} for a detailed description of these performance metrics). For precision and recall, we take into account top 100 places of each user's recommendation list. To eliminate possible effects of the probe selection, we repeat the procedure for ten independent randomly selected probe sets and present the averaged results.

\begin{table}
\centering
\begin{tabular}{rrrrr}
\toprule
                     & \multicolumn{2}{c}{Movielens} & \multicolumn{2}{c}{Netflix}\\
\cmidrule(lr){2-3} \cmidrule(l){4-5}
              method & RMSE & MAE  & RMSE & MAE\\
\midrule
     overall average & 1.12 & 0.93 & 1.09 & 0.92\\
      object average & 0.98 & 0.78 & 1.02 & 0.82\\
        user average & 1.04 & 0.83 & 1.00 & 0.80\\
multilevel spreading & 0.94 & 0.75 & 0.97 & 0.77\\
     user similarity & 0.91 & 0.72 & 0.93 & 0.72\\
   object similarity & 0.89 & 0.70 & 0.96 & 0.74\\
                 SVD & 0.85 & 0.67 & 0.87 & 0.68\\
           slope one & 0.91 & 0.71 & 0.93 & 0.73\\
  slope one weighted & 0.90 & 0.71 & 0.93 & 0.73\\
\bottomrule
\end{tabular}
\caption{Performance of algorithms for data with explicit ratings (averaged over 10 realizations).}
\label{tab:perf_with_ratings}
\end{table}

Method \emph{overall average} is used only as a benchmark; it uses the average rating in the input data as estimate for every user-object pair. For similarity-based methods, we employ the Pearson correlation coefficient which slightly outperforms cosine similarity in terms of RMSE and MAE. For \emph{SVD}, we used parameter values $D=20$, $\eta=0.001$ and $\lambda=0.1$ that result in favorable performance (see Section~{sec:SVD} for the meaning of these parameters). Note that a better founded approach would be to learn ``optimal'' values of these parameters from the data itself. This can be achieved by choosing $\{D,\eta,\lambda\}$ based on RMSE or MAE computed for a small test set of predictions (which could again be created by taking 10\% of the input data) and only then reporting the resulting method's performance computed for the probe. However, with only three free parameters to tune, the results are likely to differ little from the results obtained by our naive approach where $\{D,\eta,\lambda\}$ are chosen directly from performance observed for the probe.

While numerical performance values may seem very close to each other across all the methods (perhaps with the exception of \emph{overall average}), differences between the methods from the user's point of view are much greater than one would expect from RMSE varying at the second decimal place. For example, \emph{user average} outperforms \emph{object average} for the Netflix data set, yet in fact it has zero filtering ability. For a given user, estimated rating of all unrated objects is the same (equal to this user's average rating), this user is thus provided no useful information as to which object to select in the future. Further, the performance of \emph{object average} may seem appealing with respect to the method's simplicity, yet one can easily check that objects with the highest estimated score are likely those that received only a few ratings. This is because while a rarely viewed object may easily receive the highest possible average rating of five, a popular object inevitable receives some worse marks which result in a lower average rating. For example, top-rated movies in both tested data sets have all received less than five ratings and scored 5.0 on average. This analysis shows that RMSE and MAE, while useful and easy to understand, give only a very limited information about a method's performance.

\begin{table}
\centering
\begin{tabular}{rrrrrrr}
\toprule
                             & \multicolumn{3}{c}{Movielens} & \multicolumn{3}{c}{Netflix}\\
\cmidrule(lr){2-4}\cmidrule(l){5-7}\\
                      method & $P_{100}$ & $R_{100}$ & rank & $P_{100}$ & $R_{100}$ & rank\\
\midrule
                 global rank & 0.039 & 0.311 & 0.143 & 0.041 & 0.272 & 0.066\\
         Bayesian clustering & 0.028 & 0.276 & 0.137 & 0.045 & 0.262 & 0.069\\
                        pLSA & 0.071 & 0.575 & 0.090 & 0.071 & 0.456 & 0.050\\
                         LDA & 0.081 & 0.543 & 0.093 & 0.081 & 0.439 & 0.048\\
                       ProbS & 0.052 & 0.422 & 0.112 & 0.053 & 0.359 & 0.051\\
                       HeatS & 0.039 & 0.336 & 0.116 & 0.001 & 0.020 & 0.099\\
ProbS$+$HeatS, $\lambda=0.2$ & 0.067 & 0.548 & 0.080 & 0.062 & 0.421 & 0.046\\
\bottomrule
\end{tabular}
\caption{Performance of algorithms for data without explicit ratings (averaged over 10 realizations).}
\label{tab:perf_binary}
\end{table}

Table~\ref{tab:perf_binary} summarizes performance of methods requiring data without explicit ratings (binary data). As performance metrics we use precision $P$, recall $R$ and the relative rank (marked as rank in the table) of probe objects (if probe object $\alpha$ belonging to user $i$ appears on place $x$ of this user's recommendation list and this user has collected $k_i$ objects, the relative rank of $\alpha$ is $x/(M-k_i)$). Method \emph{global rank} corresponds to recommendation of the most popular items that have not yet been collected by a given user. The results of pLSA and LDA are obtained from $K=50$, while a slight increase in performance is observed when $K$ increases further. Results of Bayesian clustering are obtained with $(K_{\rm user}, K_{\rm item})=(70,35)$ for Movielens and $(K_{\rm user}, K_{\rm item})=(70,140)$ for Netflix, where $K_{\rm user}/K_{\rm items}$ is in a rough proportion to the corresponding ratio of users to items. Note that \emph{global rank} and Bayesian clustering are able to yield low relative rank but they fail to score in precision and recall.
           % checked, done
\section{Outlook}
\label{sec:outlook}
After reviewing extensively the past work in the field of recommender systems, we describe here a few challenges that the field of recommendation needs to tackle in the future.

To begin with, let us take the conceptual question of the possibility of effective recommendation. For a long time, we all thought that we know ourselves better than anyone else does. Without much fanfare and fuss about philosophical implications, IT experts and online businesses continue in their exploration of the part of our knowledge that resides not on in our minds, but at crossroads of communities. This in effect has violated both our self-knowledge belief and an implicit, long cherished notion: individuals are masters of themselves. Conceptually, this admitting is nothing short of a revolution. The potential of the new approach is huge: the ``extra-body'' knowledge about our preferences manifests itself in communication data and is thus much easier to analyze and decipher than the part hiding among our neurons and synapses. What Amazon or Netflix does is just scratching on the surface of the huge potential, as they only take into account a tiny fraction of information about us. Google's Gmail gains more insight about what its users do in the hope that emails reveal more than what can be obtained from data about searching, book-buying, and movie-rating, thereby matching ads more closely to our preferences and hence increasing their efficiency. Some recent online communities go even a step further than Gmail. For example, Facebook.com lets its members to create trusted relationships and keeps track of members' activities and conversations, obtaining the opportunity to infer intimate details about users' preferences. Though most of this information is implicit and not yet ready for recommendation, the huge data basis in principle can yield much more insight than hitherto seen. By letting the users to reveal more and more, the potential for inferring their future wants grows and we are still to know what the consequences will be. The induced danger of privacy violation calls for new, privacy preserving recommender systems where no sensitive data leave users' computers, yet users can enjoy the benefit of collaborative filtering.

IMDB (Internet Movie Database) and similar web sites aggregate millions of votes for a wide range of movies, and online sellers of movies use some degree of collaborative filtering to make recommendations given one's past purchasing history. However, an open reputation-sharing mechanism remains to become widespread. One can project forward to imagine innovative applications, such as ``Movies Wanted": a system where plot descriptions are collaboratively developed and voted on, to highlight movies desired by a constituency. The net effect of reputation filtering will be to bring more old, foreign, and niche movies to light, with similar effects for music and other culture. Cultural opportunities that languish for want of attention due to high search costs will reach audiences that did not know what they were missing. Many recommender systems provide suggestions based on expressed or observed preferences. But reputations could also encode other properties of media, such as ``ethicalness'' of lyrics (and indeed of the performers' lives and aims if one desires), or specific legal or reproduction rights. Licensing schemes like Creative Commons certify an artistic work as having particular legal properties; it is then feasible to provide both recommendations and direct access just within the set of freely available music.

Beyond music and movies, numerous cultural areas and experience goods are ripe for recommendation services provided by reputations. Book ratings and suggestions provide a navigation tool through humanity's ever-growing literary output---most notably from Amazon, but also from a variety of small-scale services and personal lists. Travel guidebooks aid in getting the insider view of an unfamiliar locale, but interpreted experiences of natives and previous travelers could be even better. Whether for festivals, museums, opera, or the thousands of other shared activities which enrich our social landscape, the cultural sector is fertile ground for development.

In the age of Internet and World Wide Web, ICT tools allows information to spread faster and wider than ever before, and dominates the way we form our opinions and knowledge. While there has been an undeniable progress in the information availability, a fundamental question remains elusive: do we get more diverse information than in the past? Although the ICT Revolution was expected to allow people to access ever more diversified information sources and products, one often sees that this is not the case. Popular viral videos copy the strategy of blockbuster films and target the tastes of the general audience, giving rise to global hits. On the Internet, a few sites attire a huge part of web traffic. A similar process is in action in science where disproportional attention is given to a small fraction of all new and exciting works \cite{NewmanM20091}. The problem is that search engines and recommender systems fall prey to a self-reinforcing rich-get-richer phenomenon: items that were popular in the past tend to be served to even more users in the future. The natural outcomes of such defective dynamics are the narrowing of people tastes and opinions together with a general cultural flattening. To address this issue, we need to consider the long-term impacts of information filtering systems on the information ecology and study information filtering tools that favor diversity without sacrificing their overall performance \cite{Zhou_PNAS}.

Another interesting facet of the mentioned diversity challenge is related to a concept of ``crowd-avoidance''. There are situations where the generated data naturally fits the paradigm of recommender systems, yet using a standard recommender system may result in poor outcomes. For example, when given data of user preferences for restaurants, it is natural to recommend a user a new place to eat. However, if too many users are recommended the same place, it gets crowded and nobody enjoys their meal. Similarly, when given data of industrial sectors active in various countries (which can be effectively represented by a bipartite network \cite{Pietronero2011}, so much discussed and utilized in this review), one may recommend a country a new sector on the basis of its similarity with already active sectors. However, if the country faces a strong competition from its neighbors, it may do better by choosing a less similar sector where the competition is weaker. The same happens on a smaller scale where companies routinely compete with products of other companies, yet avoiding a direct clash may be very beneficial. The concept of crowd avoidance in recommender systems could yield benefits in situations similar to those described above, where resources cannot be shared by an arbitrary number of parties due to constraints of geographic space or limited interest of customers.

The crowd avoidance phenomenon ranges from practically non-existing (in the case of e-book distribution, for example) to very strong (no two customers can share an item) where it approaches the classical assignment problem \cite{Burkard09}. The most challenging seems to be the moderate case where recommending an item to a small number of consumers does not create a problem yet (which fits well the above-described product and restaurant recommendation, for example). Note that this whole problem is in principle similar to quantum physics systems where occupation numbers are confined by constraints (such as the Pauli exclusion principle which says that two identical fermions cannot simultaneously occupy the same quantum state) or where mutual repulsion among particles sharing the same site exists. Analogies with physics can thus prove useful in studying this kind of systems.

The science of recommendation is just starting---despite impressive progresses, much remains to be understood. For further advances intuition alone is no longer enough and a multidisciplinary approach will surely bring powerful tools that may help innovative matchmakers to turn the immense potential of recommendations into real life applications.
                          % checked, done

\section*{Acknowledgments}
This work was partially supported by the EU FET-Open Grant 231200 (project
QLectives) and National Natural Science Foundation of China (Grant Nos. 11075031, 11105024, 61103109 and 60973069). CHY is partially supported by EU FET FP7 project STAMINA (FP7-265496). TZ is supported by the Research Funds for the Central Universities (UESTC).

\end{document}